\documentclass[preprint]{aastex}

\usepackage{amsmath}
\usepackage{wasysym}
\usepackage{graphicx}
\usepackage{amssymb}
\usepackage{nicefrac}

\newcommand{\myemail}{robyn@mit.edu}
\newcommand{\Nsmooth}{\ensuremath{N_s}}
\newcommand{\Nres}{\ensuremath{N_{\mathrm{p}}}}

\newcommand{\half}{\ensuremath{\frac{1}{2}}}

\newcommand{\bias}{\ensuremath{b}}
\newcommand{\gt}{\ensuremath{\Gamma_{\mathrm{true}}}}
\newcommand{\eg}{\ensuremath{E(\hat{\Gamma})}}
\newcommand{\gpbc}{\ensuremath{\hat{\Gamma}_{\mathrm{corr}}}}
\newcommand{\nmd}{\ensuremath{n_{\mathrm{max}}}}
\newcommand{\ndl}{\ensuremath{\hat{n}_{\mathrm{ul}}}}

\newcommand{\nperi}{\ensuremath{N_{\mathrm{peri}}}}

\newcommand{\sGamma}{\mathsf{\Gamma}}

\newcommand{\rn}{\ensuremath{\hat{r}_{\Nsmooth}}}
\newcommand{\rnmin}{\ensuremath{\hat{r}_{\Nsmooth,\mathrm{min}}}}
\newcommand{\Na}{\ensuremath{N_{\alpha}}}

\shorttitle{Tidal caustics in M31}
\shortauthors{Sanderson & Bertschinger}

\begin{document}

\title{Seen and unseen tidal caustics in the {A}ndromeda galaxy}

\author{R. E. Sanderson\altaffilmark{1} and E. Bertschinger}
\affil{MIT Department of Physics and Kavli Institute for Space Research, Cambridge, MA  02139}

\altaffiltext{1}{email: \myemail}

\begin{abstract}

Indirect detection of high-energy particles from dark matter interactions is a promising avenue for learning more about dark matter, but is hampered by the frequent coincidence of high-energy astrophysical sources of such particles with putative high-density regions of dark matter. We calculate the boost factor and gamma-ray flux from dark matter associated with two shell-like caustics of luminous tidal debris recently discovered around the Andromeda galaxy, under the assumption that dark matter is its own supersymmetric antiparticle.  These shell features could be a good candidate for indirect detection of dark matter via gamma rays because they are located far from the primary confusion sources at the galaxy's center, and because the shapes of the shells indicate that most of the mass has piled up near apocenter.  Using a numerical estimator specifically calibrated to estimate densities in N-body representations with sharp features and a previously determined N-body model of the shells, we find that the largest boost factors do occur in the shells but are only a few percent.  We also find  that the gamma-ray flux is an order of magnitude too low to be detected with Fermi for likely dark matter parameters, and about 2 orders of magnitude less than the signal that would have come from the dwarf galaxy that produces the shells in the N-body model.  We further show that the radial density profiles and relative radial spacing of the shells, in either dark or luminous matter, is relatively insensitive to the details of the potential of the host galaxy but depends in a predictable way on the velocity dispersion of the progenitor galaxy.

\end{abstract}

\keywords{dark matter, galaxies: individual (M31), galaxies: kinematics and dynamics, gamma rays: galaxies, methods: analytical, methods: numerical}

\section{Introduction}

The nature of the dark matter is one of the foremost questions in astrophysics.   Although most astrophysicists agree that it is probably some kind of particle \citep{bertone:2004aa}, there are to date no conclusive detections.  Countless experiments are attempting to do so, whether by detecting dark matter particles directly \citep{2004PhRvD..69a1101A,2005NuPhS.143..417K,2009arXiv0905.4273G,2009PhRvD..80k5005A,2009PhRvL.102a1301A,2009PhRvL.103n1802A,2010arXiv1003.0904F}, creating them in the laboratory \citep{2006PhRvD..74j3521B,2008ARNPS..58..293H}, or observing the standard-model byproducts of interactions between them \citep{2008Natur.456..362C,2009Natur.458..607A,2009PhRvL.102e1101A,2009PhRvL.102r1101A, 2010arXiv1001.4531F, 2010JCAP...01..031S}.  This last method, usually referred to as ``indirect detection", usually assumes that the dark matter particle is its own antiparticle, so that an interacting pair of dark matter particles self-annihilates to produce various kinds of standard model particles.  This assumption is motivated by predictions of supersymmetry that the dark matter could be the lightest supersymmetric partner (LSP) of a standard model particle, and by the cosmological result that such a particle with a mass of between 20 and 500 GeV would have been produced in the early universe in sufficient number to resolve the discrepancy between the energy density of luminous matter observed today and the total energy density of matter required to explain the gravitational history of the universe \citep{bertone:2004aa}.  

Indirect detection, because it relies on observing the products of pairwise annihilation, has a signal strength that varies as the dark matter density squared.  It is therefore most effective in regions with the very highest number density of dark matter particles.   Because dark matter interacts with luminous matter primarily through gravity, these are often the same regions where the density of luminous matter is highest, such as the centers of galaxies \citep{bergstrom:1998aa}.  Although the signal from pairwise dark matter annihilation may be highest in those regions, it also suffers from confusion with high-energy astrophysical sources like pulsars and X-ray binaries, which tend to be concentrated wherever the density of luminous matter is high \citep{2009ApJS..183...46A}.   However, in some instances the dark matter density can be high while the luminous matter density is low; for example, in the recently discovered ultra-faint dwarf galaxies orbiting the Milky Way \citep{PhysRevD.75.083526}.  Since the dark matter has a lower kinematic temperature than luminous matter, it may also have more small-scale structure than luminous matter, increasing or ``boosting" the production of standard model particles by pairwise annihilation above the level predicted for a smooth distribution \citep{peirani:2004aa,diemand:2007aa,kuhlen:2007aa,2007JCAP...05..015M,bringmann:2007aa,pieri:2007aa,vogelsberger:2007aa}.  Cases where there is no confusion between particles produced by pairwise dark matter annihilation and those produced by high-energy astrophysical sources offer a high potential for a confirmed indirect detection, provided that the signal is still detectable.

One possible scenario for indirect detection is the debris created by a merger between a larger host galaxy and a smaller progenitor galaxy on a nearly radial orbit.  \citet{hernquist:1988aa,hernquist:1989aa} showed that in such a case, the mass from the progenitor accumulates at the turning points of its orbit, producing shells of high-density material at nearly constant radius on opposing sides of the host galaxy.  The dynamics governing the formation and shape of the shells can be understood in the context of earlier work on spherically symmetric gravitational collapse.  \citet{fillmore:1984aa} and \citet{bertschinger:1985aa} demonstrated that radial infall of gravitating, cold, collisionless matter forms a series of infinite-density peaks at successive radii, known as caustics.  \citet{2006MNRAS.366.1217M} extended this case to include warm matter with various velocity dispersions: the effect of the velocity dispersion is to make the peaks of finite width and height, so they are no longer caustics in the mathematically rigorous sense but retain many of the same properties, including the possibility of extremely large local density enhancements.   There have been multiple attempts to estimate the production of gamma rays through self-annihilation from infall caustics in the dark matter halos of galaxies \citep[e.g.][]{2001PhRvD..64f3515H,2003PhLB..567....1S,2005JCAP...05..007P,2006PhRvD..73b3510N,2007JCAP...05..015M,2007PhRvD..75l3514N,2008PhRvD..77d3531N,2008PhRvD..78f3508D,2009PhRvD..79h3526A} or to otherwise determine ways in which dark matter in caustics could be detected \citep{2008PhRvD..78f3508D,2006PhRvD..73b3510N,2003PhLB..567....1S}.  Most work on the gamma-ray signal has found that caustics enhance the production from a smooth distribution by a factor of between 10 and 100.  In the case of so-called ``tidal caustics" like those seen around shell galaxies, the infall is not spherically symmetric, but the density is still clearly enhanced in the shells.

Dark matter and luminous matter alike are concentrated in tidal caustics at large distances from the bulk of the luminous matter in the host galaxy.  Shell galaxies \citep[see, e.g., ][]{malin:1983aa} are beautiful examples of the extreme case of this phenomenon, where the angular momentum of the progenitor is nearly zero.  Unfortunately, all the known shell galaxies are too far away to indirectly detect the dark matter in the shells: the flux of gamma rays is too attenuated, incoming charged particles are deflected by the Galactic magnetic field, and high-energy detectors have insufficient angular resolution to separate the shells from the host.  However, M31 also appears to have shells \citep{fardal:2007aa}, though they are not as symmetric as those in classic shell galaxies, and is close enough that the Fermi LAT can distinguish their position from that of M31's center \citep{2009arXiv0907.0626R}.  Furthermore, an N-body model of the shells already exists \citep{fardal:2006aa} that can be used to estimate whether dark matter in them could be indirectly detected.  

N-body models of dark matter distributions have been used to estimate standard-model particle fluxes for indirect detection in the Galactic halo and the factor by which dark matter substructure could increase the rate of pairwise annihilations \citep{kuhlen:2007aa}.  Both these quantities are proportional to the volume integral of the square of the dark matter density, which we will call the ``rate" for short.  The rate is estimated from an N-body representation of the dark matter distribution by substituting a Riemann sum for the volume integral and inferring the density in each Riemann volume from the N-body representation by one of many well-studied methods.  However, neither the estimation of the square of the density rather than the density itself nor the choice of a suitable set of Riemann volumes has been tested.  Likewise, the ability to recover the correct rate when the density distribution is sharply peaked has not been explored, although this is the scenario that would most likely lead to an observable signal and the reason that the M31 shells are of interest.

This paper describes tests of a number of well-known algorithms for calculating the rate and discusses the best algorithm to use in situations where the density gradient is large (Section \ref{sec:stats}).  We then present estimates, using the optimal algorithm, of both the boost factor from the M31 shells over the smooth distribution of dark matter in M31's halo and the rate at which gamma rays from pairwise annihiliation would be seen by Fermi given likely parameters of a supersymmetric dark matter candidate (Section \ref{sec:m31}). 

We find that the best way to estimate the rate from an N-body representation, whether the density is nearly uniform or has a large gradient, is with the simplest possible method: a nearest-neighbor estimator with a relatively small smoothing number to find the density, and fairly small constant Riemann cubes to perform the integral.  This result is surprising, given that so many more sophisticated density estimators exist.  We further find, using this result, that the largest boost factors from tidal debris in M31 are 2.5 percent in the most concentrated regions of the shells, and that the gamma rays from the debris are too few to be detected by Fermi for likely supersymmetric dark matter candidates: the total additional flux in gamma rays, which is model-dependent, is less than $7.4 \times 10^{-11}\ \gamma\ \textrm{cm}^{-2} \textrm{s}^{-1}$ for likely dark matter models.  

An ancillary result from our analysis of the tidal caustics, and a consequence of radial infall, is that the density profile of each shell and the radial spacing of the shells depend on the radial derivative of the gravitational potential at each shell and on the mass and size of the dwarf galaxy before infall.  This means that information about the initial qualities of the dwarf galaxy can be inferred from the shells without requiring a detailed model of the potential of the host galaxy.  Assuming that the stars in the dwarf galaxy are initially virialized (with a Maxwellian velocity distribution), the density profile can be fit with an analytic function whose width depends on these properties.  We discuss further implications of this result for recently discovered shells around other nearby galaxies.

\section{The Optimal Estimator for High Density Contrast}
\label{sec:stats}

The key to the calculation presented in this paper is the estimation of the integrated squared density,
\begin{equation}
\Gamma = \int \rho^2 dV,
\end{equation}
from an N-body representation of the dark matter mass distribution.  The particles making up the N-body representation are independent observations of the mass density function $\rho$.  Generally, probability density functions are defined as those that are everywhere positive and normalized to one \citep[as in][Chapter 4]{izenman2008}.  The mass density function sampled by the particles in the N-body representation satisfies the first of these two conditions, and dividing by the total mass to get a scaled number density will satisfy the second.  So the analysis of estimators for the probability density and its functionals applies equally to the problem at hand.  The development and characterization of estimators for this quantity is a well-studied problem in statistics, in the context of estimators for probability density distributions \citep[][and many others]{hall1987, bickel1988,birge1995, laurent1996,martinez1999,gine2008,gine2008a,tchetgen2008}.   

Density estimators studied in the literature are divided into two classes: parametric (in which a particular functional form for $\rho$ is assumed) and nonparametric (in which assumptions about the form of $\rho$ are kept to a minimum).  Nonparametric estimators are commonly used with data sets like N-body realizations, where the goal is usually to discover the form of $\rho$ and/or calculate other quantities from it \citep{izenman2008}.  Among the wide variety of nonparametric estimators available, nearest-neighbor estimators \citep{fix1989} are one of the oldest and most well-studied varieties.  The nearest-neighbor estimator uses an adaptive local smoothing length equal to the distance to the $\Nsmooth^{\textrm{th}}$ nearest particle to the location where the density is being estimated.  The density at that point is then taken to be $\Nsmooth / V_d(\Nsmooth)$, where $V_d$ is the volume in $d$ dimensions, centered on the target location, that encloses $\Nsmooth$ particles.  \citet{loftsgaarden1965} and \citet{devroye1977} showed that nearest-neighbor density estimators converge to the underlying distribution at every point as the number of particles in the realization, \Nres, goes to infinity, provided that $\Nsmooth/\Nres \to 0$ in the same limit.  They can also be considered as part of the larger class of adaptive kernel estimators \citep{moore1977} and are even more closely related when $V_d$ is replaced by a weighted sum over the $\Nsmooth$ particles \citep{mack1979}.  However, because the function they return may not be normalizable, nearest-neighbors is more suited to individual density estimates at a point than to recovery of the entire function \citep{izenman1991}.  All the estimators we test in this work are based on either the simple nearest neighbors method or one using a weighted sum, although the shape of $V_d$ varies.  We describe them in detail in Appendix \ref{appx:estimators}.

The usual measure of the quality of a nearest-neighbors estimator is its root-mean-squared (RMS) error, 
\begin{equation}
\label{rmse1}
\textrm{r.m.s.e.} \equiv \frac{1}{\gt}\sqrt{E\left[\left(\hat{\Gamma} - \gt\right)^2\right]},
\end{equation}
The RMS error compares the expectation value of the estimator, in this case the rate estimator $\hat{\Gamma}$, with the true value of the rate, \gt.  For the tests in this work, we used density distributions for which \gt\ may be calculated analytically. \citet{bickel1988} demonstrated that, given some constraints on the maximum slope of the underlying density distribution, the error of a one-dimensional integrated squared density estimator with a kernel of a constant size can converge as $\Nres^{-1/2}$; \citet{gine2008a} recently showed that a simple estimator of this type can be made adaptive using a particular rule to calculate the kernel size from the data and still converge at the same rate.  Most of the estimators we test in this work use adaptive kernels with a simpler rule than the one suggested by \citeauthor{gine2008a} for two reasons.  The first is simply conceptual and computational simplicity: rules for choosing an optimal kernel size tend to require minimizing the cross-validation function (a proxy for the RMS error) of the data, which requires an optimization program, and the resolution convergence even with the optimal kernel chosen in this way can still be slower than $\Nres^{-1/2}$.  The second is that extending the result of \citeauthor{gine2008a} to several orthogonal dimensions is not trivial \citep{wu2007}.  

As is common in the literature, we consider the RMS error in two parts: the bias and standard deviation \citep{Lindgren1976}, where
\begin{equation}
\label{rmse2}
\left(\textrm{r.m.s.e.}\right)^2 = \bias^2 + \left( \textrm{std}(\hat{\Gamma}) \right)^2.
\end{equation}
The bias, $\bias$, is the difference between the expectation value of the estimator and the true value of the parameter it is estimating.  An unbiased estimator has $\bias = 0$, one for which $\eg > \gt$ has a positive bias, and one for which $\eg < \gt$ has a negative bias.  The standard deviation indicates the size of the spread of individual estimates around the expectation value.  For this work we scale the bias, standard deviation and RMS error by a factor of $\gt$, so   
\begin{equation}
\label{biasdef}
\bias \equiv  E\left(\frac{\hat{\Gamma} - \gt}{\gt}\right) = \frac{\eg}{\gt} -1
\end{equation}
and
\begin{equation}
\textrm{std}(\hat{\Gamma}) \equiv \frac{1}{\gt} \sqrt{E\left[ \left( \hat{\Gamma} - \eg \right)^2\right]}
\end{equation}
are consistent with Equations \eqref{rmse1} and \eqref{rmse2}.  

We used numerical experiments to assess the bias, standard deviation, and RMS error of the various estimators, so we must be clear about how these values are calculated numerically.  For each experiment, $10^4$ random realizations of the density distribution of interest comprise one sample.  The expectation value of a quantity is then defined as the mean of that quantity over the set of all random realizations.  The random realizations are subject to Poisson fluctuations, so this number of realizations corresponds to sampling error of about one percent.  We take 20 samples of the expectation value, so the relative error on the mean from these 20 samples is about 0.2 percent. 

\begin{figure}[htd]
\epsscale{1}
\plotone{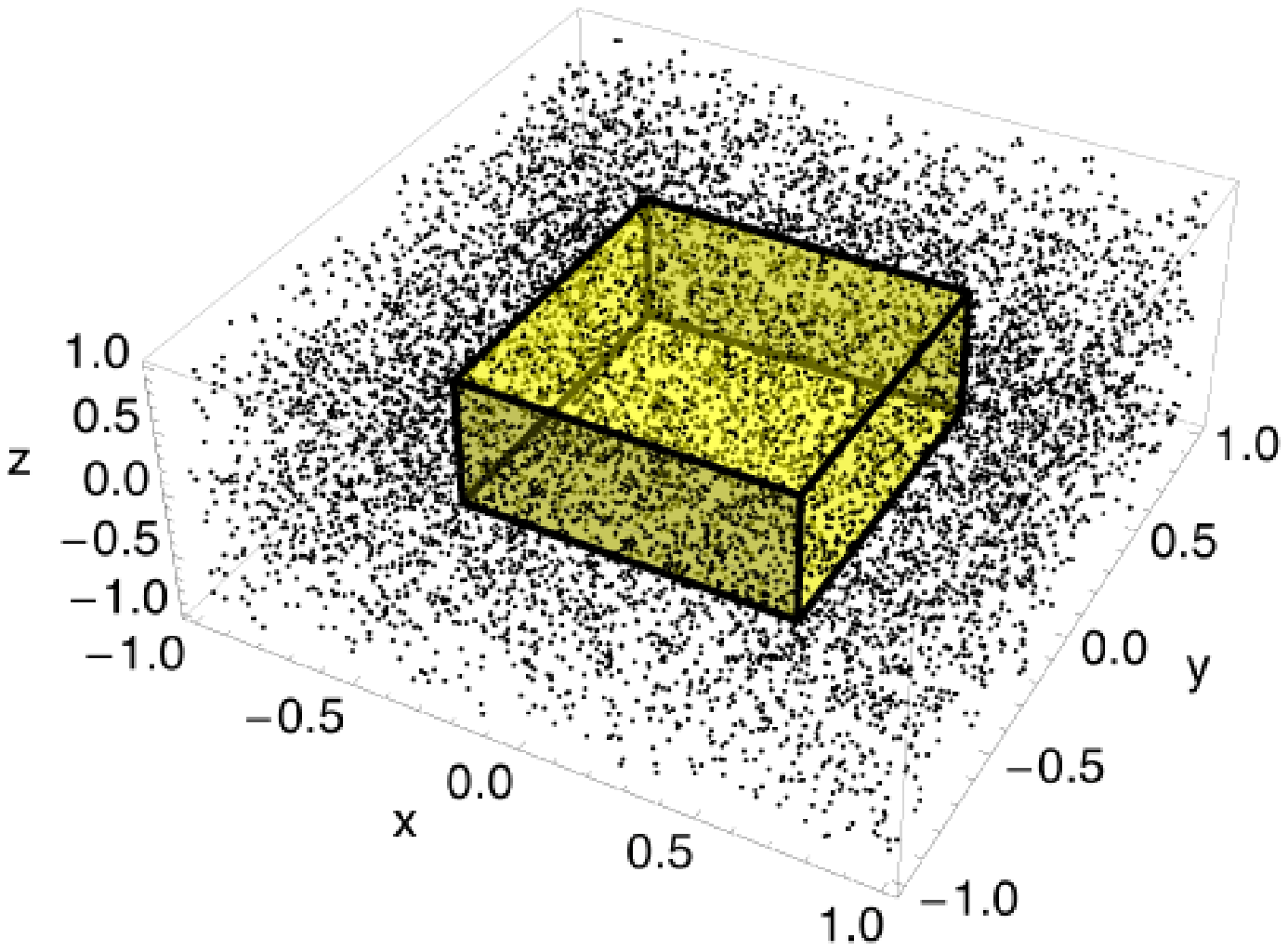}
\caption{(Color online) A three-dimensional random N-body realization of the uniform distribution.  All the points (particles in the realization) in the full volume are used to estimate the density; to avoid edge effects, the volume tessellation and Riemann sum are performed over the volume indicated by the shaded (yellow in electronic edition) box.  \label{EstimationRegion}}
\end{figure}

The number of particles in each random realization (shown as points in an example in Figure \ref{EstimationRegion}) is drawn from a Poisson distribution with a specified mean, denoted in the following sections as $\Nres$.  This precaution keeps the number of particles in the density distribution, and in the subset of that distribution used for the volume integral (the shaded box in Figure \ref{EstimationRegion}), purely Poisson; the error associated with using a fixed number of particles depends on $\Nres$, so we must eliminate it if we wish to establish how the estimators behave as $\Nres$ varies.

The method for estimating the rate has two distinct parts: how to determine the number and placement of the Riemann volumes making up the sum, and how to estimate the density in each Riemann volume.  We tested five different rate estimators that together use three different well-known density estimation methods and two different ways of assigning Riemann volumes (adaptive and constant).  The rate estimators are described in detail in Appendix \ref{appx:estimators} and briefly summarized in Table \ref{tbl:variousRateEstimators}.  

We first evaluate and, if possible, eliminate the bias.  We constructed rate estimators using density estimators that have a very small or zero bias when used to estimate the density, but we demonstrate in this section that they do not always produce unbiased estimates of the rate without further correction.  We wish to reduce the bias of the estimators when it is possible to do so without increasing their standard deviations.  Bias resulting from the statistics of Poisson point processes, here referred to as ``Poisson bias,'' can be eliminated this way---analytically for some of our estimators, and numerically for the others---once it has been measured using random realizations of the uniform density distribution (Section \ref{sec:pb}).  We examined the RMS error of rate estimates for the uniform density distribution, after correcting for the Poisson bias, to separate the contribution of Poisson processes to the overall RMS error of each estimator from additional error that arises when the density is not uniform (Section \ref{sec:UniformStdDev}).

The second step in determining the best estimator is to determine the RMS error in the case where the density distribution has sharp features with high contrast (like shells).  Discreteness effects will then introduce additional bias that depends on the resolution, the smoothing number, and the scale of the high-contrast features (Section \ref{sec:rb}).  To understand when and how this bias contributes, we tested each estimator using random realizations of a simple caustic density distribution that has an analytic expression for $\gt$ (Section \ref{sec:ub}).  Finally, we compared the RMS errors of the estimators for the high-contrast density distribution to determine which one to use in the calculation of the gamma-ray flux (Section \ref{sec:bestPerformance}).

\subsection{Eliminating Poisson Bias}
\label{sec:pb}

\begin{figure}[htd]
\plotone{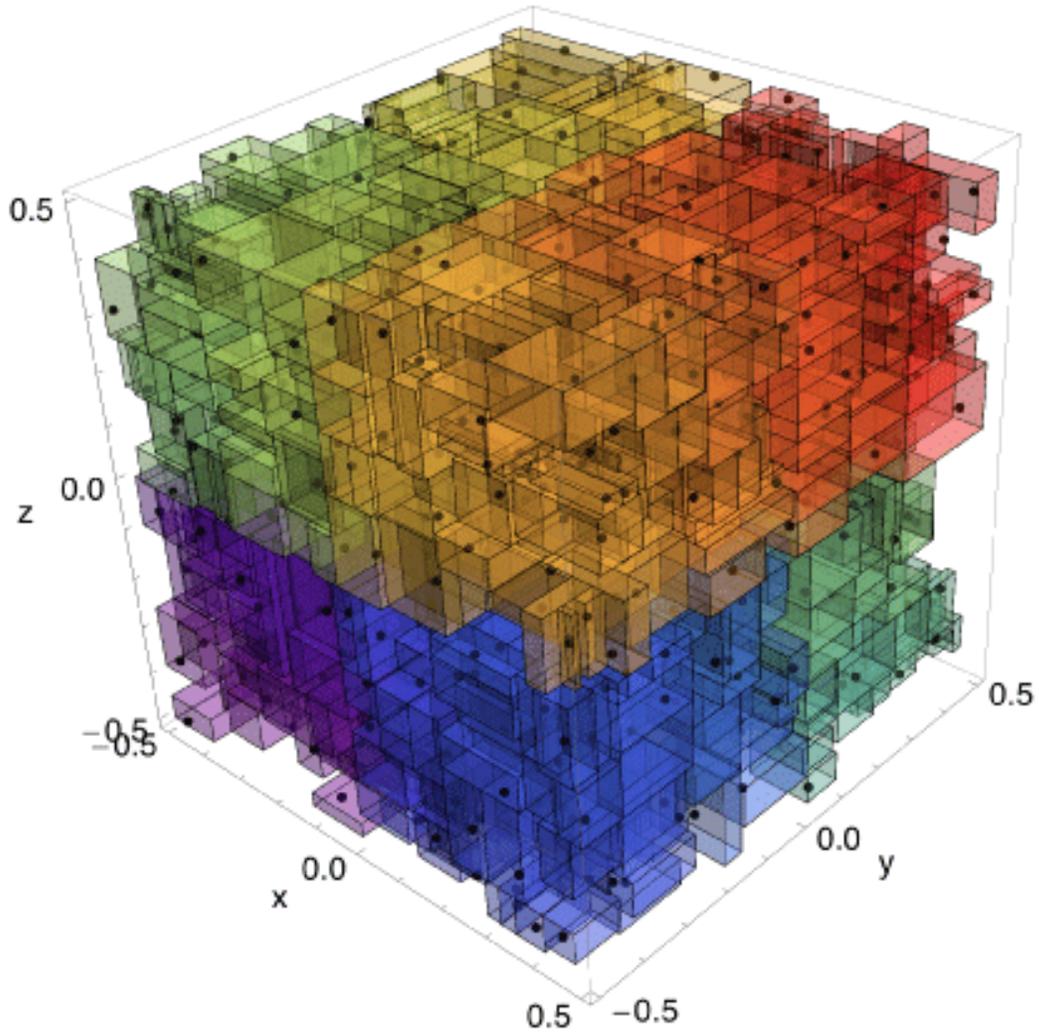}
\caption{(Color online) Example of adaptive volume tessellation referred to in the text, produced by the space-filling tree in the FiEstAS estimator \citep{ascasibar:2005aa}.  The shading (colors from red to blue in electronic edition) indicates the way in which the tree divides the space (the spatial distribution of the inorder traversal). For clarity, not all the boxes with edges on the boundaries are shown. \label{treeboxes}}
\end{figure}

Using Poisson statistics, it is possible to construct a rate estimator $\hat{\Gamma}_u$ that is unbiased for a uniform density distribution; that is, one for which $E(\hat{\Gamma}_u) = \gt$ (Appendix \ref{appx:estimators}, Equation \eqref{estimatorU}).  It uses a constant Riemann volume $dV$ for the integral and the distance to the $\Nsmooth^{\mathrm{th}}$ nearest neighboring particle, denoted $r_{Ns}$, to estimate the density.  However, this estimator is not necessarily unbiased for non-uniform distributions.  If the density is location-dependent, the integration volume $dV$ must small enough to accurately sample the density gradient everywhere in the distribution.  This choice of $dV$ can be impractically small for distributions with high density contrast.  One solution is to choose $dV$ adaptively based on the local density of particles (Figure \ref{treeboxes}), using smaller boxes in higher-density regions, but this method introduces bias because the box dimensions, like $r_{Ns}$, are then subject to Poisson statistics.  We chose $\hat{\Gamma}_n$ (Appendix \ref{appx:estimators}, Equation \eqref{estimatorN}) to isolate the contribution from choosing $dV$ adaptively.

Furthermore, the simple nearest-neighbor estimator is not the only option.  Many other density estimation algorithms exist, often optimized for much better performance.  However, even for these algorithms it is usually the case that $E(\hat{n})^2 \ne E(\widehat{n^2})$, especially since the bias and standard deviation are usually optimized for the first moment (the density) at the expense of the higher moments (including density-squared).  We have chosen two other methods besides nearest-neighbor, both well-studied as density estimators, to see whether a good density estimator also makes a good rate estimator ($\hat{\Gamma}_f$, Equation \eqref{rateEstimatorF}, and $\hat{\Gamma}_s$, Equation \eqref{rateEstimatorS}).  We also include a rate estimator based on one of these density estimators that has been used in the literature to calculate the rate ($\hat{\Gamma}_d$, Equation \eqref{rateEstimatorD}).  All the rate estimators are summarized in Table \ref{tbl:variousRateEstimators}.

\begin{deluxetable}{clp{2.4in}l}
\tablecaption{Various estimators of the gamma-ray emissivity $\Gamma$.    \label{tbl:variousRateEstimators}}
\tablehead{\colhead{Estimator} & \colhead{Definition} & \colhead{Long name} & \colhead{See equation(s)}}
\startdata
 $\hat{\Gamma}_u$ & $= dV \sum_{i=1}^{N_V} \widehat{n^2}_{u,i}$ & Unbiased nearest-neighbors estimator with uniform box size & \ref{unbiasedDensitySquared}, \ref{estimatorU}\\
 $\hat{\Gamma}_n$ & $ \propto \sum_{i=1}^{\Nres} \widehat{n^2}_{u,i} dV_i$ & Unbiased nearest-neighbors estimator with adaptive box size & \ref{unbiasedDensitySquared}, \ref{estimatorN} \\
$\hat{\Gamma}_f$ & $\propto \sum_{i=1}^{\Nres} (\hat{n}_{f,i})^2 dV_i $ & FiEstAS estimator \citep{ascasibar:2005aa} with adaptive box size & \ref{densityEstimatorF}, \ref{rateEstimatorF} \\
$\hat{\Gamma}_s$ & $\propto \sum_{i=1}^{\Nres} (\hat{n}_{s}(r_i))^2 dV_i$ & Epanechikov kernel density estimator with adaptive box size & \ref{densityEstimatorS}, \ref{EpanechikovKernel}, \ref{rateEstimatorS}\\
$\hat{\Gamma}_d$ & $\propto \sum_{i=1}^{\Nres} \hat{n}_{s}(r_i) $ & Method from \citet{diemand:2007aa} & \ref{rateEstimatorD}\\
\enddata
\tablecomments{See Appendix \ref{appx:estimators} for longer descriptions of the various estimators.}
\end{deluxetable}%

\begin{figure}[htbp]
\begin{center}
\plotone{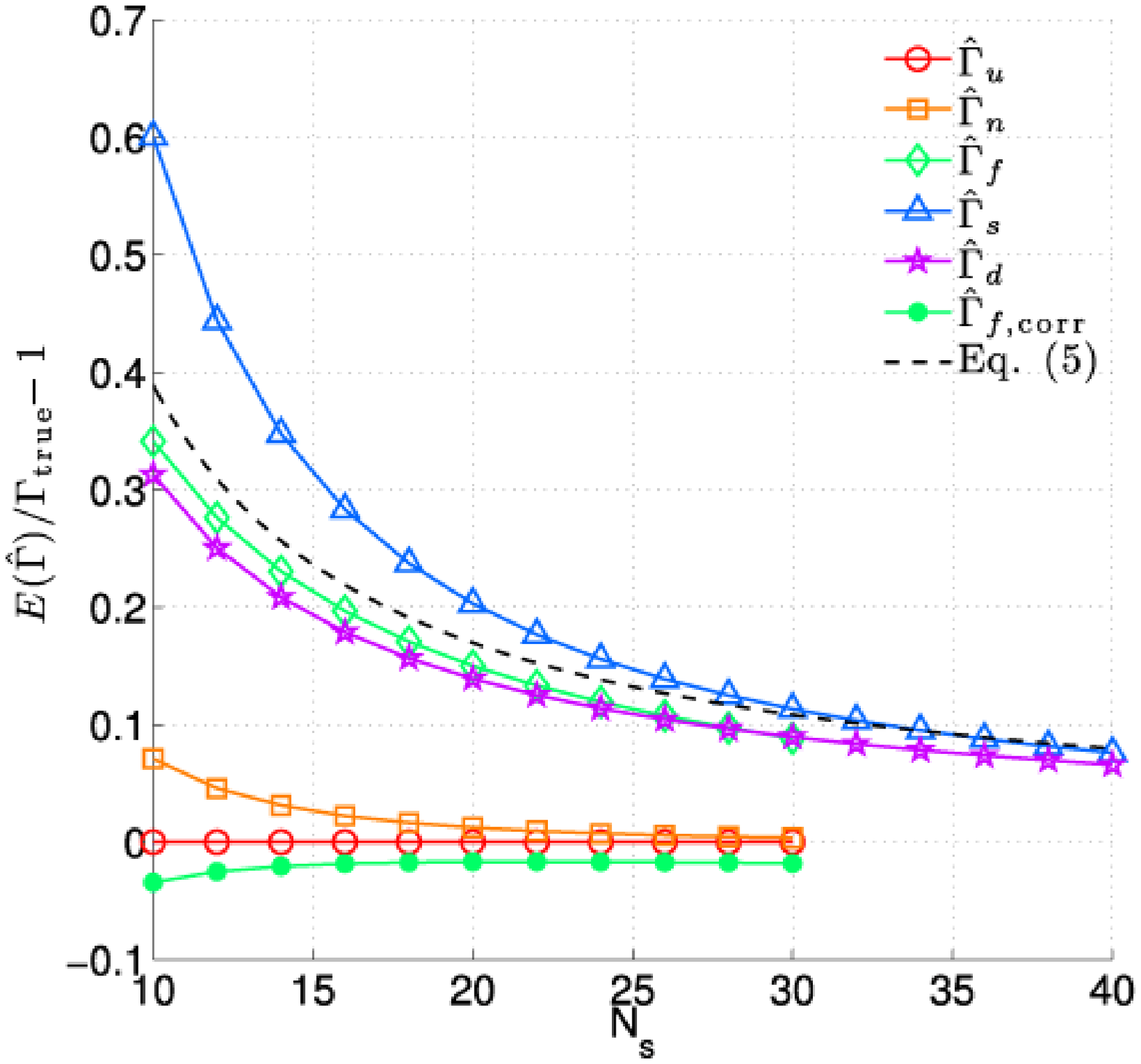}
\caption{(Color online as indicated in square brackets) All the rate estimators except for $\hat{\Gamma}_u$ ([red] circles) require slightly different but similarly shaped bias corrections.  Equation \eqref{biasfitfunction} is an adequate ansatz for the \Nsmooth-dependence of the correction: an example corresponding to Equation \eqref{naivebias} is shown as the [black] dashed line and the [green] filled points show the effect of using Equation \eqref{naivebias} to correct $\hat{\Gamma}_f$.  Fit parameters are listed in Table \ref{tbl:biasfits}; Table \ref{tbl:variousRateEstimators} explains the label abbreviations.  \label{biasfits}}
\end{center}
\end{figure}

We calculated the expectation value and standard deviation of each rate estimator using random realizations of a uniform distribution, as described above.  Equations \eqref{rmse2} and \eqref{biasdef} then give the RMS error and bias in terms of these quantities.  As expected, the analytically unbiased estimator $\hat{\Gamma}_u$ has a numerically confirmed bias of zero (orange squares in Figure \ref{biasfits}).  Furthermore, using an adaptive Riemann volume does change the bias (green diamonds in Figure \ref{biasfits}).  The more complicated estimators require even more substantial correction (cyan triangles, blue pentagrams, and purple hexagrams in Figure \ref{biasfits}).  Interestingly, all the bias curves have the same overall shape.  We use the multiplicative factor that transforms the biased density-squared estimator into the unbiased one,
\begin{equation}
\label{naivebias}
\frac{E[(\hat{n}_b)^2]}{E(\widehat{n^2}_u)} = \frac{\Nsmooth^2}{(\Nsmooth-1)(\Nsmooth- 2)}, 
\end{equation}
generalized to the form
\begin{equation}
\label{biasfitfunction}
\bias(\Nsmooth) + 1 = \frac{\bias_1 \Nsmooth^2}{(\Nsmooth - \bias_2)(\Nsmooth - \bias_3)},
\end{equation}
as an ansatz for the shape of each bias curve.  This parameterization evokes the introduction of an effective $\Nsmooth$, but can also adjust for the statistical effect of choosing adaptive Riemann volumes, and is a good fit to all the bias curves (Figure \ref{biasfits}).

Fitting $E(\hat{\Gamma}_n)/\gt$ to a function of the form in Equation \eqref{biasfitfunction} determined $\bias_n$, the bias from using adaptive Riemann volumes.  The results for $\hat{\Gamma}_f$, $\hat{\Gamma}_s$, and $\hat{\Gamma}_d$ were fit using the same function to determine $\bias_f$, $\bias_s$, and $\bias_d$ respectively, to identify bias from using a particular density estimation method.  The fitted parameters for each estimator are summarized in Table \ref{tbl:biasfits}.  We used the fits to correct for the Poisson bias in all the remaining work discussed in this paper.

\begin{deluxetable}{llll}
\tablecolumns{4}
\tablewidth{0pt}
\tablecaption{Best-fit bias corrections. \label{tbl:biasfits}}
\tablehead{\colhead{Estimator} & \colhead{$\bias_1$} & \colhead{$\bias_2$} & \colhead{$\bias_3$}}
\startdata
u & 1 & 2 & 1 \\
n & $1.0001 \pm 0.0004$ & $2.81 \pm 0.01$ & $-2.99 \pm 0.03$ \\
f & $0.991 \pm 0.004$ & $1.41 \pm 0.03$ & $1.41 \pm 0.03$  \\
s & $0.971 \pm 0.003$ & $-0.3 \pm 0.2$ & $4.1 \pm 0.1$ \\
d & $1.00024 \pm 0.00005$ & $0.60 \pm 0.01$ & $1.89 \pm 0.01$ \\
\enddata
\tablecomments{Error ranges indicate the 95-percent confidence level.  The form of the bias correction is given in Equation \eqref{biasfitfunction}. See Appendix \ref{appx:estimators} and Table \ref{tbl:variousRateEstimators} for descriptions of the various estimators.}
\end{deluxetable}%

\begin{figure}[htbp]
\begin{center}
\plotone{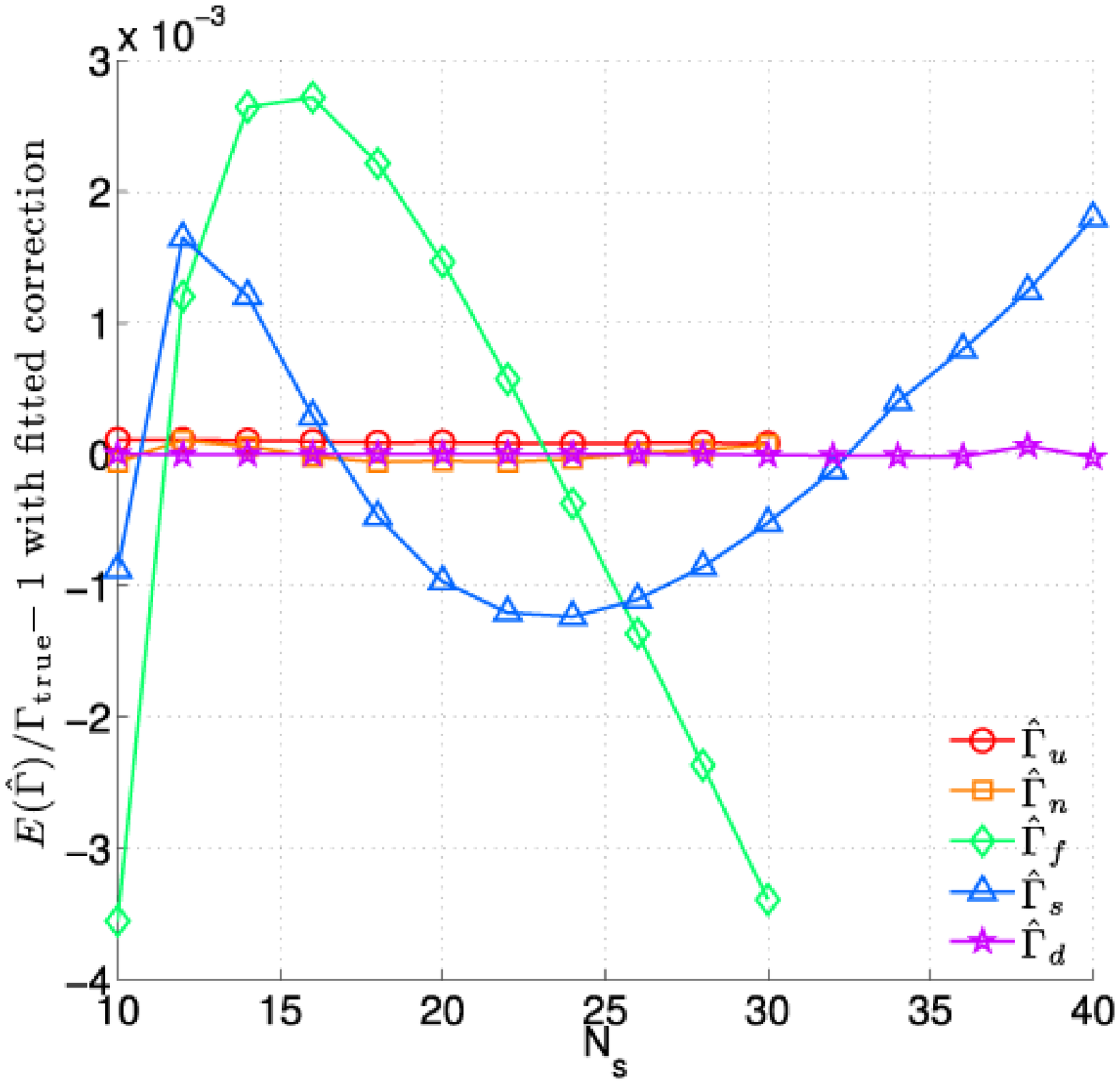}
\caption{(Color online) Once estimators are corrected for Poisson bias using the fitting formula in Equation \eqref{biasfitfunction}, the corrected versions all have a bias much smaller than the typical standard deviation of a few percent.  The residual bias is comparable to that of $\hat{\Gamma}_u$.  Table \ref{tbl:variousRateEstimators} explains the label abbreviations. \label{biascorrected}}
\end{center}
\end{figure}

\subsection{Performance of Estimators on the Uniform Distribution}
\label{sec:UniformStdDev}

\begin{figure}[htb]
\plotone{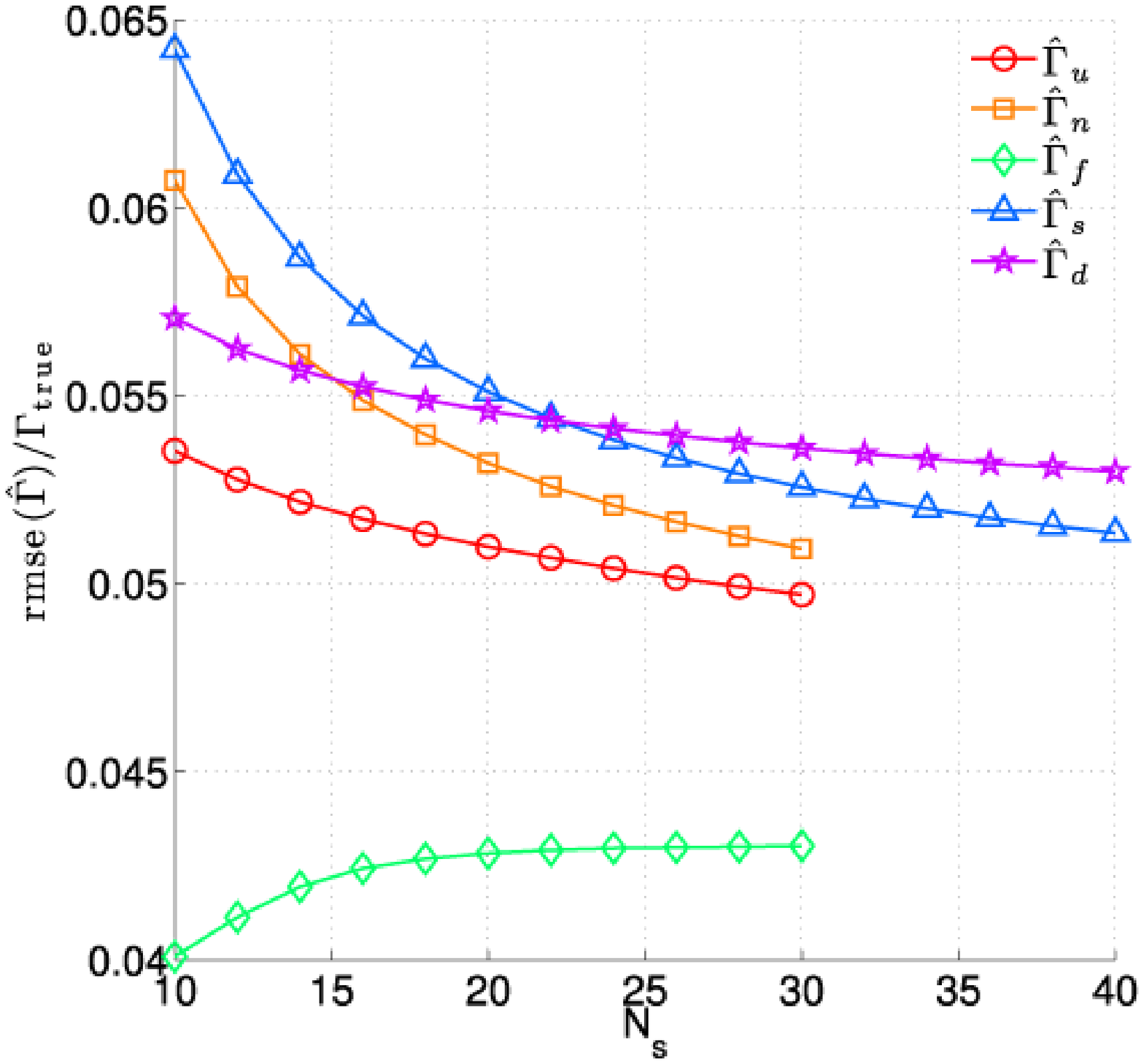}
\caption{(Color online as indicated in square brackets) The RMS error of the corrected estimators is around five percent, and declines with increasing \Nsmooth.  $\hat{\Gamma}_f$ ([green] diamonds) and $\hat{\Gamma}_u$ ([red] circles) have the smallest RMS error at low \Nsmooth.  Introducing an adaptive Riemann volume ([orange] squares) increases the RMS error significantly at low \Nsmooth.  The RMS error starts at a higher value and scales more slowly with $\Nsmooth$  for the kernel-based estimators ([blue] triangles and [purple] stars) than for the nearest-neighbors estimators, an effect of the weighting function used to estimate the density.  Table \ref{tbl:variousRateEstimators} explains the label abbreviations.  \label{Neff}}
\end{figure}

All the estimators can be corrected to measure $\gt$ for a uniform density with an accuracy of better than one percent, although the $f$ and $s$ estimators have higher-order uncorrected behavior (Figure \ref{biascorrected}).  The typical standard deviation of any of the estimators is larger than this residual bias by about an order of magnitude, so it will dominate the RMS error (compare Figure \ref{Neff} to Figure \ref{biascorrected}).  We want the estimator with the best combination of small RMS error and small \Nsmooth: although the value of \Nsmooth\ is not as important in the uniform-density case, it limits the sensitivity of the estimator to small-scale density fluctuations if the density is not uniform.  We will discuss this further in Section \ref{sec:rb}.

It is not clear that reducing the bias will automatically reduce the RMS error, since correcting for the bias can change the standard deviation of an estimator.  In our case, for a given $\Nsmooth$, the bias correction simply multiplies the uncorrected estimator by a constant factor, so the expectation values are related by
\begin{equation}
E(\gpbc) = E(\hat{\Gamma}_{\mathrm{uncorr}})/(\bias(\Nsmooth) + 1)
\end{equation}
and the standard deviations are related by the same factor:
\begin{equation}
\textrm{std}(\gpbc) = \textrm{std}(\hat{\Gamma}_{\mathrm{uncorr}}/(\bias(\Nsmooth) + 1)) = \textrm{std}(\hat{\Gamma}_{\mathrm{uncorr}})/(\bias(\Nsmooth) + 1)
\end{equation}
From this last expression, we see that if the uncorrected estimator \emph{underestimates} $\gt$, correcting it will increase the standard deviation.  However, all the uncorrected estimators overestimate \gt\ (Figure \ref{biasfits}), so correcting the bias will also reduce the standard deviation.  $\bias(\Nsmooth) + 1$ is always close to unity and never larger than 2, so the change to the standard deviation is slight.  

For a given $\Nsmooth$, the RMS error of the nearest-neighbor-style estimators is generally smaller than that of the kernel-based estimators and decreases faster with \Nsmooth\  (Figure \ref{Neff}).  In a kernel-based estimator, because each particle within the smoothing radius is individually weighted, the estimator must know the location of every one of the $\Nsmooth$ particles used in the density estimate, not just the $\Nsmooth$th one, and each weight is less than 1.  $\Nsmooth$ in the kernel-based estimators must therefore be much larger to get the same RMS error as in a nearest-neighbors estimator with a given $\Nsmooth$, as can be clearly seen in Figure \ref{Neff}.   To achieve the same RMS error as $\hat{\Gamma}_u$ at $\Nsmooth = 10$, $\hat{\Gamma}_s$ must use $\Nsmooth\sim 25$.  The RMS error of $\hat{\Gamma}_d$ converges so slowly that although it starts out with slightly better performance than $\hat{\Gamma}_n$, it only begins to compete with $\hat{\Gamma}_u$ at $\Nsmooth > 30$.  Using such a large $\Nsmooth$ is a built-in disadvantage for these estimators if one hopes to retain sensitivity to small-scale density fluctuations, and also significantly increases the computational load.  For this reason we decided not to test the kernel-based estimators on systems with high density contrast, since the nearest-neighbors estimators are better suited to our needs.

Using an adaptive Riemann volume with the spherical nearest-neighbor estimator, as is done in $\hat{\Gamma}_n$, increases the RMS error at low \Nsmooth.  At the same time, $\hat{\Gamma}_f$ uses the same adaptive Riemann volume and achieves a lower RMS error.  This unusual behavior may be caused by the different shapes of the density estimation volume (spherical) and Riemann volume (orthohedral) used in $\hat{\Gamma}_n$; we discuss this possibility in Section \ref{sec:ub}.  The RMS error of $\hat{\Gamma}_n$ may be smaller relative to that of the other estimators in the case of high density contrast because the adaptive Riemann volumes can resolve the density gradient better, so we tested it, along with $\hat{\Gamma}_u$ and $\hat{\Gamma}_f$, on samples with high density contrast.  These tests are described in the next few sections.

\subsection{Additional Bias for Systems with High Density Contrast}
\label{sec:rb}

In regions of high density contrast, like caustics, the maximum resolvable density is limited by the minimum nearest-neighbor distance expected for a given smoothing number \Nsmooth\ and number of particles \Nres.  In the limit that \Nsmooth\ and \Nres\ are both large, the expectation value of the minimum nearest-neighbor distance scales as
\begin{equation}
E(\rnmin)  \propto \left( \frac{\Nsmooth }{\Nres }\right)^{1/3}, \qquad \qquad \Nsmooth, \Nres \gg 1.
\end{equation}
This scaling is derived by calculating the first order statistic of the probability distribution of nearest-neighbor distances for a three-dimensional Poisson point process (for a longer explanation, please see Appendix \ref{appx:rnmin}).  The scaling with $\Nres$ is valid for $\Nres \gtrsim 10^2$, but the scaling with $\Nsmooth$ only approaches the asymptotic limit for values much too large to be practical ($\Nsmooth \gtrsim 10^3$).  For reasonable values of \Nsmooth, the power-law index must be determined numerically as discussed in Appendix \ref{appx:rnmin}:
\begin{equation}
\label{rnminscaling}
E(\rnmin)  \propto \frac{\Nsmooth^{\gamma}}{\Nres^{1/3}}, \quad \gamma = 0.51 \pm 0.06 \qquad \qquad \Nres > 10^2, \quad 10\lesssim\Nsmooth\lesssim 45
\end{equation}
The corresponding maximum density then scales as
\begin{equation}
\label{nmaxscaling}
\nmd \propto \frac{N_s}{r_{\Nsmooth,\mathrm{min}}^3} \propto \frac{\Nres}{\Nsmooth^{3\gamma-1}}.
\end{equation}
For \Nsmooth\ in the range of interest, $3\gamma - 1 \approx 1/2$.  As expected, using more particles or a smaller smoothing number increases the sensitivity to small-scale fluctuations and the maximum density.  The upper limit on the density introduces bias into the density estimation that also depends on $\Nres$ and $\Nsmooth$ in the combination given by Equation \eqref{nmaxscaling}.  The estimator will perform normally as long as the local density is less than $\nmd$, but returns $\nmd$ for densities larger than $\nmd$.  Given a density estimator $\hat{n}$, the undersampling-limited density estimator $\ndl$ that incorporates this effect can be written
\begin{equation}
\ndl =
\left\{
\begin{array}{ll}
  \hat{n} & E(\hat{n}) < \nmd  \\
   \nmd &     E(\hat{n})  > \nmd 
\end{array}
\right.
\end{equation}
The upper limit on the density changes the way the corresponding rate estimator works, since now the piecewise function $\ndl$ separates regions of the Riemann sum where the density is less than the upper limit from regions where the density is too high to be resolved.  So given a bias-free uniform-density estimator $\hat{\Gamma}$, the corresponding density-limited estimator $\hat{\Gamma}_{\mathrm{dl}}$ is
\begin{equation}
\hat{\Gamma}_{\mathrm{dl}} = \left. \hat{\Gamma}_i \right|_{n<\nmd} + \frac{\nmd^2 V_{n>\nmd}}{\bias + 1}
\end{equation}
where $\bias$ is a factor with the form of Equation \eqref{biasfitfunction} and the appropriate fitted constants from Table \ref{tbl:biasfits}.  There is therefore an undersampling bias, $\bias_{us}$, in the rate estimator that depends on both $\Nres$ and $\Nsmooth$.  The $\Nres$- and $\Nsmooth$-dependence enter two ways: in the criterion for separating the Riemann sum and directly in the rate calculation for one of the terms:
\begin{eqnarray}
\label{eq:undersamplingBias}
\bias_{us} &\equiv& \frac{\hat{\Gamma}_{\mathrm{dl}}}{\gt} - 1   \nonumber \\
&=& \frac{\left. \eg \right|_{n<\nmd}}{\gt} + \frac{\nmd^2 V_{n>\nmd}}{\gt(\bias + 1)} - 1, \nonumber \\
\end{eqnarray}
Each of the first two terms is less than or equal to 1 because they are both evaluated over subsets of the full integration volume.  Additionally, their sum must be less than or equal to 1 because the limiting density is less than or equal to the density in all the Riemann volumes in that sum.  So $\bias_{us} \le 0$. In the limit where the realization is fully resolved, the bias should be zero since $V_{n>\nmd} =0$, independent of \Nsmooth\ and \Nres. 

In general, the $\Nres$- and $\Nsmooth$-dependence in Equation \eqref{eq:undersamplingBias} is complicated since some unknown fraction of the total volume is under-resolved. To determine the undersampling bias, we tested $\hat{\Gamma}_u$, $\hat{\Gamma}_n$, and $\hat{\Gamma}_f$, including their respective corrections for Poisson bias, on N-body realizations of a one-dimensional caustic for which $\gt$ can be calculated analytically.  The caustic has an adjustable sharpness represented in terms of a velocity dispersion $\sigma$: the smaller $\sigma$ is, the narrower and taller the peak in the density.  The caustic density as a function of position and time may be expressed in terms of Bessel functions:
\begin{eqnarray}
\rho(x,t) &=& \frac{\rho_0}{\sqrt{2\pi\sigma^2t^2}} \sqrt{\frac{\left|x-x_c\right|}{t}}\ e^{-\left(x-x_c\right)^2/4\sigma^2t^2} \nonumber \\
&\times&\mathcal{B}\left[\frac{(x-x_c)^2}{4 \sigma^2 t^2}\right]
\label{analyticCaustic1}
\end{eqnarray}
with
\begin{equation}
\mathcal{B}(u) = 
\left\{
\begin{array}{cc}
\frac{\pi}{\sqrt{2}} \left[ \mathcal{I}_{-1/4}(u) + \mathcal{I}_{1/4}(u) \right]  & x \le x_c  \\
\frac{\pi}{\sqrt{2}} \left[ \mathcal{I}_{-1/4}(u) - \mathcal{I}_{1/4}(u) \right] &  x > x_c 
\end{array}
\right.
\label{analyticCaustic2}
\end{equation}
where $x_c = 1/4\alpha t$ is the position of the caustic and $\mathcal{I}$ is a modified Bessel function of the first kind.  Table \ref{tbl:parameters} briefly explains the parameters $\rho_0$, $\alpha$, and $\sigma$ and the dimensions for the integration volume, and gives the values used in our tests where applicable.   We derive this result and explain the parameters more fully in Appendix \ref{appx:oneDcaustic}.  Equation \eqref{analyticCaustic1} is given in terms of the mass density $\rho$, which is easily related to the number density $n$.  When we construct random realizations of the caustic, we hold the normalization of the mass density $\rho_0$ and the size of the integration volume $V$ constant as $\Nres$ changes by setting the particle mass $m_p = \rho_0 V/\Nres$, so that realizations with different $\Nres$ will have the same \gt.  

Given the values in Table \ref{tbl:parameters}, $\gt$ can be calculated by performing a single numerical integral (Appendix \ref{appx:oneDcaustic}).  We can use this simple model to vary the density contrast and the scale of the density variations, simply by generating N-body representations of the caustic described by Equation \eqref{analyticCaustic1} for different $\sigma$. 

\begin{figure}[htb]
\plotone{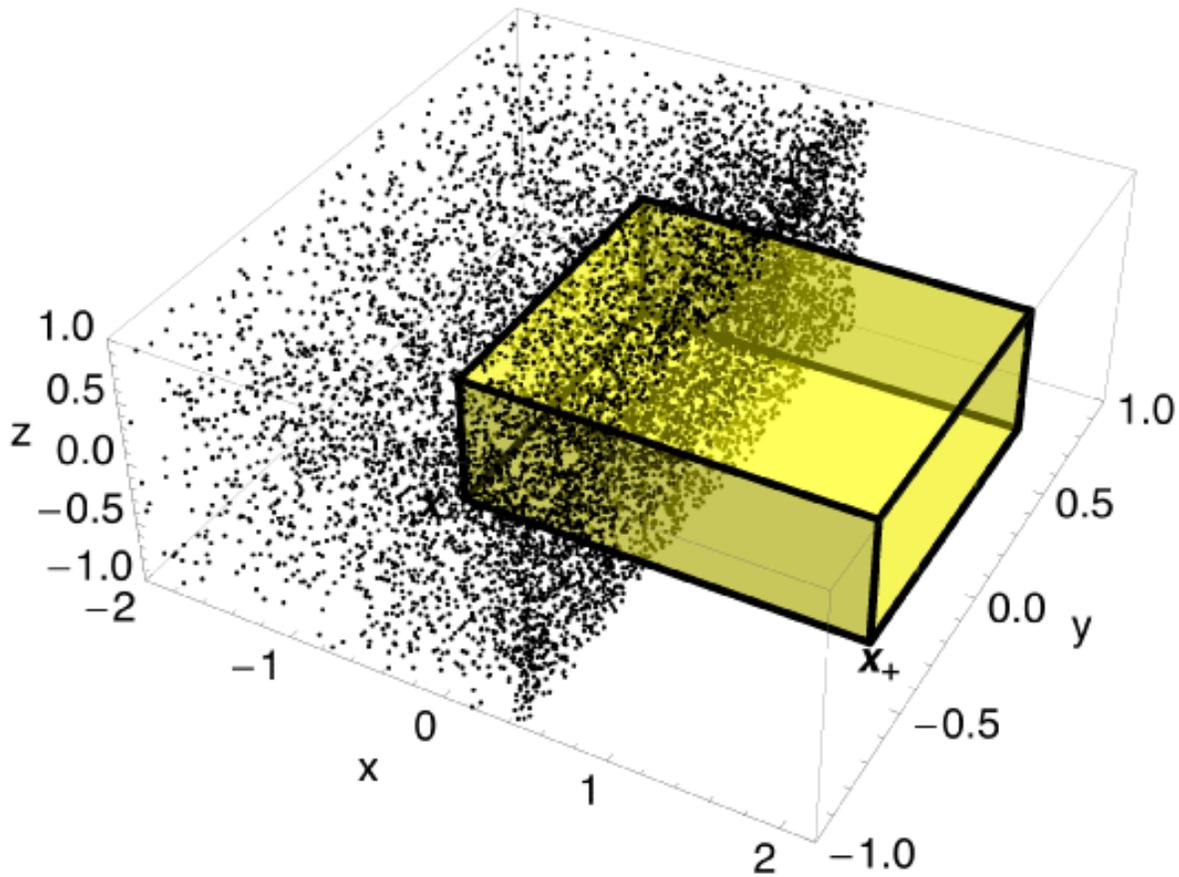}
\caption{(Color online) An example of a random N-body realization of the analytic one-dimensional caustic used to test estimators for undersampling bias.  The points are the locations of the particles in the realization. The shaded (yellow in electronic edition) box indicates the integration volume, positioned to include the complete caustic shape and avoid edge effects. For this sample, $\Nres = 10^4$, $\sigma = 0.01$, and the dimensions of the box are given in Table \ref{tbl:parameters}.  \label{NBodyCaustic}}
\end{figure}

To make a random N-body realization of the caustic, $\Nres$ particles are initially distributed uniformly in their initial three-dimensional positions $\vec{q}$.  Then a displacement function $x(q_x)$ is applied to the $q_x$ coordinate to generate the sample (the set of blue points in Figure \ref{NBodyCaustic}).  The general form of $x(q_x)$ is given in Equation \eqref{eq:qtox-with-disp}.  The form for a particular sample is determined by setting the parameters $\alpha$ and $t$, which also set the location of the peak density of the caustic, $x_c$, and by setting $\sigma$, the width of the normal distribution from which the random components of the particles' initial velocities are drawn.  The values of $\alpha$, $t$, and $\sigma$ we used are summarized in Table \ref{tbl:parameters}.  The locations of the particles in $y$ and $z$ remain uniform.  

It is important to choose the initial range of $q_x$ so that the corresponding range of $x$ values defined by the mapping $x(q_x)$ is larger than the integration range in $x$, since otherwise the system will be incompletely sampled in $x$.  In practice, we determined the range in $q_x$ by choosing a range in $x$ that is larger than the integration range and then inversely mapping it to $q_x$ under the assumption that the random velocity contribution is zero (otherwise the mapping is not invertible).  This method does not work for random velocities comparable in magnitude to the bulk velocity, and we adjusted our method of calculating $\gt$ to account for the incomplete sampling in these cases, as noted in Appendix \ref{appx:oneDcaustic}.

To test the estimators, we used them to calculate the rate for a set of samples with a given density contrast and resolution by integrating density-squared over the integration volume (shown as a yellow box in Figure \ref{NBodyCaustic}).  The integration volume is smaller than the dimensions of the realization to avoid unwanted edge effects, but extends well beyond the edge of the caustic, which is the feature of interest.  Since the undersampling bias depends on both the resolution and smoothing number, we varied both these parameters for each set of samples with a given $\sigma$.  Table \ref{tbl:calibrationRange} summarizes the ranges and step sizes we used to explore this parameter space.  Because the parameter space for these tests was so much larger than in the uniform-density case, we used 5000 random realizations at each combination of contrast and resolution, so the expected level of sampling fluctuations is about 1.5 percent.

\begin{deluxetable}{rccp{4in}}
\tablewidth{0pt}
\tablecaption{Parameters for the analytic one-dimensional caustic.\label{tbl:parameters}}
\tablehead{\colhead{Parameter} & \colhead{Value} & \colhead{Dimension} & \colhead{Notes}}
\startdata
$\rho_0$ & $625$ & [M]/[L]${}^3$ & Mass density of the sample at $t=0$.  Particle mass is adjusted to keep the same mass density at varying resolution. \\
$\sigma$ & varies & [L]/[T] & Sharpness parameter. A smaller $\sigma$ makes a sharper caustic. \\
$\alpha$ & 1/2 & 1/[L][T] & Describes the displacement function used to generate the caustic.  See Appendix \ref{appx:oneDcaustic}, Equations \eqref{eq:qtox-with-disp} and \eqref{eq:single-integral-alpha}.\\
$t$ & 1 & [T] & Corresponds to $x_c = 0.5$.  See Appendix \ref{appx:oneDcaustic}, Equation \eqref{eq:xcDefined}.\\
$L_y$,$L_z$ & 0.5 & [L] & The rate is integrated from $-L_{y,z}$ to $L_{y,z}$, where $y$ and $z$ are the dimensions parallel to the caustic. Avoids edge effects, as illustrated in  Figure \ref{NBodyCaustic}.  \\
$[x_{-}, x_{+}]$ & [-0.5,2]  & [L] & Limits of the rate integration in the direction perpendicular to the caustic, chosen so that the rate is integrated across the caustic.  Shown in Figure \ref{NBodyCaustic}. \\
\enddata
\tablecomments{These quantities are defined in more detail in Appendix \ref{appx:oneDcaustic}.  The units are given as dimensions only since they may be scaled as needed.} 
\end{deluxetable}%

\begin{deluxetable}{rp{3in}}
\tablewidth{0pt}
\tablecaption{Parameter space for testing undersampling bias.\label{tbl:calibrationRange}}
\tablehead{\colhead{Parameter} & \colhead{Values Tested}}
\startdata
Estimator & \{u,n,f\}, corrected for Poisson bias where appropriate using Equation \eqref{biasfitfunction} and the appropriate coefficients in Table \ref{tbl:biasfits}. \\
\Nsmooth & $10\ldots30$, in steps of 2\\
$\log_{10}\Nres$ & $3\ldots 4.5$, in steps of 0.25\\
$\log_{10}\sigma$ & $-3\ldots 0.5$, in steps of 0.25\\
\enddata
\end{deluxetable}%

\subsection{Understanding the Undersampling Bias}
\label{sec:ub}

We draw conclusions about the behavior of the undersampling bias with various $\Nsmooth$ and $\Nres$ using the results of our tests at various levels of density contrast, represented in our model caustic by the parameter $\sigma$.  

\begin{figure}[htbp]
\begin{center}
\plottwo{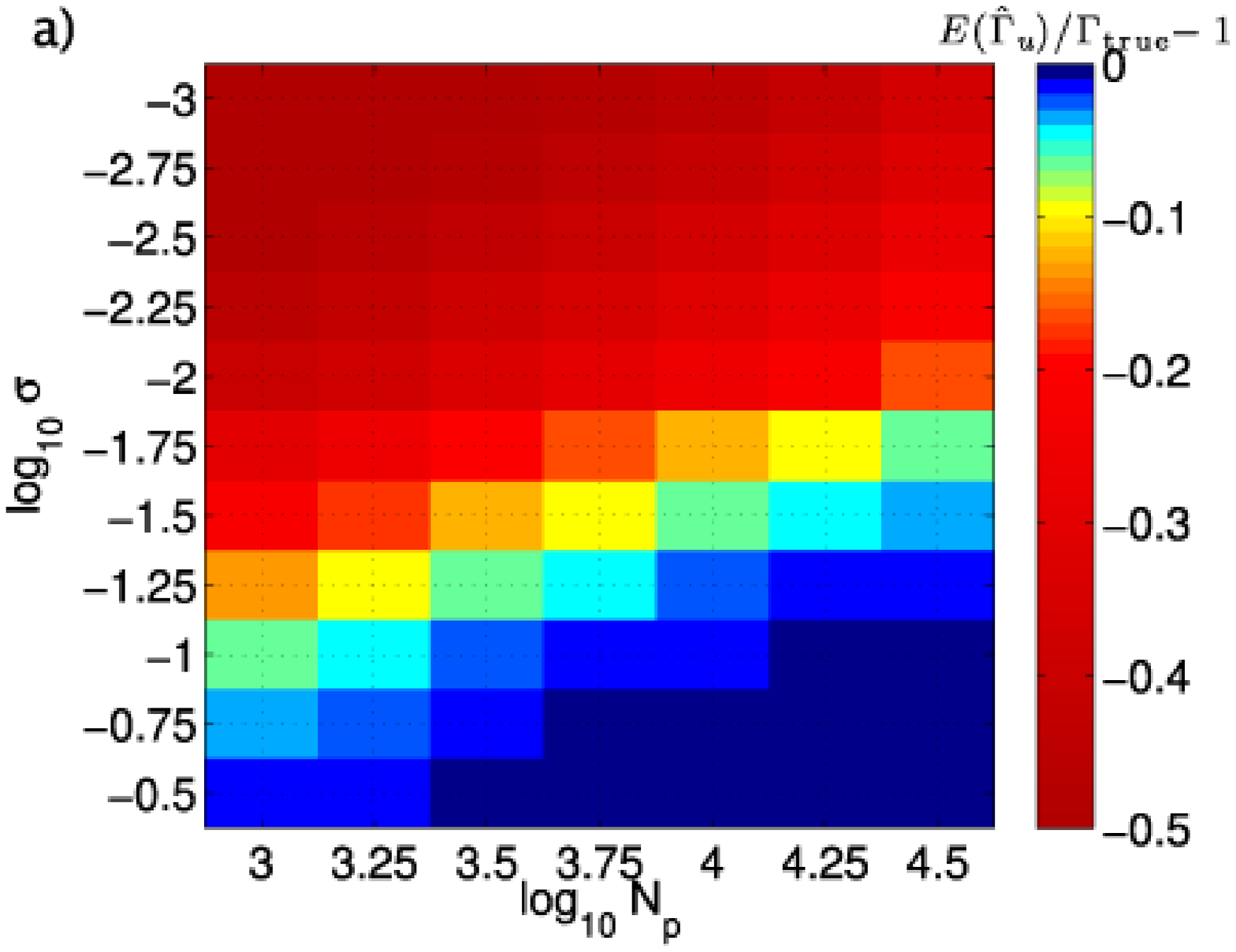}{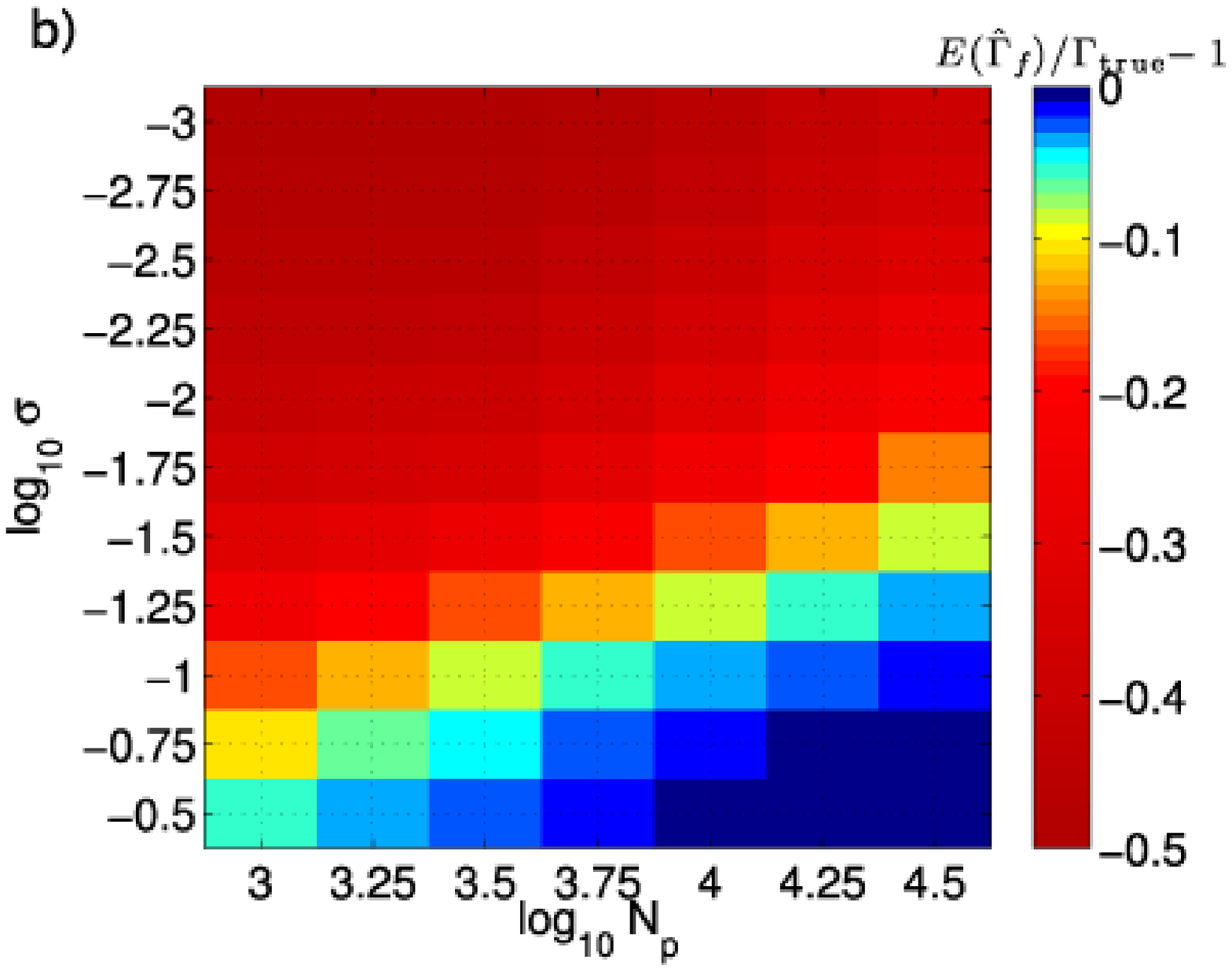}
\caption{(Color online) Using a larger number of particles resolves sharper caustics in all three cases: $\hat{\Gamma}_u$ (a), $\hat{\Gamma}_f$ (b), and $\hat{\Gamma}_n$ (not shown, but very similar to the two others).  Regions colored white (dark or bright blue in the electronic version) are considered fully resolved because the bias is less than the typical standard deviation of about 2 percent.  In the electronic version, different colors indicate the magnitude of the undersampling bias $\bias_{us}$: blue indicates $|\bias_{us}| < 0.02$, aqua $|\bias_{us}| \sim 0.05$, yellow $|\bias_{us}| \sim 0.1$, and red $|\bias_{us}| > 0.2$.  \label{BiasMapU}}
\end{center}
\end{figure}

As is expected, using higher resolution (a larger $\Nres$) leads to better rate estimates of sharper features (Figure \ref{BiasMapU}).  Even the highest-resolution realizations we tested could only resolve moderately sharp features. The estimators using adaptive Riemann volumes (for example, $\hat{\Gamma}_f$ shown in the right panel of Figure \ref{BiasMapU}) appear to require more particles to obtain the same bias as the constant-Riemann-volume case; this is partly due to the fact that in the adaptive scheme there is exactly one volume per particle, but the same number of constant Riemann volumes (about $10^4$) is used regardless of resolution. 

\begin{figure}[htbp]
\begin{center}
\plottwo{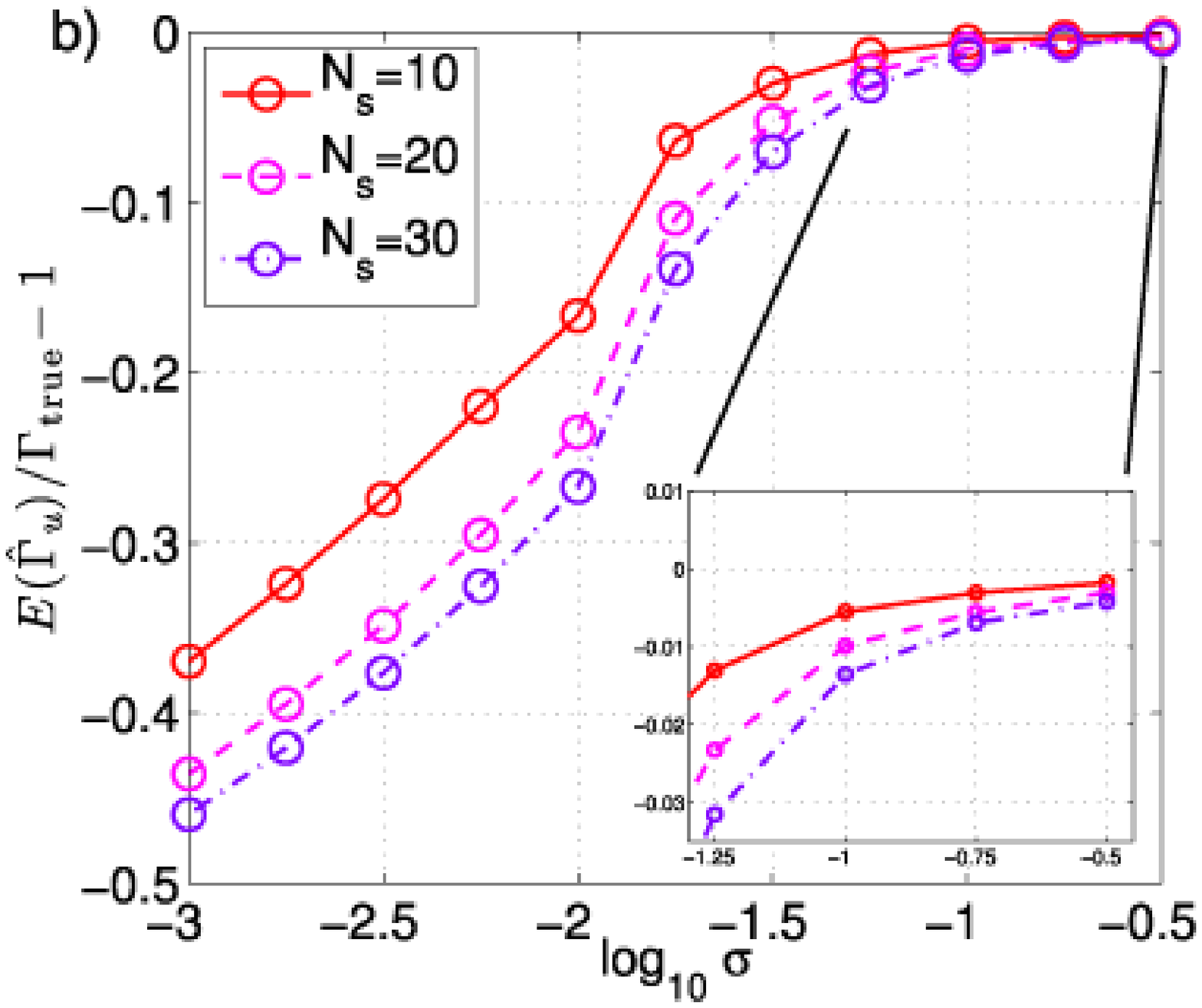}{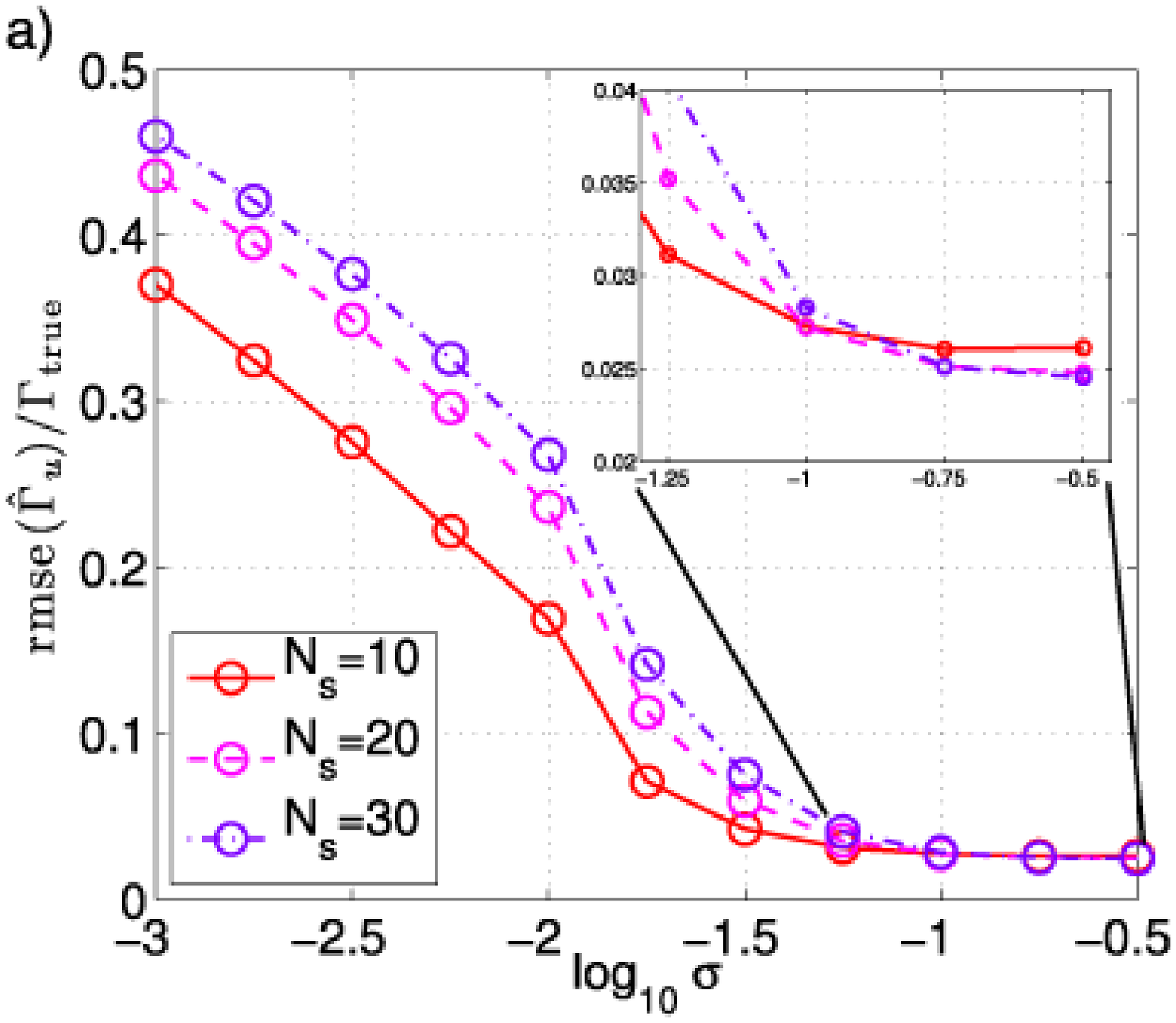}
\caption{(Color online) Using a lower value of $\Nsmooth$ leads to lower bias whether or not the realization fully resolves all the features in the underlying distribution (a, main figure and inset).  The increase in the standard deviation, and therefore RMS error, from using a smaller $\Nsmooth$ is small compared to the improvement in the bias when there are under-resolved features (b).  If the realization is fully resolved, the increase in RMS error is detectable but extremely small (b, inset). \label{NsmoothBehavior}}
\end{center}
\end{figure}

We found that the undersampling bias depends strongly on $\Nsmooth$ when the features of interest are marginally or under-resolved, and only weakly when they are fully resolved (Figure \ref{NsmoothBehavior}, left panel).  As expected, using a smaller $\Nsmooth$ leads to a lower bias because the size of the volume used for density estimation scales with $\Nsmooth$ as shown in Equation \eqref{rnminscaling}, so that a larger $\Nsmooth$ blurs out more small-scale features.  A lower $\Nsmooth$ also leads to a larger standard deviation of the density estimates, which increases the total RMS error, but this effect is small compared to the improvement in the bias in the under-resolved regime and only very slightly affects the fully-resolved case (Figure \ref{NsmoothBehavior}, right panel).  

\begin{figure}[htbp]
\begin{center}
\plotone{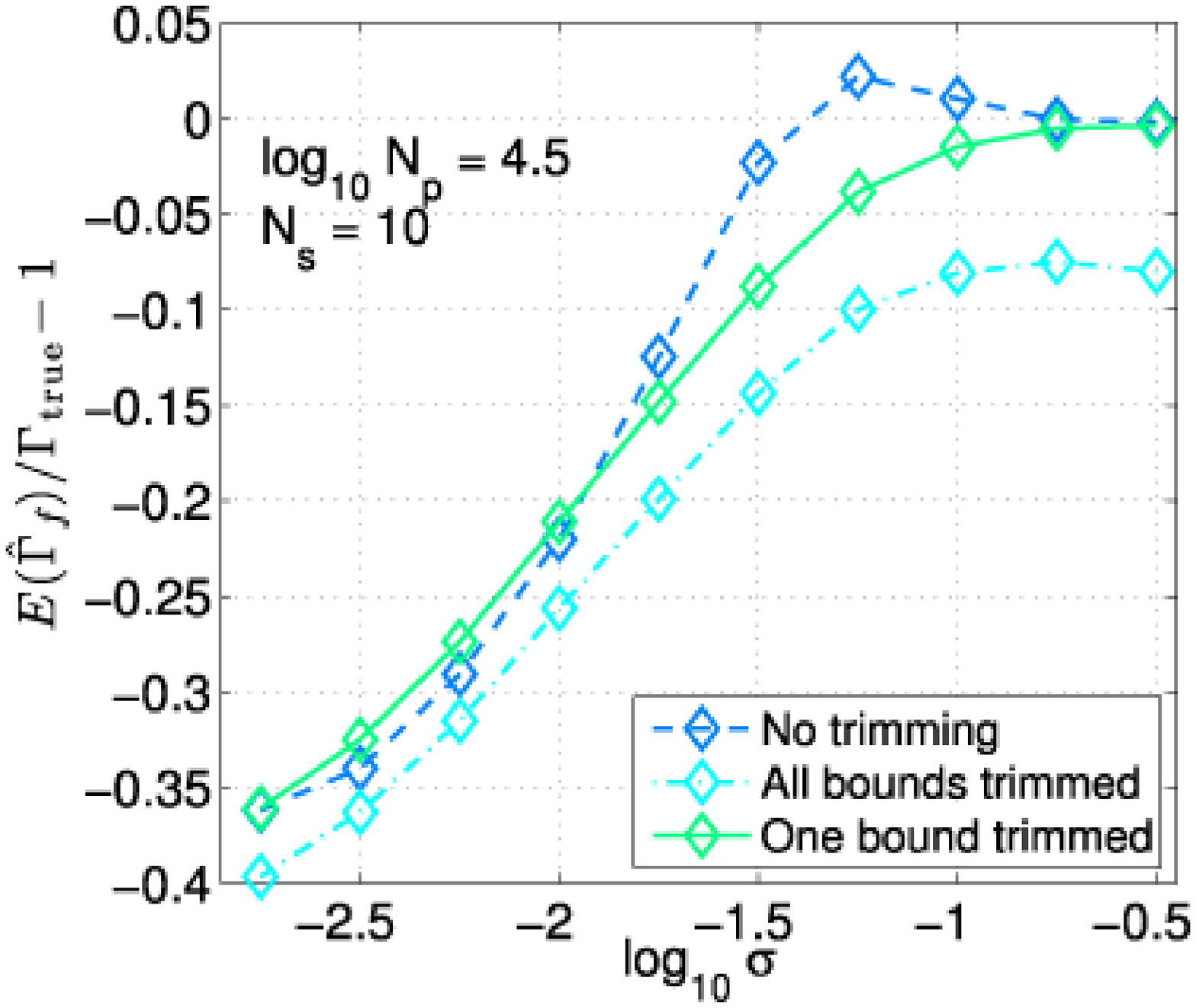}
\caption{(Color online as indicated in square brackets) The adaptive Riemann volume algorithm is sensitive to the treatment of the boundaries of the integration volume relative to the geometry of the density distribution.  Thanks to the location of the caustic parallel to one boundary of the integration volume, the rate is overestimated (solid [blue] lines) at marginal resolutions unless the Riemann volumes on that boundary are trimmed to fit the face of the caustic (dot-dashed [green] lines), but trimming all the boundaries in the same manner underestimates the rate (dashed [cyan] lines). \label{adaptiveBounds}}
\end{center}
\end{figure}

We found that the algorithm used to determine the Riemann volumes adaptively was sensitive to the way in which the boundaries of the integration volume were treated.  Boundaries closest to the sharp edge of the caustic (which in our test cases is parallel to one face of the integration volume) must be trimmed as described in Section 2.2 of \citet{ascasibar:2005aa} to avoid artificially overestimating the rate: without trimming, the Riemann volumes on the face of the caustic next to the boundary are artificially elongated into the region ahead of the sharp edge where the density is effectively zero.   The density estimate in those volumes will then be artificially high because the density is assumed to be constant over the entire Riemann volume, leading to rate estimates with positive bias when the distribution is marginally resolved (Figure \ref{adaptiveBounds}, solid [blue] lines).  Such "bleed-over" still occurs with a constant Riemann volume but is much less significant because the box size does not depend on the local density.   

However, applying the trimming algorithm uniformly to all the Riemann volumes with one or more faces on a boundary of the integration volume artificially \emph{underestimates} the rate by significantly reducing the total integration volume on the trailing edge of the caustic, where the density is small but still nonzero (Figure \ref{adaptiveBounds}, dashed [cyan] lines).  In this system, restricting the trimming only to the face nearest the caustic edge results in the correct bias behavior (Figure \ref{adaptiveBounds}, dot-dashed [green] lines).  Because the choice of how to treat the boundaries appears to depend on the particular geometry of the system in question, it may be difficult to extrapolate the performance of estimators that use the adaptive Riemann volumes from our test case to systems with arbitrary geometry.  In particular, it is not immediately clear how this method would extend to the shells in the dynamical model of M31, which have spherical edges near several boundaries of the integration volume. 

\subsection{Performance of Estimators on the Non-uniform Distribution}
\label{sec:bestPerformance}

\begin{figure}[htbp]
\plotone{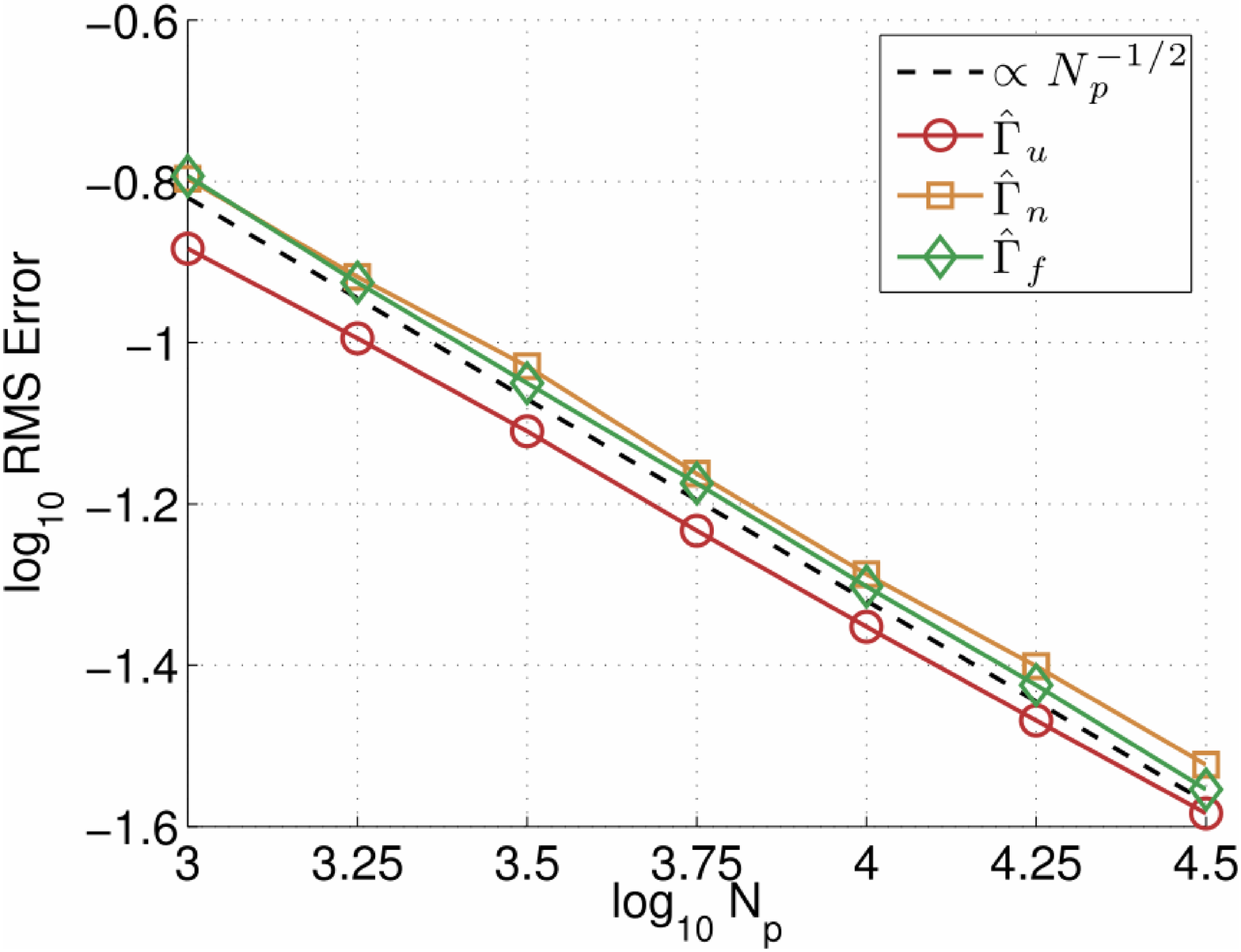}
\caption{(Color online) All three estimators tested on non-uniform distributions achieve a convergence rate of $N_p^{-1/2}$ (dashed line) as predicted by prior analytical work.  Tests with $\Nsmooth = 10$ and $\log_{10}\sigma = -0.75$ are shown.  \label{fig:convergenceRate}}
\end{figure}

The best estimator; that is, the one with the smallest RMS error, has the best combination of undersampling bias near zero and small standard deviation for the smallest $\sigma$.  In the resolution-limited regime the RMS error is dominated by the bias.  If the caustic is fully resolved, the bias is zero and the standard deviation, which is constant for a given \Nres, dominates the RMS error.  We have already established that a smaller $\Nsmooth$ improves performance in the under-resolved regime without substantially increasing the RMS error for resolved distributions.  Figure \ref{fig:convergenceRate} shows that all three estimators achieve the convergence rate of $N_p^{-1/2}$ predicted by \citet{bickel1988} and \citet{gine2008a} for sharpnesses that are fully resolved at the highest resolution.  

\begin{figure}[htb]
\plotone{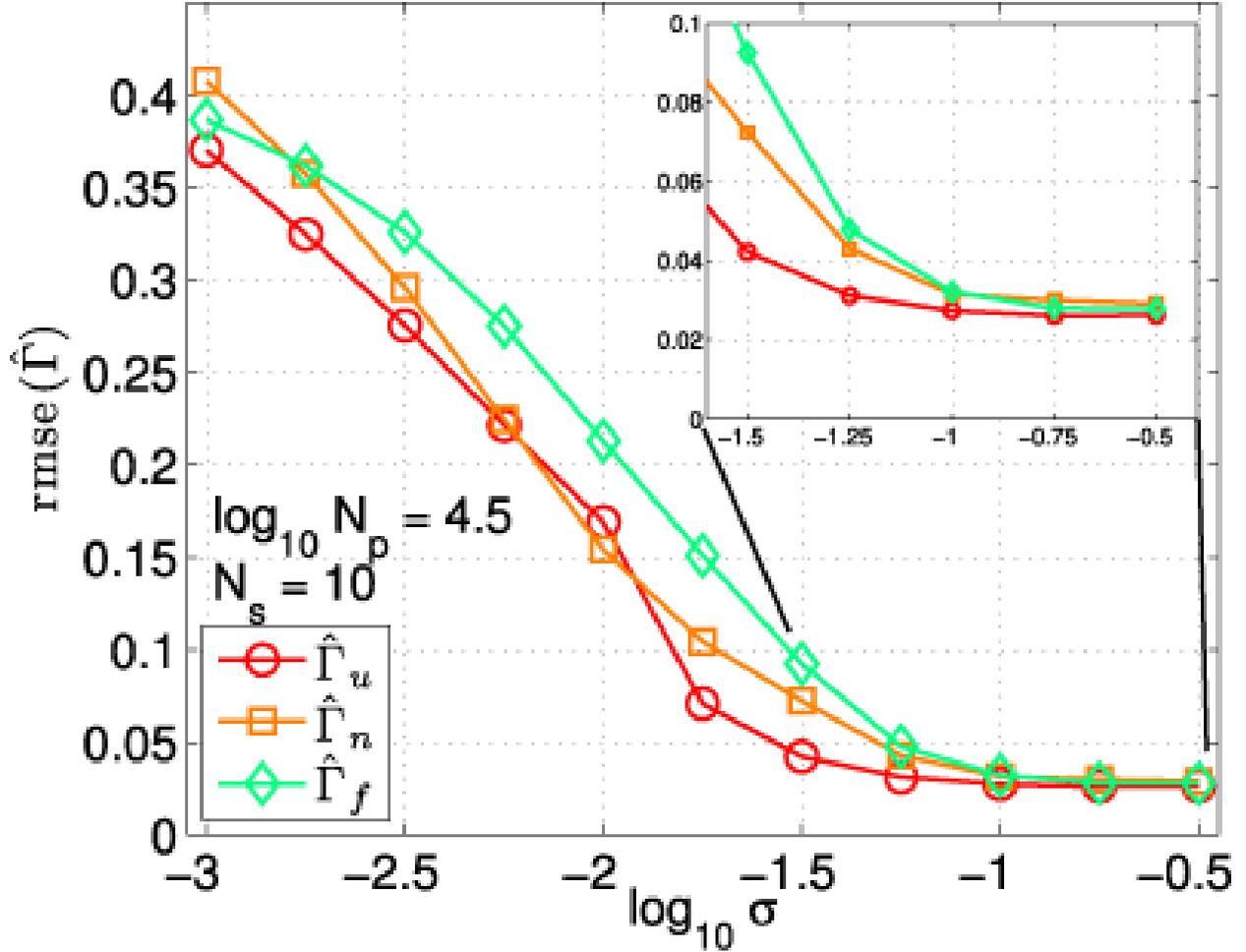}
\caption{(Color online as indicated in square brackets) All the estimators have very similar performance.  $\hat{\Gamma}_u$ ([red] circles) converges faster in the marginally-resolved regime than the estimators that use adaptive Riemann volumes ([orange] squares and [green] diamonds).  In the fully resolved regime the RMS errors of the three are nearly identical (inset). \label{rmserror}}
\end{figure}

Comparing the RMS errors for sharper and sharper caustics shows that all three tested estimators have very similar performance (Figure \ref{rmserror}).  The nearest-neighbors estimator with constant Riemann volume ([red] circles in Figure \ref{rmserror}) converges slightly faster than the other two but once the distribution is resolved they are nearly indistinguishable from one another (Figure \ref{rmserror}, inset).  The FiEstAS method, thanks to the space-filling tree organizing the particles, is faster than the nearest-neighbors estimators, so if the distribution is known to be completely resolved (the regime shown in the inset of Figure \ref{rmserror}), then this method can be used without loss of performance to take advantage of its greater speed.  However, in situations where parts of the distribution may be under-resolved and creating a higher-resolution realization is not possible, $\hat{\Gamma}_u$ should be used to take advantage of its ability to resolve slightly sharper features with fewer particles.

\clearpage
\section{Calculation of the Boost Factor and Gamma-ray Flux}
\label{sec:m31}

In this section we describe the N-body model of the M31 tidal debris (Section \ref{sec:nbm}) and present derivations of formulae for the boost factor and gamma-ray flux (Section \ref{sec:calc}) in terms of the numerically estimated rate.  We then describe how we used the results of the bias calibrations described in Section \ref{sec:stats} to correct our numerical estimates of the boost factor and flux (Section \ref{sec:calib}), and present the results (Sections \ref{sec:boost} and \ref{sec:signal}).

\subsection{N-body Model}
\label{sec:nbm}

For this work we use the N-body model of the tidal shell system in \citet{fardal:2007aa}.  The model uses a Plummer sphere as the progenitor of the tidal debris, orbiting in a static, 3-component representation of M31's potential: a spherical halo and bulge, and an axisymmetric exponential disk.  To construct a model of the potential, the parameters of the halo, bulge, and disk were first fit to a rotation curve of M31 from several combined sources, not including the tidal shells and the associated tidal stream \citep[see][for details]{geehan:2006aa}.  Then the orbit of the center of mass of the progenitor satellite was fit to the three-dimensional position data and radial velocity measurements available for the stream \citep{fardal:2006aa}.  Finally, the mass and size of the progenitor were constrained with an N-body model of the stream using the previously determined orbit \citep{fardal:2007aa}.  Dynamical friction and the response of M31 to the merger are ignored since the mass ratio of the progenitor to M31 is approximately 1/500.    \citet{fardal:2007aa} emphasize that the N-body model is not the result of a full exploration of this many-dimensional parameter space, but for this work we are only interested in the end configuration of the debris, which acceptably matches the available observations.

We further make the assumption that there is a comparable mass of dark matter associated with the stellar tidal debris.  This assumption is probably generous, since the dark matter in galaxies is thought to be much more diffuse than the stellar matter and much of it will have been stripped away by tides before the progenitor of this tidal debris even reaches the starting point of the simulation.   However, for a first estimate we consider this assumption sufficient, though it is not a strict upper limit for reasons we will discuss at the end of this paper.

\subsection{Formulae}
\label{sec:calc}

In this section we derive expressions for the boost factor and gamma-ray flux in terms of the estimated rate from the N-body representation.

\subsubsection{Boost Factor}

To calculate the boost factor, we must assume a halo model because dark matter in the shell can interact with dark matter in the halo.  Denoting the halo dark matter with $h$ and the shell dark matter with $s$, there are three terms in the total rate:
\begin{eqnarray}
\Gamma &=& \int n_{\mathrm{tot}}^{2} dV \nonumber \\
&=& \int (n_h + n_s)^{2} dV \nonumber \\
&=& \int (n_h^{2} + 2 n_h n_s + n_s^{2}) dV, \nonumber \\
&\equiv&\Gamma_{hh} + \Gamma_{sh} + \Gamma_{ss}
\label{eq:totalrate}
\end{eqnarray}
Equation \eqref{eq:totalrate} shows that the boost factor $\beta$ depends on the halo dark matter density:
\begin{eqnarray}
\beta &\equiv& \frac{\Gamma_{\mathrm{tot}} - \Gamma_{hh}}{\Gamma_{hh}} =   \frac{1}{\Gamma_{hh}}  \int n_s ( 2n_h + n_s) dV.\nonumber \\
\label{fluxContributions}
\end{eqnarray}
If $n_h \gtrsim n_s$, the additional emissivity from the tidal debris is dominated by the first term in equation \eqref{fluxContributions} and the boost factor scales linearly with the dark matter density in both the halo and the shell.

For the halo, we use the same density distribution that was used in the dynamical model of the shells \citep{geehan:2006aa} with the addition of a small core with size $r_{\mathrm{core}}$ of half the size of one Riemann volume.   The core eliminates the infinite-density cusp at $r=0$. This halo is spherically symmetric and of Navarro-Frenk-White form \citep{navarro:1996aa, 1997ApJ...490..493N}:
\begin{equation}
\label{eq:analyticHaloDensity}
n_{h}(r) = \frac{n_{h,0}}{[(r+r_{\mathrm{core}})/r_h][1+(r + r_{\mathrm{core}})/r_h]^2}
\end{equation}
where $n_{h,0}  \equiv \rho_{h,0}/m_p = 3.67 \times 10^3\ \mathrm{kpc}^{-3} $ and $r_h = 7.63$ kpc are determined by \citeauthor{geehan:2006aa} by fitting a mass model to a set of measurements of dynamical tracers of M31's halo. We divide the fitted mass density $\rho_{h,0}$ by $m_p$, the mass of the simulation particles, to get consistent number densities $n_h$ and $n_s$.  

\subsubsection{Gamma-ray Flux}
We use the notation of \citet[hereafter FPS]{fornengo:2004aa} to present the calculation of the gamma-ray flux.   The differential flux of photons in an infinitesimal band of photon energy $E_{\gamma}$, $d\Phi_{\gamma}/dE_{\gamma}$, can be factored into a contribution from ``particle physics" that specifies the spectrum of the radiation and an achromatic contribution $\Phi^{\mathrm{cosmo}}$ from ``cosmology"---the shape, size, density, and distance of the dark matter---that sets the normalization, as in eq. 1 of FPS:
\begin{equation}
\frac{d\Phi_{\gamma}}{dE_{\gamma}} = %
\frac{d\Phi^{\mathrm{SUSY}}}{dE_{\gamma}} %
\Phi^{\mathrm{cosmo}}
\end{equation}
The rate at which gamma rays would be detected by the Fermi LAT is
\begin{equation}
\label{eq:phigamma}
\frac{dN_{\gamma}}{dt} = %
\Phi^{\mathrm{cosmo}} %
\int_{E_{\mathrm{th}}}^{m_{\chi}} \frac{d\Phi^{\mathrm{SUSY}}(E_{\gamma})}{dE_{\gamma}} A_{\mathrm{eff}}(E_{\gamma}) dE_{\gamma}
\end{equation}
where $A_{\mathrm{eff}}(E_{\gamma})$ is the effective area of the detector, $E_{\mathrm{th}}$ is the lowest detectable energy, and $m_{\chi}$ is the dark matter mass.  For the Fermi LAT, $E_{\mathrm{th}}$ is 30 MeV, and above 1 GeV the effective area for diffuse events is roughly independent of energy to within about 10 percent of the mean value over the energy range of integration \citep{2009arXiv0907.0626R}.  Furthermore, the supersymmetric calculations of the particle physics contribution in \citet{2004JCAP...07..008G}, which we use for this work, take $E_{\mathrm{th}} = 1$ GeV.  So for the remainder of this work we will use $E_{\mathrm{th}} = 1$ GeV and assume $A_{\mathrm{eff}}$ is independent of energy.  With this simplification we can work with the total flux for now and later multiply it by $A_{\mathrm{eff}}$ to get the detection rate in the LAT.

Also following FPS, the particle-physics contribution is
\begin{equation}
\label{eq:dphisusydE}
\frac{d\Phi^{\mathrm{SUSY}}}{dE_{\gamma}} 
= \frac{\left< \sigma v\right>}{2 m_{\chi}^{2}} \frac{dN_{\gamma}}{dE_{\gamma}}
\end{equation}
where $\left< \sigma v\right>$ is the velocity-averaged cross section and $N_{\gamma}$ is the yield, called $Q_{\gamma}$ in \cite{peirani:2004aa} and FPS.  We have moved the constant factor $1/4\pi$ to be part of $\Phi^{\mathrm{cosmo}}$ because it is most easily understood as part of the attenuation of the gamma-ray flux over distance.

To evaluate $N_{\gamma}$, we use a subset of the benchmark models of \citet{battaglia-2004-33} to span the space of supersymmetric WIMP candidates.  \citet{2004JCAP...07..008G} have calculated the gamma-ray yields above 1 GeV for 10 of the 12 models in \citet{battaglia-2004-33}: models $A'$ through $L'$ with the exception of models $E'$ and $F'$.   We use the values given in the table for the number of photons in the continuum emission times the cross section, $N_{\gamma, \mathrm{cont.}} \left<\sigma v\right>$, listed in Table 1 of \citet{2004JCAP...07..008G}.  The continuum emission is not as diagnostic as the line emission at $E_{\gamma}=m_{\chi}/2$, but the branching ratio for line emission is smaller by a factor of $10^3$.  

Compared to Equations \eqref{eq:phigamma} and \eqref{eq:dphisusydE}, we find that
\begin{equation}
\label{eq:phisusy}
\Phi^{\mathrm{SUSY}} = \int_{E_{\mathrm{th}}}^{m_{\chi}} \frac{d\Phi^{\mathrm{SUSY}}}{dE_{\gamma}} dE_{\gamma} = \frac{N_{\gamma, \mathrm{cont.}} \left<\sigma v\right>}{2 m_{\chi}^2} 
\end{equation}
For the models in \citet{2004JCAP...07..008G}, $N_{\gamma, \mathrm{cont.}} \left<\sigma v\right>$ is in the range $10^{-29}$---$10^{-24}$ cm${}^3$ s${}^{-1}$, and $m_{\chi}$ is generally given in GeV.  So a useful scaling of this formula in typical units is
\begin{eqnarray}
\label{eq:PhiSUSYfiducial}
\Phi^{\mathrm{SUSY}} &=& 1.54 \times 10^{-8} \textrm{cm}^4 \textrm{kpc}^{-1}  \textrm{s}^{-1} \textrm{GeV}^{-2} \nonumber \\
&\times & \left(\frac{N_{\gamma\ \mathrm{cont.}} (\sigma v)}{10^{-29} \textrm{cm}^{-3} \textrm{s}^{-1}} \right)  \left( \frac{m_{\chi}}{1 \textrm{GeV}}\right)^{-2}.\nonumber 
\\
\end{eqnarray}

In addition to the benchmarks, we also use the most optimistic value of $\Phi^{\mathrm{SUSY}}$ from FPS.  According to their Figure 8, $\Phi^{\mathrm{SUSY}}\lesssim 10^{-8}$ for all the models explored, and the maximum occurs for $m_{\chi} \approx 40$ GeV or $N_{\gamma, \mathrm{cont.}} \left<\sigma v\right> \approx 10^{-25}$ cm${}^{3}$ s${}^{-1}$.  We use these values to represent the most optimistic estimate of the flux.

The astrophysical factor $\Phi^{\mathrm{cosmo}}$ depends on the square of the dark matter mass density $\rho_{\chi}$:
\begin{equation}
\label{eq:phiCosmoMass}
\Phi^{\mathrm{cosmo}} = \frac{1}{4\pi d^{2}} \int_{\mathrm{obj}} dV \rho^{2}_{\chi}(x,y, z) 
\end{equation}
where $x,y,z$ indicate physical distances in a coordinate system centered on the object, and the integral is over the total volume of the object.  The factor $1/4\pi d^{2}$ accounts for the attenuation in flux over the distance $d$ from the object to the observer.  The key element in calculating the astrophysical contribution to the flux is thus determining the integral-density-squared $\int \rho^{2} dV$.  We calculate this quantity from the simulation results, which consist of the locations of the $\Nres$ simulation particles, each with mass $m_{p}$.  A numerical density estimator gives the number density $n_{p}$ of the simulation particles as a function of position, which is related to the mass density $\rho_{\chi}$ by mass conservation:
\begin{equation}
\rho_{\chi} = m_{\chi} n_{\chi} = m_{p} n_{p}
\end{equation}
so that
\begin{equation}
\int \rho_{\chi}^{2} dV = m_{p}^{2} \int n_{p}^{2} dV
\end{equation}
and
\begin{equation}
\label{eq:phiCosmoNumber}
\Phi^{\mathrm{cosmo}} = \frac{m_p^2}{4\pi d^2} \int n_{p}^{2} dV \equiv  \frac{m_p^2}{4\pi d^2} E(\hat{\Gamma}_u).
\end{equation}

It is useful to rewrite the expression \eqref{eq:phiCosmoNumber} with the units and values used in the N-body representation.  For the simulation of \cite{fardal:2007aa}, $m_{p} = 1.68 \times 10^{4} M_{\astrosun}$ and $d = 785$ kpc.  $E(\hat{\Gamma})$ is calculated in units of $\textrm{kpc}^{-3}$.  So a useful version of \eqref{eq:phiCosmoNumber} for this work is
\begin{eqnarray}
\label{eq:phiCosmoFiducial}
\Phi^{\mathrm{cosmo}} &=& (1.87 \times 10^{-14} \textrm{ GeV}^{2} \textrm{ kpc cm}^{-6}) \nonumber \\
&\times& \left(\frac{m_{p}}{10^{4} M_{\astrosun}}\right)^{2} \left(\frac{d}{785\ \mathrm{kpc}}\right)^{-2}\nonumber \\
&\times& \left(\frac{E(\hat{\Gamma})}{\textrm{kpc}^{-3}}\right) \nonumber \\
\end{eqnarray}

The total flux of gamma rays for a given model of dark matter and density distribution, scaled to typical values in the problem, is obtained by combining equations \eqref{eq:PhiSUSYfiducial} and \eqref{eq:phiCosmoFiducial}:
\begin{eqnarray}
\label{eq:phitotal}
\Phi_{\gamma} &=& \Phi_{\gamma,0}  \left( \frac{N_{\gamma}(\sigma v)}{10^{-29} \textrm{cm}^{3} \textrm{s}^{-1} } \right) \nonumber \\ %
&\times& \left( \frac{m_{\chi}}{1 \textrm{GeV}} \right)^{-2}  \left(\frac{m_{p}}{10^{4} M_{\astrosun}}\right)^{2} \nonumber \\
&\times& \left(\frac{d}{785\ \mathrm{kpc}}\right)^{-2}  \left(\frac{E(\hat{\Gamma})}{\textrm{kpc}^{-3}}\right),
\end{eqnarray}
where $\Phi_{\gamma,0} = 2.88 \times 10^{-22}$ cm${}^{-2}$ s${}^{-1} = 9.09\times10^{-11}$ m${}^{-2}$ yr${}^{-1}$.  The effective area of the Fermi LAT varies between 0.7--0.85 square meter above 1 GeV \citep{2009arXiv0907.0626R}.

\subsection{Calibration of the Rate Estimate}
\label{sec:calib}

\begin{figure}[htbp]
\begin{center}
\plotone{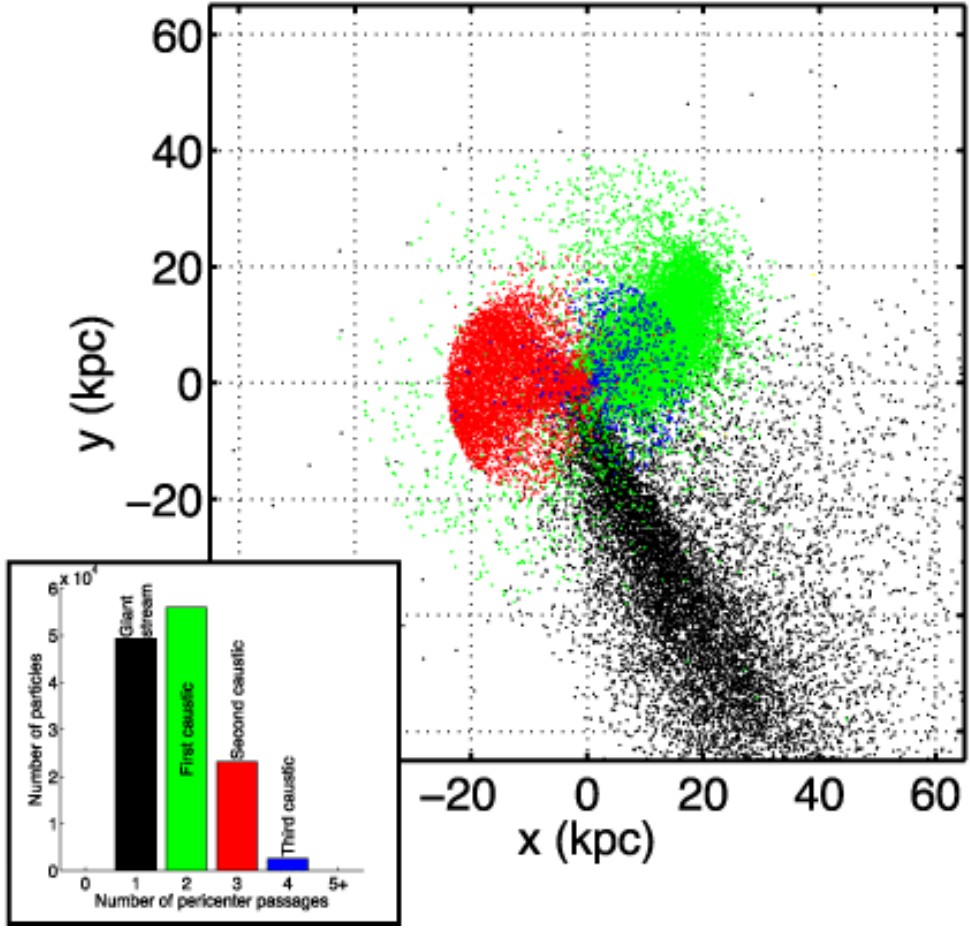}
\caption{(Color online) Sorting particles by the number of pericenter passages, $\nperi$ (the inset gives the shading [color in electronic version] scheme used here), easily identifies the main dynamical structures in the tidal debris.  This $x$--$y$ projection shows the N-body model as it would be seen from Earth, with $x$ and $y$ measured relative to M31's center and aligned with the east and north directions on the sky, respectively.  In this projection the edge of the younger (medium gray, red online) shell is sharper than that of the older (dark gray, green online) shell, but both are nearly spherical relative to M31's center. \label{colorShellsxy}}
\end{center}
\end{figure}

\begin{figure}[htbp]
\begin{center}
\plotone{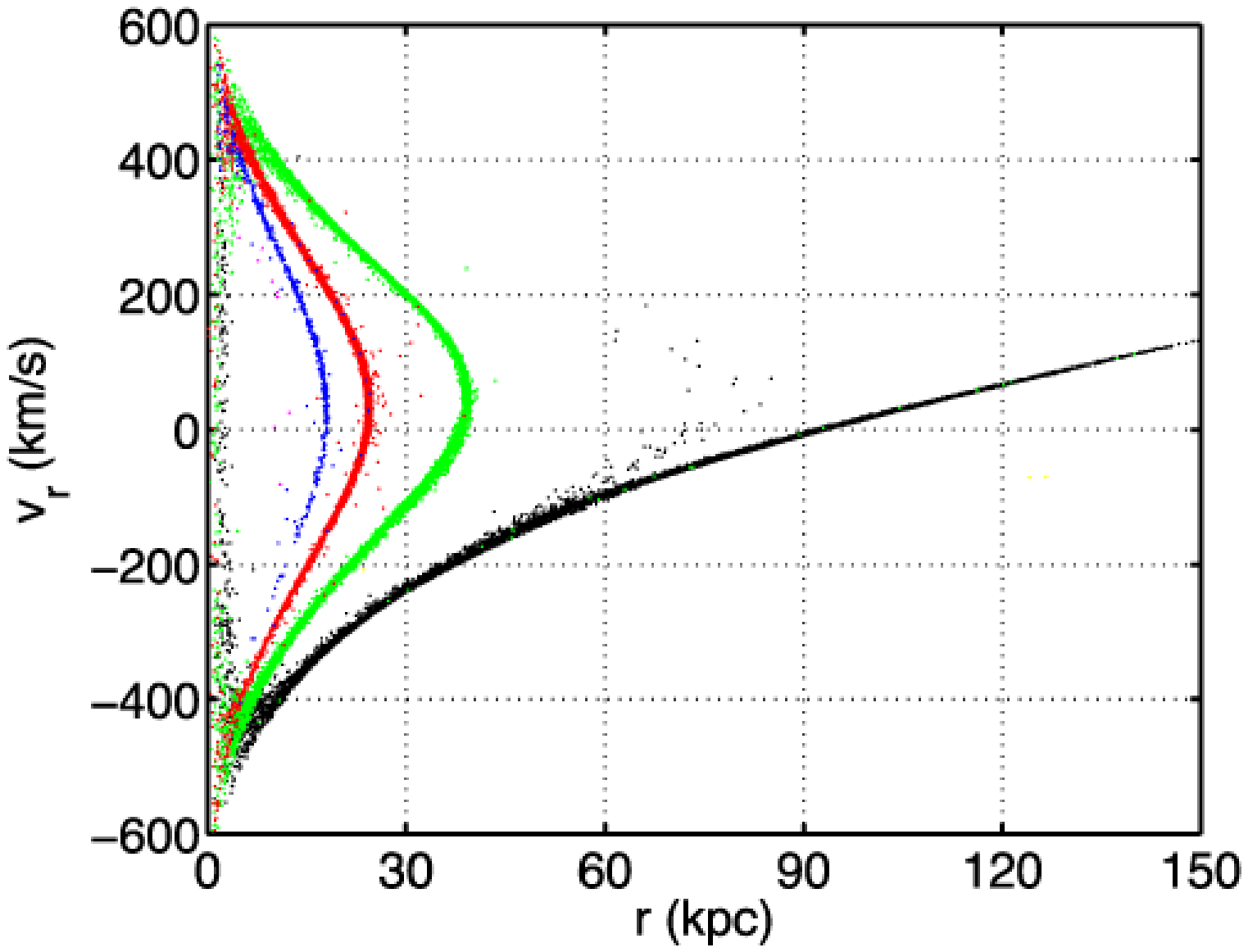}
\caption{(Color online) Looking at the projected phase space of $r$ and $v_r$, measured relative to M31's center, lets us determine the order in which the shells formed.  In the self-similar model, the outermost caustic forms first: we see here that this corresponds to the bound particles with lowest $\nperi$ (shading/colors are the same as in Figure \ref{colorShellsxy}).  The youngest caustic contains the particles that have undergone the largest number of orbits.  \label{colorShellsrvr}}
\end{center}
\end{figure}

To calibrate the result from the density-squared calculation, we must estimate the number of particles in each shell (and thereby $\Nres$) and the $\sigma$ that gives the best approximation to each caustic's shape.  Since the shells are located at the apocenters of the particles' orbits, we can unambiguously determine which particles are in which shell by counting the number of pericenter passages, $\nperi$, that each particle has experienced (Figure \ref{colorShellsxy}). There are three shells and one tidal stream in the system, each composed of particles with a different $\nperi$.  Examining the system in a projection of phase space in the $r$-$v_r$ plane, shown in Figure \ref{colorShellsrvr}, shows the order in which the shells were formed.  The first caustic to form corresponds to the first (outermost) winding in phase space with the largest apocenter.  Particles making up this caustic at a given moment have the lowest $\nperi$ because this caustic marks the location of the first turnaround point for bound particles.  The most recently formed caustic has the highest number of pericenters and smallest apocenter distance, and has not yet been fully filled: the innermost winding is not yet complete because only a few particles have had time to complete three full orbits.  We chose the two oldest shells for analysis because they contain most of the mass.  We will refer to the oldest caustic, whose constituent particles are shown in green and have undergone two pericenter passages, as caustic 1 or shell 1.  The second caustic, whose constituent particles are shown in red and have undergone three pericenter passages, will be referred to as caustic 2 or shell 2.

\begin{figure}[htbp]
\begin{center}
\plotone{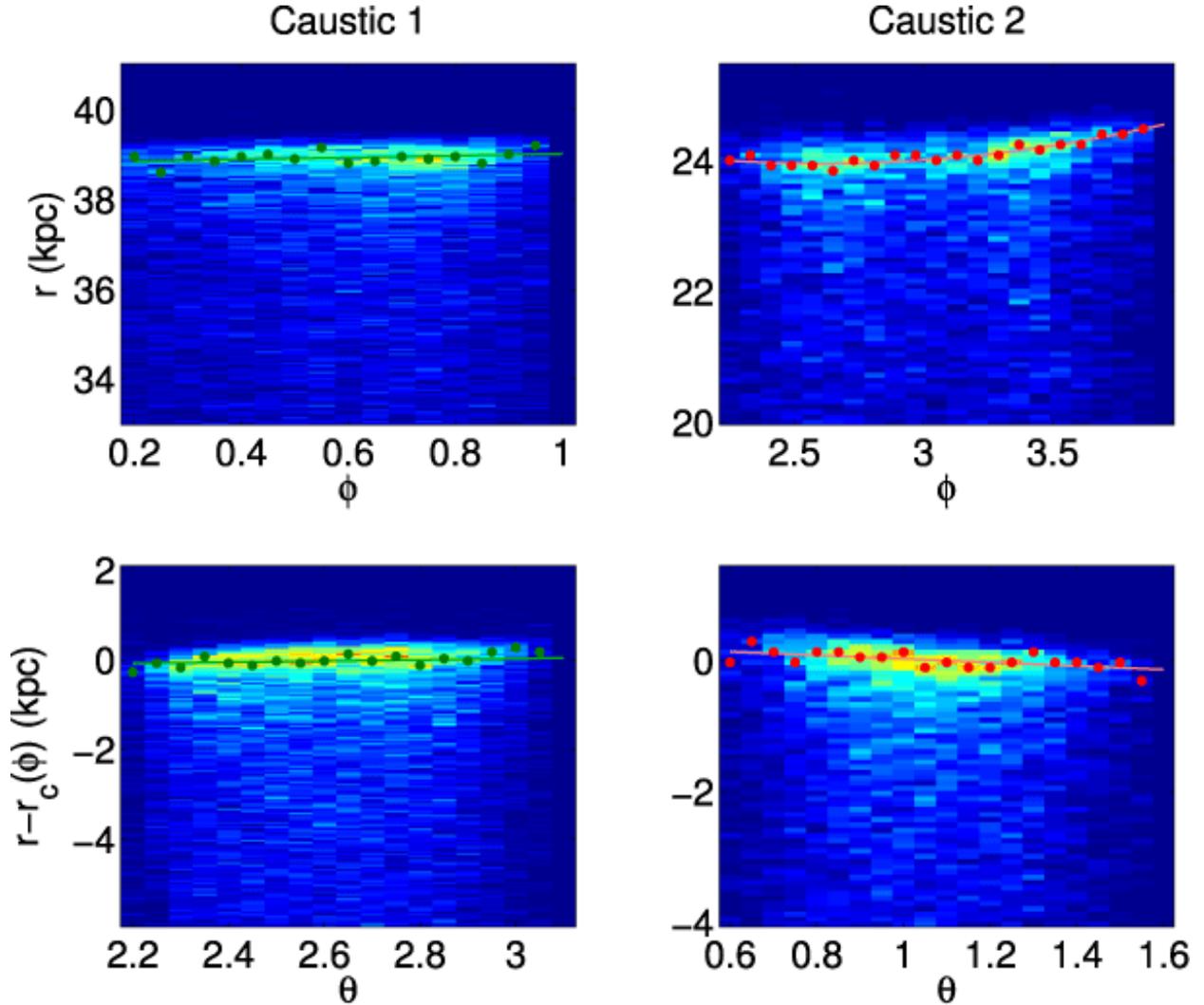}
\caption{(Color online) Each caustic surface was corrected for asphericity by fitting the position of the peak bin in $r$ as a function of angle, first in the $\phi$ direction (top row) and then in the $\theta$ direction (bottom row).  The points show the peak bin location and the lines indicate the best fit for the position of the caustic surface.  In the background is the two-dimensional binned density map of the caustic.  A linear fit was sufficient for all but the $\phi$-dependence of Caustic 2 (top right panel) which used a quadratic fit.  \label{caustic2phi}}
\end{center}
\end{figure}

Since the satellite galaxy's orbit is nearly radial, the caustics are nearly spherical.  However, slight systematic deviations from spherical symmetry, in projection along $r$, can slightly increase the perceived width of the caustic and cause $\sigma$ to be overestimated.  In order to use our analytical caustic to estimate $\sigma$ for each shell in M31 and determine the undersampling correction to the result from the density estimator, we had to correct for the slight asphericity in each shell.  To do this we determined the caustic radius of each shell in projection along the $\phi$ direction by binning the particles in $r$ and $\phi$ and finding the r-bin with the highest number of counts for each slice in $\phi$.  The bins were chosen as small as possible for resolution while still being able to identify the peak $r$ bin for each $\phi$.  Once the peak $r$ was obtained as a function of $\phi$, we fit this set of points with a polynomial $r_c(\phi)$ and calculated $dr = r-r_c(\phi)$ for each particle to correct for the asphericity of the shell in $\phi$.  The polynomial was of the lowest order possible, since higher orders introduce more spurious spread at the edges of the fitted region.  In all but one case a linear fit was sufficient.  The process was repeated with the corrected particle radii, $dr$, in the $\theta$ direction to find and subtract $dr_c(\theta)$.  Figure \ref{caustic2phi} illustrates this process.  

Correcting for the asphericity in this manner determines the radius $r_c(\theta,\phi)$ of the caustic surface in two steps, so that 
\begin{equation}
\label{rcFitFunction}
r_c(\theta,\phi) \equiv r_c(\phi) + dr_c(\theta).
\end{equation}
The bin widths in $r$ used for caustics 1 and 2 in this procedure limit the accuracy of $r_c(\theta,\phi)$ to $0.05$ and $0.08$ kpc, respectively.  However, fitting the radial profile with the analytical caustic determines the caustic radius more accurately, assuming that $r_c(\theta,\phi)$ has sufficiently corrected the asphericity.

After correcting the caustics for asphericity, we binned the particles in $x \equiv r-r_c(\theta,\phi)$ to construct a radial density profile.  We used the Wand rule \citep{Wand96data-basedchoice} to choose a starting bin width and fit Equation \eqref{eq:generalDP} to the density profile to get baseline parameters, then decreased the bin width until the fit parameters converged.  The final bin width was 0.1 kpc for both caustics, larger than the bins used to correct for the asphericity, indicating that $r_c(\theta,\phi)$ adequately represents the caustic surface.  Using smaller bins produced no appreciable change in the fitted parameters.

Having determined the ideal bin width, we fit the model caustic of Equation \eqref{eq:generalDP} to each density profile.   The model has a total of four parameters: the caustic width $\delta r$, $x_c$, the initial phase space density $f_0$, and the phase-space curvature $\kappa$, that are discussed in detail in Appendix \ref{appx:oneDcaustic}.   As noted there, this density profile is universal for caustics created by quasi-radial infall of initially virialized material.  We allowed all the parameters except $\kappa$ to vary in the fit, but checked the fitted values of $x_c$, $\delta r$ and $f_0$ by comparing them to estimates.  For $x_c$ this is trivial; it should be close to zero after the asphericity correction.  We used mass conservation to estimate $f_0$, since the number of particles in the caustic and its geometry are both known, and energy conservation to estimate $\delta r$, using Equation \eqref{eq:EstimatingDeltaR} and about 10-20 percent of the particles in each shell just around the caustic peak.  For reference, the covariance term in Equation \eqref{eq:energyConsvVariances} was two orders of magnitude smaller than the other terms.  

For each caustic, we determined $\kappa$ by fitting a parabola to the phase-space profile near the peak.  The portion of data to fit was determined by the range over which the parabola was a good fit to the phase-space profile.  As a check, we also estimated $\kappa$ using the local gravity at each caustic, using Equation \eqref{eq:kappaFromGravity}.  Both are shown in the table and agree with each other.  Keeping the value of $\kappa$ constant, we then fit the density profile of the caustic, comparing the fitted values of $f_0$ and $x_c$ with the estimated or expected values.  The results of the fitting are shown in Table \ref{tbl:shellfits}.

\begin{figure}[htbp]
\begin{center}
\plotone{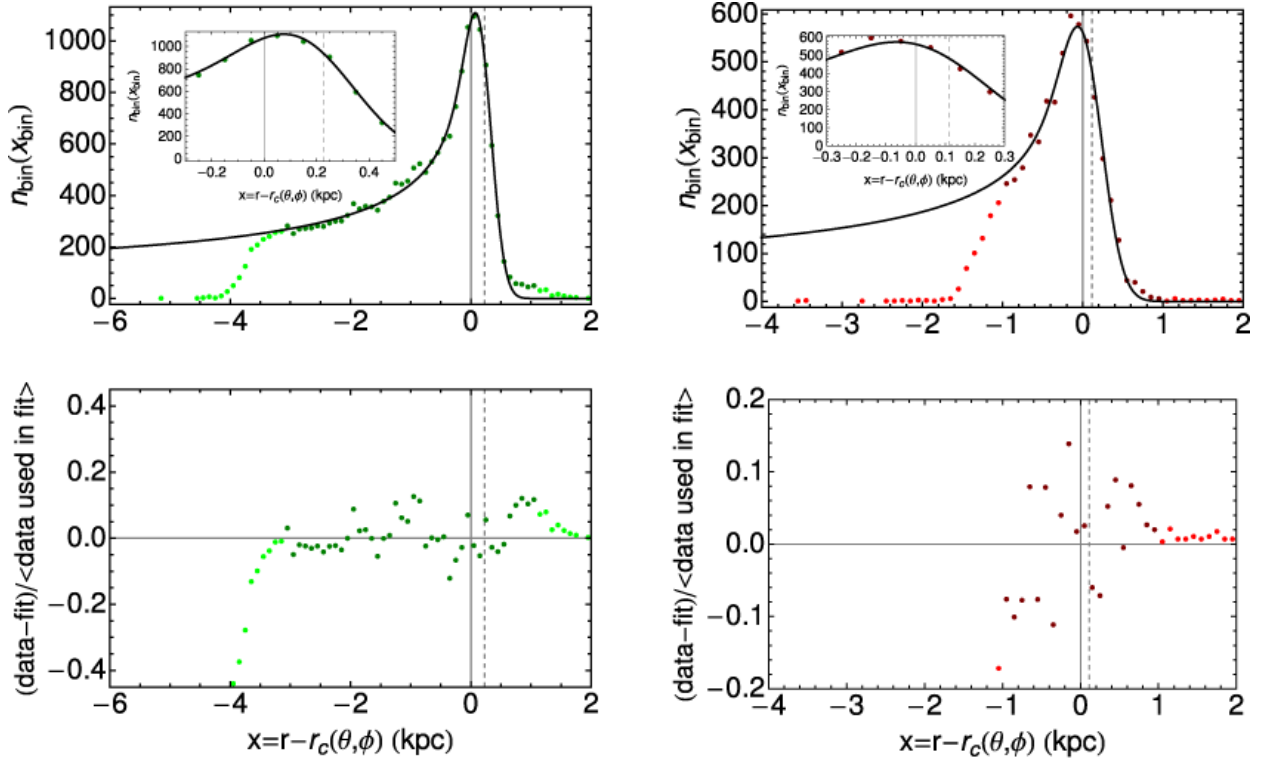}
\caption{(Color online) The radial density profiles of caustics 1 (left upper panel; green in electronic edition) and 2 (right upper panel; red in electronic edition) can be fit surprisingly well with the functional form in Equation \eqref{analyticCaustic1} after they have been corrected for asphericity with the process illustrated in Figure \ref{caustic2phi}.  The insets in the upper panels show how well the function fits the region right around the peak of each caustic, which is the most important region for determining the true width.   The residuals (lower panels) from the data used in the fit (shown in a darker shade) are evenly scattered around zero in each case, indicating that the peak's relative height is accurately determined.  The fit parameters are given in Table \ref{tbl:shellfits}.  \label{shellfits}}
\end{center}
\end{figure}

As expected, the model fits the data well all the way through the peak in both cases (Figure \ref{shellfits}).  The abrupt drop in density behind the caustic (at negative $x$) is an artifact of selecting the particles via their phase-space profile.  In both cases $x_c$, the correction to $r_c$ from the fit (dashed line), is within a few bins of the value from the asphericity correction (solid line at zero).  $x_c$ is not positioned exactly at the peak of the caustic because the satellite is not totally cold (see Appendix \ref{appx:oneDcaustic} for a more detailed explanation).  The widths of the two caustics are close, but not identical, reflecting the slight difference in the initial velocity dispersions of the particles creating them.  The estimate of $\delta r$ for caustic 1 is much closer than that for caustic 2; this could be because the second caustic is at about half the distance of the first and is thus more affected by the non-spherical portions of the potential.  The spread of energies with $r$ in caustic 2 is certainly both larger and less symmetric than in caustic 1.  Both estimates are slightly high thanks to the simplifying assumptions described in the appendix.

\begin{deluxetable}{cllp{1in}}
\tablewidth{0pt}
\tablecaption{Density profile parameters for caustics 1 and 2.\label{tbl:shellfits}}
\tablehead{\colhead{Parameter} & \colhead{Value for caustic 1} &  \colhead{Value for caustic 2} & \colhead{Notes}}
\startdata
$N_p$ & 19779 & 6524 & number of particles in part of caustic used in fit \\
$r_{c}(\theta,\phi)$ (kpc) & $38.4 + 0.21\theta + 0.14 \phi$ & $26.0 -1.8\theta+0.34 \theta^2 + 0.3\phi$ & as defined in Equation \eqref{rcFitFunction}\\
$x_c$ (kpc) & $0.228 \pm 0.005$ & $0.113 \pm 0.009$ & from fit to density profile after asphericity corrections\\
$\delta r$, estimated (kpc) & 0.23 & 0.39 & using Equation \eqref{eq:EstimatingDeltaR} \\
 $\delta r$, fitted (kpc) & 0.201 $\pm 0.005$ & 0.232 $\pm 0.009$ & from fit to density profile \\
 $\kappa$, estimated\tablenotemark{a} & $4.6 \times 10^{-4}$   & $ 2.4 \times 10^{-4}$  & estimated using Equation \eqref{eq:kappaFromGravity}\\
 $\kappa$, fitted\tablenotemark{a} & $(4.2 \pm 0.2) \times 10^{-4}$  & $(2.31 \pm 0.04)  \times 10^{-4}$ & from fit to phase-space profile\\
 $f_0$, estimated\tablenotemark{b} & $0.37$ & $0.096$ & estimated using mass conservation\\
$f_0$, fitted\tablenotemark{b} & $(0.33 \pm 0.01)$ &  $(0.086 \pm 0.003)$ & from fit to density profile \\
\enddata
\tablecomments{Error ranges on fitted parameters indicate the 95 percent confidence interval.}
\tablenotetext{a}{Units are kpc (km s${}^{-1}$)${}^{-2}$.}
\tablenotetext{b}{Units are kpc${}^{-3}$ (km s${}^{-1}$)${}^{-1}$ sr${}^{-1}$.}
\end{deluxetable}

To finish calibrating the rate calculation, we compared the characteristic widths $\delta r$ and \Nres\ of the two caustics with those used in the numerical experiments to determine whether the rate estimate suffered from significant undersampling bias.  The numerical experiments used $t=1$ Myr (we can choose the units of time and length freely).  Caustic 1 has $\delta r = 0.20$ kpc and Caustic 2 has $\delta r = 0.23$ kpc, which correspond to test distributions with $\log_{10} \sigma = -0.70$ and $\log_{10} \sigma = -0.63$, respectively.  According to Table \ref{tbl:shellfits}, Caustic 1 has $\log_{10} \Nres = 4.3$ and Caustic 2 has $\log_{10} \Nres = 3.8$.  From the results of the tests in Section \ref{sec:bestPerformance}, we find that for $\log_{10} \Nres = 4.25$, the RMS error at $\log_{10} \sigma = -0.75$ is 4.0 percent using $\Nsmooth=10$ with the FiEstAS estimator and 3.4 percent with the uniform estimator.  For $\log_{10} \Nres = 3.75$, the RMS error at $\log_{10} \sigma = -0.75$ is 5.8 percent using $\Nsmooth=10$ and the uniform estimator.  All these errors are dominated by the standard deviation, not the undersampling bias.

\subsection{Boost factor}
\label{sec:boost}

Based on the fits in the previous section, we expect the shells to be fully resolved in the simulation.  So we are free to choose any of the three estimators we tested to calculate the contribution to the total rate from interactions between shell particles, $\Gamma_{ss}$.  We chose to use $\hat{\Gamma}_f$ for convenience, since choosing a constant Riemann volume whose size relative to the shells' thickness is consistent with our tests would require at least $10^7$ Riemann volumes to fill the simulation volume, whereas with adaptive Riemann volumes the large low-density portions require much less computation time.  $\Gamma_{sh}$, which represents interactions between dark matter from the shell and dark matter in the halo, was calculated using the density estimator $\hat{n}_f$ (Equation \eqref{densityEstimatorF}) to estimate $n_s$, and evaluating Equation \eqref{eq:analyticHaloDensity} for $n_h$ at the same points where $n_s$ is estimated.  The core radius was set to half the size of the Riemann volume enclosing the origin; in practice about 0.1 kpc.  $\Gamma_{hh}$ was calculated analytically by integrating Equation \eqref{eq:analyticHaloDensity} over the simulation volume.   We find that the boost factor $\beta = 2.4 \times 10^{-3}$ for $\Gamma$ integrated over the entire simulation volume, independent of the particle physics model.

By far the largest contribution to $\Gamma_{hh}$ comes from the very center of the halo.  To get a more realistic estimate of the boost factor we recalculated it with this region excised, which would certainly be done for a real observation given that astrophysical gamma rays appear to come from the disk.  We excluded a square region 1.35 kpc (0.2 degrees) on a side, centered on M31's center.  This choice of exclusion region corresponds to about twice the resolution limit of the Fermi LAT, and is intended to be equivalent to excising the central 4 pixels about M31's center.   This technique increases the boost factor by only a small amount, to $\beta = 0.0027$ or 0.27 percent.  

\begin{figure}[htbp]
\begin{center}
\plotone{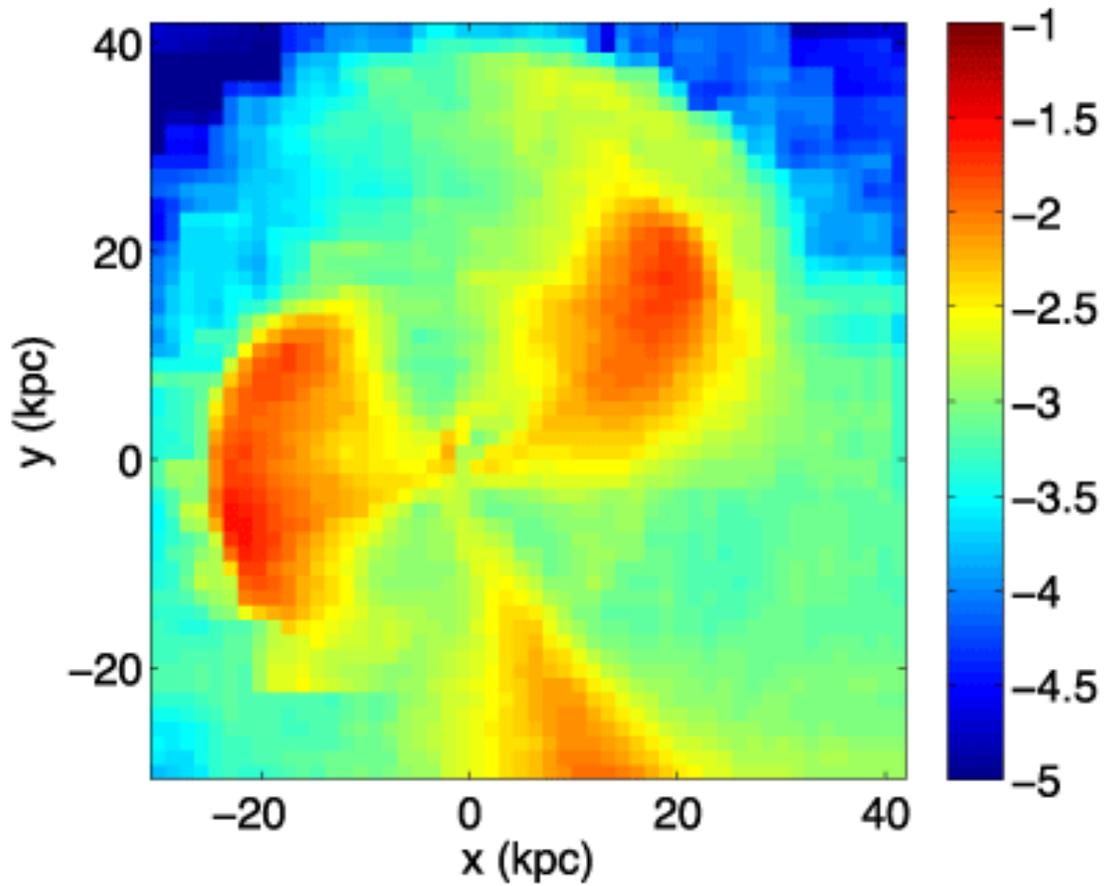}
\caption{(Color online) The largest boost factors, shown as the lightest shades (red in the electronic edition), are from the edges of the two shells.  The contrast in this figure is independent of the parameters for the particle physics model (summarized as $\Phi^{\mathrm{SUSY}}$). \label{boostFactor}}
\end{center}
\end{figure}

Finally, we mapped the spatial variation of the boost factor by using the space-filling tree in the FiEstAS algorithm.  As expected, the largest boost factors come from the edges of the two shells containing most of the mass, as shown in Figure \ref{boostFactor}; the maximum is 2.5 percent, independent of the particle physics model.

\subsection{Astrophysical factor}

The ``astrophysical factor" is the quantity $\Phi^{\mathrm{cosmo}}$ defined by Equations \eqref{eq:phiCosmoMass}, \eqref{eq:phiCosmoNumber}, and \eqref{eq:phiCosmoFiducial}.  For a given model of dark matter, comparing values of $\Phi^{\mathrm{cosmo}}$ gives the relative strength of different mass distributions as sources of high-energy particles via self-annihilation.  As described in Section \ref{sec:boost}, we used $\hat{\Gamma}_f$ to calculate $\Phi^{\mathrm{cosmo}}$ for the shells, both for interactions between two dark matter particles in the shell material and interactions between dark matter in the shell and dark matter in the halo (Table \ref{tbl:PhiCosmos}).  For comparison we also calculated $\Phi^{\mathrm{cosmo}}$ for the dwarf galaxy used in the N-body model.  The dwarf is represented as a Plummer sphere with mass $M_P = 2.2 \times 10^9\ M_{\astrosun}$ and scale radius $b = 1.03$ kpc.  Integrating the density-squared of the dwarf over volume shows that
\begin{equation}
\Gamma_{\mathrm{dwarf}} = \frac{1}{b^3} \left(\frac{3 M_P}{8 m_p} \right)^2,
\end{equation}
to be used with Equation \ref{eq:phiCosmoFiducial}.  We also compared these values with one calculated by \cite{PhysRevD.75.083526} from measurements of the mass and mass profile of the Ursa Minor dwarf galaxy in the Milky Way, scaled as if it were located in M31.  Ursa Minor has approximately the same mass as M31's faint satellites AndIX and AndXII \citep{2010MNRAS.tmp.1119C}.  We find that the shell signal is comparable in magnitude to the signal from this type of dwarf galaxy at the same distance, and two orders of magnitude less than the signal from the original Plummer sphere (whose mass is about ten times that estimated for Ursa Minor and its analogues in M31).

\begin{table}[htdp]
\begin{tabular}{|c|c|}
\tableline
Object & $\Phi^{\mathrm{cosmo}}$, Gev${}^2$ kpc cm${}^{-6}$ \\
\tableline
shell-shell (using $\Gamma_{ss}$)& $9.9 \times 10^{-9}$\\ 
shell-halo (using $\Gamma_{sh}$)& $8.5 \times 10^{-7}$\\
Plummer dwarf (using $\Gamma_{\mathrm{dwarf}}$) & $4.1 \times 10^{-5}$ \\
``Ursa Minor" \citep[based on][]{PhysRevD.75.083526} & $6.9 \times 10^{-7}$ \\
\tableline
\end{tabular}
\caption{Values of the astrophysical factor $\Phi^{\mathrm{cosmo}}$ for various configurations of the tidal debris, calculated using Equation \eqref{eq:phiCosmoFiducial}.\label{tbl:PhiCosmos}}
\end{table}

\subsection{Gamma-ray Signal}
\label{sec:signal}
Beyond calculating $\Gamma$ and $\beta$, we used Equation \eqref{eq:phitotal} to estimate the flux of gamma rays in the Fermi LAT for the various benchmarks in \citet{2004JCAP...07..008G}.  The results are shown in Table \ref{totalFluxes} along with the parameters used to calculate $\Phi^{\mathrm{SUSY}}$ in each case.  

\begin{figure}[htbp]
\begin{center}
\plotone{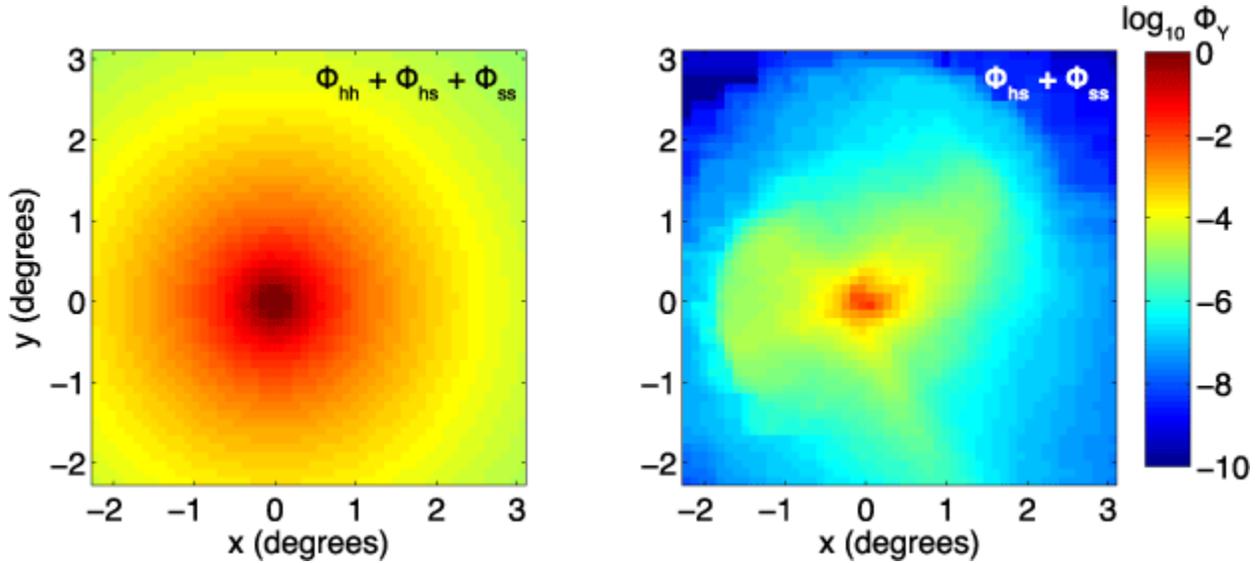}
\caption{(color online) Gamma-ray emission from the dark halo, though faint, dominates over emission from the shell even at large radius (left panel).  If $\Phi_{\gamma,\ hh}$ is removed, the remaining emission from the shell is too faint to detect with Fermi even for the most optimistic parameters in the set of benchmarks (right panel).  In these images the pixels are 0.1 degree on a side to imitate the approximate degree resolution of Fermi at the energy scale of interest, and the zero of the degree scale is centered on M31's center.  \label{countMap}}
\end{center}
\end{figure}

By using the three-dimensional representation of the density-squared field from the shell constructed with the FiEstAS estimator as a piecewise definition of the rate, we can integrate the rate along the line of sight and across pixels of arbitrary size such that the sum of the flux in all the pixels equals the total flux in Table \ref{totalFluxes}.  With this technique we made two-dimensional maps of the expected gamma-ray emission using the most optimistic value of $\Phi^{\mathrm{SUSY}}$ (labeled as ``Upper Limit" in Table \ref{totalFluxes}).  Using this value, even the center of M31's halo is only barely detectable by Fermi, if the halo is shaped as we assumed for the dynamical model and the amount of dark matter in the dwarf galaxy is comparable to the amount of luminous matter, although the emission from the dark halo completely dominates over that from the shell (Figure \ref{countMap}, left panel).  The tidal structure is at least an order of magnitude too faint to be detected even if the halo component is fitted and removed (Figure \ref{countMap}, right panel).  Because of interactions between halo and tidal dark matter, the signal from the tidal debris scales linearly with the mass of its progenitor as long as $\rho_h > \rho_s$ (Equation \eqref{fluxContributions}) so the ratio of dark matter to luminous matter in the dwarf galaxy would need to be several orders of magnitude larger, even after tidal stripping, for the tidal debris to be detectable with Fermi or for the scaling of the boost factor to become quadratic in the shell dark matter density.  For a smaller progenitor such a ratio might be plausible, but dwarf galaxies with masses of $10^9 M_{\odot}$ or higher tend to have comparable masses of luminous and dark matter in their centers based on our understanding of the Tully-Fisher relation at those masses \citep{2006ApJ...653..240G}.

\begin{deluxetable}{rlllllllllll}
\tabletypesize{\scriptsize}
\tablecaption{Contributions to the flux of gamma rays above 1 GeV from WIMP self-annihilation, for various MSSM benchmarks. \label{totalFluxes}}
\tablehead{
\colhead{\nodata} & \colhead{A'} & \colhead{B'} & \colhead{C'} & \colhead{D'} & \colhead{G'} & \colhead{H'} & \colhead{I'} & \colhead{J'} & \colhead{K'} & \colhead{L'} & \colhead{Upper Limit}}
\startdata
$m_{\chi}$, GeV\tablenotemark{a} & 242.8 & 94.9 & 158.1 & 212.4 & 148.0 & 388.4 & 138.1 & 309.1 & 554.2 & 181.0 & 40 \\

$N_{\gamma} (\sigma v)$\tablenotemark{a} & 120 & 782 & 195 & 63.6 & 1032 & 86.5 & 6303 & 930 & $7.08 \times 10^4$ & $1.87 \times 10^4$ & $1.30 \times 10^4$\\

$\Phi^{\mathrm{SUSY}}$\tablenotemark{b} & $3.14$ & $134$ & $12.0$ & $2.18$ & $72.7$ & $0.885$ & $510.$ & $15.0$ & $356.$ & $882.$ & $1.26 \times 10^4$ \\
\tableline
\tableline
  $\Phi_{\gamma,\mathrm{hh, all}}$\tablenotemark{c} \tablenotemark{h} & 3.38 & 144. & 13.0 & 2.34 & 78.3 & 0.953 & 550. & 16.2 & 383. & 950. & 13530. \\
  $\Phi_{\gamma,\mathrm{sh, all}}$\tablenotemark{d} & 0.00780 & 0.341 & 0.0307 & 0.00554 & 0.185 & 0.00225 & 1.30 & 0.0382 & 0.906 & 2.25 & 32.0 \\
  $\Phi_{\gamma,\mathrm{ss, all}}$\tablenotemark{e} & $<10^{-3}$ & 0.00424 & $<10^{-3}$ & $<10^{-3}$ & 0.0023 & $<10^{-3}$  & 0.0161 & $<10^{-3}$ & 0.0112 & 0.0279 & 0.398 \\
  $\Phi_{\gamma,\mathrm{addl, all}}$\tablenotemark{f} & 0.00810 & 0.345 & 0.0310 & 0.00561 & 0.187 & 0.00228 & 1.32 & 0.0387 & 0.917 & 2.28 & 32.4 \\
  \tableline
  $\Phi_{\gamma,\mathrm{total, all}}$\tablenotemark{g} & 3.39 & 145. & 13.0 & 2.35 & 78.5 & 0.955 & 551. & 16.2 & 384. & 953. & 13560. \\
  \tableline
  \tableline
  $\Phi_{\gamma,\mathrm{hh, nc}}\tablenotemark{i}$ & 1.92 & 81.8 & 7.35 & 1.33 & 44.4 & 0.5402 & 311. & 9.17 & 217. & 539. & 7672. \\
  $\Phi_{\gamma,\mathrm{sh, nc}}$ & 0.00509 & 0.217 & 0.0195 & 0.00353 & 0.118 & 0.00144 & 0.827 & 0.0244 & 0.577 & 1.43 & 20.4 \\
  $\Phi_{\gamma,\mathrm{ss, nc}}$ & $<10^{-3}$ & 0.00362 &  $<10^{-3}$  & $<10^{-3}$ &  0.00196 & $<10^{-3}$ & 0.0138 & $<10^{-3}$ & 0.00961 & 0.0238 & 0.339 \\
  $\Phi_{\gamma,\mathrm{addl, nc}}$ & 0.00518 & 0.221 & 0.0198 & 0.00359 & 0.120 & 0.00146 & 0.841 & 0.0248 & 0.586 & 1.46 & 20.7 \\ 
   \tableline
  $\Phi_{\gamma,\mathrm{total, nc}}$ & 1.92 & 82.0 & 7.37 & 1.33 & 44.5 & 0.542 & 312. & 9.19 & 218. & 540. & 7693.\\
\tableline
\enddata
\tablecomments{ The subscripts $hh$, $sh$, and $ss$ refer to the various terms in Equation \eqref{eq:totalrate}. $N_{\gamma} (\sigma v)$ has units $10^{-29}$ cm${}^3$ s${}^{-1}$.  $\Phi_{\mathrm{SUSY}}$ has units $10^{-11}$ cm${}^4$ kpc${}^{-1}$ s${}^{-1}$ GeV${}^{-2}$.  All $\Phi_{\gamma}$ have units $10^{-14}\ \gamma$ cm${}^{-2}$ s${}^{-1}$.  For reference, the Fermi \emph{point source} sensitivity for photons with $E>100$ MeV is on the order of $10^{-9}\ \gamma$ cm${}^{-2}$ s${}^{-1}$ \citep{2009arXiv0907.0626R}.}
\tablenotetext{a}{Taken from Table 1 of \citet{2004JCAP...07..008G} except for the rightmost column, which is based on Figure 8 of FPS as described in the text.}
\tablenotetext{b}{Calculated analytically using Equation \eqref{eq:PhiSUSYfiducial}}
\tablenotetext{c}{Calculated analytically using the same NFW halo as for the dynamical model.}
\tablenotetext{d}{Calculated by constructing a numerical estimate for $n_{\mathrm{shell}}$ in each integration volume element, then evaluating $n_{\mathrm{halo}}$ analytically at the center of that volume element and assuming its value is constant over the entire element.}
\tablenotetext{e}{Calculated numerically as described in the text.}
\tablenotetext{f}{$\Phi_{\gamma,\ \mathrm{addl}} \equiv \Phi_{\gamma,\ \mathrm{hs}} + \Phi_{\gamma,\ \mathrm{ss}}$.}
\tablenotetext{g}{$\Phi_{\gamma,\ \mathrm{total}} \equiv \Phi_{\gamma,\ \mathrm{hh}} + \Phi_{\gamma,\ \mathrm{hs}} + \Phi_{\gamma,\ \mathrm{ss}}$.}
\tablenotetext{h}{\emph{all}: integration is over entire line of sight and covers the region shown in Figure \ref{countMap} in x and y}
\tablenotetext{i}{\emph{nc}: a central region is excluded from the calculation, as described in the text. }
\end{deluxetable}

\section{Conclusions}
\label{sec:concl}

We find that unless all the features in a given density distribution are known to be fully resolved, the best way to estimate the volume integral of the square of the density (the ``rate") from an N-body realization is to use the simple nearest-neighbors estimator with a constant Riemann volume.  If the realization completely resolves even the sharpest features, all three estimators we tested should agree on the result.  The simplest method for estimating the rate works best for this problem because the other, more complicated algorithms are optimized for density estimation, not rate estimation, and because estimators using adaptive Riemann volumes appear to require slightly more particles to resolve features of a given sharpness.   In any case the estimator should be calibrated for Poisson bias as we describe in Section \ref{sec:pb}.  We also find that the improvement in the standard deviation achieved by increasing the smoothing number is smaller than the increased bias from blurring more small-scale structure for $\Nsmooth > 10$.  The correct calibration of the estimator for Poisson bias can change the estimated result by up to 10 percent for reasonable values of $\Nsmooth$.  The correct calibration for undersampling bias can change the result by a factor of 2 or more if the simulation is under-resolved; rather than attempt to correct for it, it is better to ensure that the N-body realization has sufficient resolution for the small-scale features to be resolved.

Using a calibrated estimator and a sufficiently resolved N-body realization, we calculated the boost factor and signal in gamma rays from tidal debris in M31 that displays high-contrast features.  Although we find as expected that the largest boosts come from the shell edges, they only increase the total signal by at most 2.5 percent over the signal from a self-consistent smooth halo.  Likewise, the total gamma-ray flux from the shells is three orders of magnitude lower than emission from the dark halo, and too low to be detected by Fermi for likely dark matter parameters (Table \ref{totalFluxes}).  The total signal is comparable to that predicted for an ultra-faint satellite of M31.

\section{Future Work}
\label{sec:future}

The existence of shell features around M31 provides many avenues other than indirect detection for learning about the nature, dynamics, and distribution of dark matter.  The very existence of the shells demands that the dwarf galaxy that created them must have had very low angular momentum relative to M31 because the pericenter distance is so small.  Whereas high-angular-momentum systems like that of the Sagittarius dwarf galaxy in the Milky Way are useful for constraining the shape of dark halos because the tidal debris explores a large range in angle, low-angular-momentum systems like the M31 shells and giant stream probe M31's potential over a large range in radius, and are best suited for constraining the degeneracy between the different mass components of the host galaxy.  They also act as a sensitive probe of the mass profile of the progenitor, since the combination of relatively cold initial conditions in the dwarf and a small pericenter distance acts as a kind of spectrometer, spreading the mass of the satellite galaxy out in space according to its total energy.  Because the shells' relative orientations are a good limit on the projected angular momentum of the progenitor, variations in the initial position and velocity of the center of mass are not likely to be degenerate with variations in the shape and phase space distribution of the debris, although this is still being tested.  This makes the shapes and phase space distributions of the shells extremely sensitive to the initial phase space distribution of the progenitor satellite, and can place limits on the cuspiness of the mass profile of the dwarf.  

The analytical caustic used in this work can also be used as a model for caustics that form under the much more complicated equations of motion responsible for quasi-radial gravitational infall.  As shown in Appendix \ref{appx:oneDcaustic}, the resulting form is identical to that obtained by \citet{2006MNRAS.366.1217M} in their analysis of those caustics with intuitive identifications of the normalization, caustic location and distance from the caustic surface, and requires no numerical integration to obtain the complete profile so it may be easily used for fitting.  Although our model is less general (it does not predict the relative locations of caustics) it is consistent with the more general case, and more tractable if only the universal density profile is desired.  The height and width of each caustic are sensitive to the initial phase space distribution of material in the caustic, while the profile depends on the potential of the host galaxy only through the gravitational force at the location each caustic---a complete mass model is not necessary.  In light of recent discoveries of shells around many more nearby galaxies besides Andromeda \citep{2010arXiv1003.4860M}, this technique may provide a way to constrain the properties of luminous matter in dwarf galaxies by examining the tidal debris they produce, as will be discussed in an upcoming paper (Sanderson, in prep.).

Although the M31 tidal debris is probably not a candidate for indirect detection, we hope that our discussion of how to estimate such signals from N-body realizations will improve those estimates in future work.  We also hope it may inspire attempts to develop optimized estimators for this quantity, similar to the way that optimized density estimators have been developed, since no optimization other than simple bias correction was applied to the estimators used in this work, and connect the astrophysics community with the body of statistical literature on such estimators.  N-body modeling will undoubtedly prove an indispensable component of the prediction and interpretation of direct and indirect detections of dark matter.

\section{Acknowledgements}

The authors acknowledge support from NASA grant NNG06GG99G.  The numerical experiments in the paper were performed using the  MIT Kavli Institute computing cluster, which is supported in part by the Kavli Foundation.  The authors thank Paul Hsi for maintaining, troubleshooting, and upgrading the cluster.  RES thanks Will Farr for the use of his N-body integrator and for many helpful conversations.

\clearpage

\appendix

\section{Rate Estimators}
\label{appx:estimators}

Here we describe the five rate estimators tested in this work.  They are based on three types of density estimators (spherical and Cartesian nearest-neighbor algorithms and a spherical smoothed kernel method) and two methods for determining Riemann volumes (constant size and adaptive tessellation). 

\subsection{Nearest Neighbor}

An N-body representation of a continuous number density distribution $n(\vec{x})$ is a Poisson point process with a spatially varying mean.  As such, all estimators (e.g., $\hat{n}$) of the density and its higher moments ($n^2$, $n^3$, etc.) obey the statistics of point processes.  In the case of a uniform distribution, these are simply the well known Poisson statistics, and the simplest estimator calculates the density in terms of the distance to the $\Nsmooth^{\mathrm{th}}$ particle, called the nearest neighbor distance.  For a given value of the smoothing number $\Nsmooth$ there is a particular nearest neighbor distance $r_{Ns}$ for each particle, and the density near the particle is estimated using
\begin{equation}
\label{biasedDensity}
\hat{n}_{b} = \frac{3\Nsmooth}{4\pi r_{Ns}^3},
\end{equation}
However, if we compute $E(\hat{n}_b)$ by integrating over the probability distribution describing the distances between particles, we find that
\begin{equation}
\label{biasedDensityExpectation}
E(\hat{n}_b) = \frac{\Nsmooth}{\Nsmooth-1} n, 
\end{equation}
indicating that the estimator is biased since $E(\hat{n}_b)\ne n$.  This is a result of the random fluctuations in the distance $r_{Ns}$ from particle to particle, which obey Poisson statistics.  Even in cases where the density is not uniform a similar effect is present.  

The Poisson bias of the estimator \eqref{biasedDensity} can be easily eliminated in the case of the uniform distribution by noting that $E(\hat{n}_b)$ differs from $n$ by a constant factor only.  Dividing by this factor produces the estimator
\begin{equation}
\label{eq:unbiased-nearest-neighbor}
\hat{n}_u = \frac{3(\Nsmooth-1)}{4\pi r_{Ns}^3},
\end{equation}
which has $E(\hat{n}) = n$.  

We wish to construct a minimally biased estimator for the rate, $\Gamma = \int n^2 dV$.  It is well known that in a Poisson distribution $E(\hat{n}_u^2)$, the expectation value of the square of the unbiased density estimator in equation \eqref{eq:unbiased-nearest-neighbor}, is not equal to $n^2$; still, an unbiased estimator for $n^2$ in the case of the uniform distribution does exist:
\begin{equation}
\label{unbiasedDensitySquared}
\widehat{n^2}_u = (\Nsmooth-1)(\Nsmooth-2)\left(\frac{3}{4\pi r_{Ns}^3}\right)^2
\end{equation}
with $\Nsmooth$ and $r_{Ns}$ defined as before.  Then $E(\widehat{n^2}_u) = n^2$.

Using $\widehat{n^2}$, we can construct an unbiased estimator for $\Gamma$ by using a Riemann sum over $N_V$ identical volumes $dV$ to approximate the volume integral, so that
\begin{equation}
\label{estimatorU}
\hat{\Gamma}_u = dV \sum_{i=1}^{N_V} \widehat{n^2}_{u,i}
\end{equation}
where $\widehat{n^2}_{u,i}$ is given by evaluating Equation \eqref{unbiasedDensitySquared} at the center of subvolume $i$, and the total volume $V = N_V dV$.  Using Equation \eqref{estimatorU}, $E(\hat{\Gamma_u}) = \gt$.  This estimator provides a useful check that the code is functioning properly. 

We also tested the same density estimation method with an adaptive Riemann volume, in which each particle occupies its own box ($N_V = \Nres$).  Each particle's Riemann volume contains all the space closer to that particle than any other.  The size of such a Riemann volume is also affected by Poisson statistics, so this rate estimator will not be unbiased even if Equation \eqref{unbiasedDensitySquared} is used to calculate the density-squared.  For simplicity, we represent the $\Nsmooth$ dependence of the additional bias from the adaptive box size as a prefactor, to be determined numerically, and use Equation \eqref{unbiasedDensitySquared} to estimate the density-squared:
\begin{equation}
\label{estimatorN}
\hat{\Gamma}_n = \frac{1}{1+\bias_n(\Nsmooth)} \sum_{i=1}^{\Nres} \widehat{n^2}_{u,i} dV_i
\end{equation}

\subsection{FiEstAS}
A variation on the nearest-neighbor estimator is implemented by Ascasibar and Binney (2005) in their algorithm FiEstAS.  We refer to this estimator as $f$.  It too uses the $\Nsmooth$th nearest neighbor, but instead of a spherical volume considers the volume of the Cartesian box enclosing \Nsmooth particles when calculating the density, so that for a particle $i$:
\begin{equation}
\label{densityEstimatorF}
\hat{n}_{f,i} = \frac{\Nsmooth}{V_{\Nsmooth,i}}
\end{equation}
Conveniently, the construction of the tree used in calculating $dV_i$ also chooses the Riemann volume adaptively in the same manner as for the estimator $n$.  Now there are two contributions to the bias: the Poisson bias from using $(\hat{n}_{f,i})^2$ to estimate the density-squared and the Poisson bias from determining the adaptive Riemann volumes.  For simplicity, we represent the $\Nsmooth$-dependence of both contributions with a single prefactor, so the rate estimator is
\begin{equation}
\label{rateEstimatorF}
\hat{\Gamma}_f \equiv \frac{1}{1+\bias_f(\Nsmooth)}  \sum_{i=1}^{\Nres} (\hat{n}_{f,i})^2 dV_i
\end{equation}

\subsection{Kernel-based}
We also tested two kernel-based rate estimators.  Kernel-based density estimators use a weighted sum to smooth over the $\Nsmooth$ nearest particles, so that the estimated density at location $\vec{x}$ is
\begin{equation}
\label{densityEstimatorS}
\hat{n}_{\mathrm{s}}(\vec{x}) =  \sum_{j=1}^{\Nsmooth} W(\vec{x}_j - \vec{x}, \vec{h})
\end{equation}
The smoothing vector $\vec{h}$ is a generalized nearest-neighbor distance.  The vector has length $| \vec{x}_{\Nsmooth} - \vec{x} |$.  For a one-dimensional spherical kernel $\vec{h} \equiv r_N \hat{r}$ and the kernel function $W$ is nonzero when $| \vec{x}_{j} - \vec{x} | < r_N$.  For a three-dimensional kernel, $\vec{h} \equiv \vec{x}_{\Nsmooth} - \vec{x}$ and $W$ is nonzero when $|\vec{x}_j - \vec{x}| <  |\vec{h}|$. 

The type of kernel chosen can have a significant effect on the bias and variance.  \citet{sharma:2006aa} have tested the bias and variance of density estimators with a variety of kernels on uniform density distributions, and we use their notation here.  The one-dimensional kernel can be written in the form
\begin{equation}
W(\vec{r},\vec{h}) = \frac{f W(u)}{V_h}
\end{equation}
where $\vec{r}$ is the distance from the target location, $u$ is the scaled distance
\begin{equation}
u = r/h,
\end{equation}
$f$ is the kernel normalization
\begin{equation}
\frac{1}{f} = \int_0^1 W(u) 4\pi u^2 du,
\end{equation}
and $V_h$ is the volume enclosed by the smoothing length $\vec{h}$. 

\citeauthor{sharma:2006aa} found that the Epanechnikov kernel
\begin{equation}
\label{EpanechikovKernel}
W(u) = \left\{
\begin{array}{cc}
1 - u^2 & 0 \le u \le 1\\
0 & \textrm{otherwise}
\end{array} \right.
\end{equation}
has the smallest bias and variance in estimating the density in the case of a uniform distribution.  We use this density estimator and an adaptive Riemann volume for the rate estimator $s$, and again collect the $\Nsmooth$-dependence of the bias in a prefactor:
\begin{equation}
\label{rateEstimatorS}
\hat{\Gamma}_s \equiv \frac{1}{1+\bias_s(\Nsmooth)} \sum_{i=1}^{\Nres} \left[\hat{n}_s(r_i)\right]^2 dV_i
\end{equation}
Again, $1/(1+\bias_s)$ is the bias when using a constant Riemann volume.

\citet{diemand:2007aa} used an adaptation of the kernel-based method to estimate the rate from simulations of the Milky Way's dark halo and halo substructure.  Starting with Equation \eqref{rateEstimatorS}, they make the substitution $\hat{n}_s(r_i) dV_i = 1$ (there is one particle per Riemann volume).  We examine this variation of the kernel-based method, referred to as estimator $d$.  We include a bias-correcting prefactor that is equal to 1 in \citeauthor{diemand:2007aa}:
\begin{equation}
\label{rateEstimatorD}
\hat{\Gamma}_d \equiv \frac{1}{1+\bias_d(\Nsmooth)} \sum_{i=1}^{\Nres} \hat{n}_{s}(r_i) 
\end{equation}
From a Poisson-statistics standpoint, the substitution implicitly assumes that $E(\hat{n})^2 = E(\widehat{n^2})$, yet the estimator itself is linear instead of quadratic in the density.  For this reason it is expected to behave differently than the rate estimator \eqref{rateEstimatorS}.

\section{Calculation of the Minimum Nearest-neighbor Distance}
\label{appx:rnmin}

In this appendix we derive an expression for the expectation value of the minimum nearest-neighbor distance, $E(\rnmin)$, in the case of a uniform density distribution of particles.  The scaling of this value with the smoothing number $\Nsmooth$ and the number $\Nres$ of particles in the simulation subsequently determines the maximum density that can be both represented in an N-body realization with \Nres\ particles and calculated using the nearest-neighbor estimator with \Nsmooth\ nearest neighbors.  $E(\rnmin)$ is the first order statistic of the estimator \rn, which is related to the nearest-neighbors density estimator $\hat{n}$ by
\begin{equation}
\rn = \left(\frac{4\pi \hat{n}}{3 \Nsmooth}\right)^{-1/3}.
\end{equation}

We start with the PDF of the nearest-neighbor distance,
\begin{equation}
p_{\rn}(\rho) d\rho = \frac{\exp(-4\pi n \rho^3/3)}{(\Nsmooth-1)!} \left(\frac{4\pi n \rho^3}{3}\right)^{\Nsmooth-1} d\left(\frac{4\pi n \rho^3}{3}\right),
\end{equation}
which can be derived from directly integrating over the joint PDF for the nearest $\Nsmooth$ particles.  The PDF for each particle is Poisson.  To calculate the first order statistic we also need the CDF of the nearest neighbor distance, 
\begin{equation}
P_{\rn}(\mu) = \frac{1}{(\Nsmooth-1)!} \left[ \sGamma(\Nsmooth) - \sGamma\left(\Nsmooth, 4\pi n \mu^3/3\right)\right] = 1 - \frac{\sGamma\left(\Nsmooth, 4\pi n \mu^3/3\right)}{\sGamma(\Nsmooth)},
\end{equation}
where $\sGamma(N)$ and $\sGamma(N,x)$ are the complete and incomplete gamma functions, respectively.

The PDF of \rnmin\ is that of the first order statistic of the PDF of the nearest-neighbor distance:
\begin{equation}
p_{\rnmin}(\nu) d\nu = \Nres \left\{ 1- P_{\rn}(\nu)\right\}^{\Nres-1} p_{\rn}(\nu) d\nu.
\end{equation}
Substituting the expressions for the PDF and CDF of the nearest-neighbor distance,
\begin{equation}
p_{\rnmin}(\mu) d\mu =\frac{\Nres}{(\Nsmooth-1)!} \exp\left(-4\pi n \mu^3/3\right) \left(\frac{4\pi n \mu^3}{3}\right)^{\Nsmooth-1} \left( \frac{\sGamma\left(\Nsmooth, 4\pi n \mu^3/3\right)}{\sGamma(\Nsmooth)} \right)^{\Nres-1} d\left(\frac{4\pi n \mu^3}{3}\right).
\end{equation}
Changing variables to $y = 4 \pi n \mu^3 / 3$ gives us a simpler expression:
\begin{equation}
p_{\rnmin}(y) dy =\frac{\Nres}{(\Nsmooth-1)!} e^{-y} y^{\Nsmooth - 1} \left[\frac{\sGamma(\Nsmooth,y)}{\sGamma(\Nsmooth)}\right]^{\Nres - 1} dy.
\end{equation}
$y$ represents the average number of particles in a sphere of radius $\mu$.

The expectation value of $\rnmin$ is then determined by weighted integration over its PDF:
\begin{equation}
\label{ernmin}
E(\rnmin) = \left( \frac{3}{4\pi n}\right)^{1/3} \frac{\Nres}{[(\Nsmooth - 1)!]^{\Nres}}  \int_0^{\infty} e^{-y} y^{\Nsmooth -2/3} \left[\sGamma\left(\Nsmooth, y \right)\right]^{\Nres-1} dy,
\end{equation}
using the definition of $\mu$ in terms of $y$ and the fact that $p_{\rnmin}(y) dy = p_{\rnmin}(\mu) d\mu$.

We now use two approximations for the gamma function.  The first is the asymptotic expansion for the incomplete gamma function at large $y$:
\begin{equation}
\sGamma(\Nsmooth, y) \approx e^{-y} \left[ y^{\Nsmooth-1}  + \mathcal{O}\left(y^{\Nsmooth-2}\right)\right];
\end{equation}
the second is Stirling's approximation for the complete gamma function at large argument $N$:
\begin{equation}
\sGamma(N+1) = N! \approx N^{N} e^{-N}.
\end{equation}

Using the asymptotic expansion, we may rewrite the integral:
\begin{equation}
E(\rnmin) = \left( \frac{3}{4\pi n}\right)^{1/3} \frac{\Nres}{[(\Nsmooth - 1)!]^{\Nres}} \int_0^{\infty} e^{-\Nres y} y^{\Nres(\Nsmooth - 1) + 1/3} dy
\end{equation}

Making the change of variables $t=\Nres y$, we find
\begin{equation}
E(\rnmin) = \left( \frac{3}{4\pi n}\right)^{1/3} \frac{\Nres}{[(\Nsmooth - 1)!]^{\Nres} \Nres^{\Nres(\Nsmooth-1)+4/3}} \int_0^{\infty} e^{-t} t^{\Nres(\Nsmooth - 1) + 1/3} dt
\end{equation}
which, using the definition of the complete gamma function, is
\begin{equation}
E(\rnmin) = \left( \frac{3}{4\pi n}\right)^{1/3} \frac{ \sGamma[\Nres(\Nsmooth - 1) + 4/3]}{[(\Nsmooth - 1)!]^{\Nres} \Nres^{\Nres(\Nsmooth-1)+1/3}} = \left( \frac{3}{4\pi n}\right)^{1/3} \frac{ [\Nres(\Nsmooth - 1) + 1/3]!}{[(\Nsmooth - 1)!]^{\Nres} \Nres^{\Nres(\Nsmooth-1)+1/3}} 
\end{equation}

Now we use Stirling's approximation to simplify the factorials in the ratio.  We define $\Na \equiv \Nres(\Nsmooth - 1)$ to keep things shorter:
\begin{eqnarray}
(\Na + 1/3)! & \approx & (\Na + 1/3)^{\Na + 1/3} e^{-(\Na + 1/3)} \\
\left[(\Nsmooth - 1)!\right]^{\Nres} & \approx & \left[ (\Nsmooth - 1)^{\Nsmooth - 1} e^{-(\Nsmooth-1)} \right]^{\Nres} = (\Nsmooth-1)^{\Na} e^{-\Na}
\end{eqnarray}

Substituting these two approximations back into the expression for $E(\rnmin)$, we find
\begin{equation}
E(\rnmin) = \left( \frac{3}{4\pi n}\right)^{1/3} \frac{ (\Na + 1/3)^{\Na + 1/3} e^{-(\Na + 1/3)} }{ \Nres^{\Na + 1/3} (\Nsmooth-1)^{\Na} e^{-\Na} } = \left( \frac{3(\Nsmooth - 1)}{4\pi n e}\right)^{1/3} \left(\frac{\Na + 1/3}{\Na} \right)^{\Na + 1/3}
\end{equation}
In the limit $\Na \gg 1/3$, the second term approaches $e^{1/3}$, so to leading order, $E(\rnmin) \propto \Nsmooth^{1/3}$:
\begin{equation}
E(\rnmin) \approx \left( \frac{3(\Nsmooth - 1)}{4\pi n }\right)^{1/3}
\end{equation}
The \emph{minimum} expected distance to the $\Nsmooth^{\mathrm{th}}$ particle is roughly equal to the \emph{average} expected distance to the $(\Nsmooth -1)^{\mathrm{th}}$ particle, in the limit where both $\Nsmooth$ and \Nres\ are much greater than 1.  

\begin{figure}[htbp]
\begin{center}
\plottwo{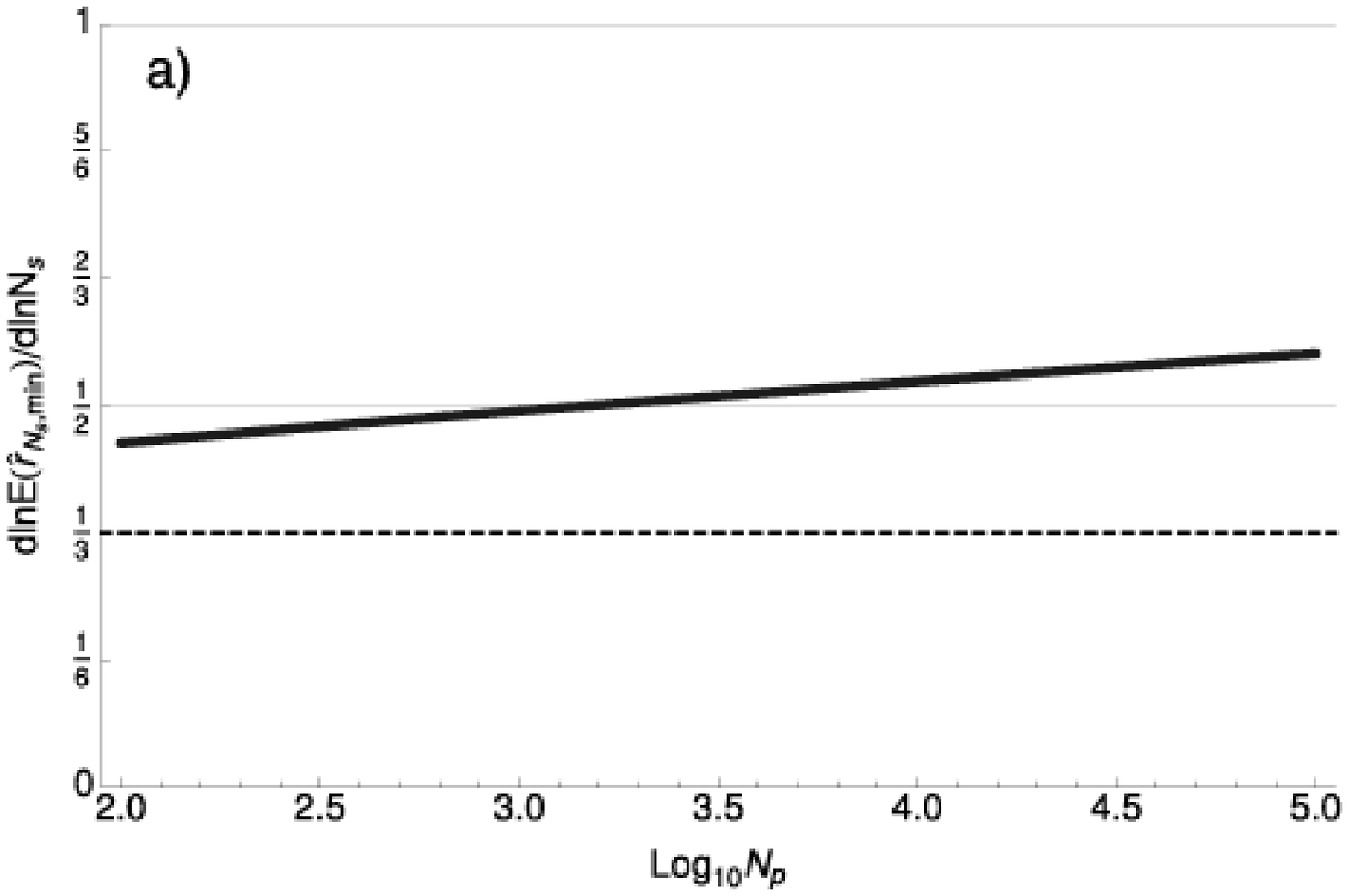}{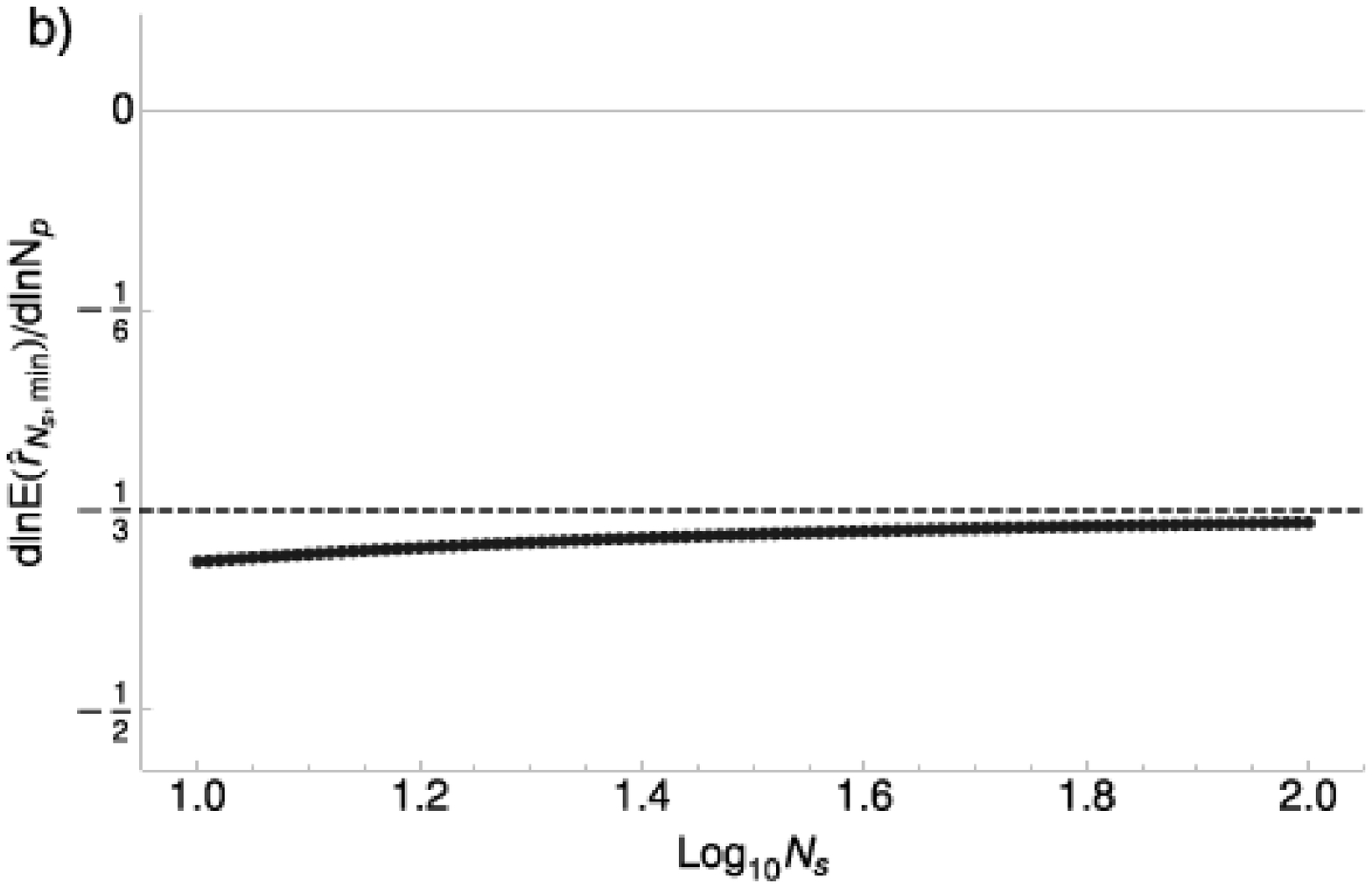}
\caption{The \Nsmooth-dependence of $E(\rnmin)$ (a, thick line) is not close to the asymptotic prediction (dashed line) in the range of interest, and varies somewhat with \Nres.  We take an average slope of $1/2$ (solid thin line) for the power law index in \Nsmooth. However, the \Nres-dependence of $E(\rnmin)$ (b) is close to and approaches the asymptotic prediction (dashed line), so we use the asymptotic slope of $-1/3$ for the \Nres\ scaling relation. \label{fig:rnminfits}}
\end{center}
\end{figure}

Although the criterion $\Nres \gg 1$ is generally satisfied, we are interested in values of \Nsmooth\ that do \emph{not} satisfy the criterion $\Nsmooth \gg 1$: in fact, we wish to use the smallest value of \Nsmooth\ possible while retaining a good RMS error.  We must therefore estimate the scaling with \Nsmooth in our region of interest by tabulating values of the integral in Equation \eqref{ernmin} at various values of \Nsmooth\ and \Nres.  The scaling in the region of interest can then be approximated by fitting the values in the region of interest to a power law whose index is a free parameter.  We find that for $2 \le \log_{10} \Nres \le 5$ and $10 \le \Nsmooth \le 50$, the \Nsmooth-dependence roughly obeys a power law $\rnmin \propto \Nsmooth^{\gamma}$ with index $\gamma = 0.51 \pm 0.06$, where $\gamma$ depends slightly on \Nres.  We confirm that the scaling of $E(\rnmin)$ with \Nres is suitably consistent with the prediction (Figure \ref{fig:rnminfits}).

\section{An Analytic One-dimensional Caustic}
\label{appx:oneDcaustic}
We would like to test the bias and variance of three common density estimation schemes in the context of high density contrast.  The three-dimensional test distribution has a one-dimensional caustic in ($x$) and is uniform in the other two dimensions ($y$ and $z$).  Below we present the derivation of Equation \eqref{analyticCaustic1} and discuss the calculation of $\Gamma$.

\subsection{Density Profile}
 Following the notation of \citet{shandarin:1989aa}, we define the Eulerian coordinates $\vec{x} \equiv (x,y,z)$ in terms of the Lagrangian coordinates or ``initial conditions" $\vec{q} \equiv (q_x, q_y, q_z)$ for a collisionless nongravitating ensemble:
\begin{equation}
\label{eq:eul-as-lag}
\vec{x}(\vec{q},t) = \vec{q} + \vec{v}_0(\vec{q}) t
\end{equation}
where $\vec{v}_0$ is the initial velocity vector of the particles.  

Conservation of mass states that:
\begin{equation}
\rho(\vec{x},t) d^3\vec{x} = \sum_{\vec{q} \to \vec{x}} \rho_0 d^3 \vec{q}
\end{equation}
where $\vec{x}(\vec{q},t)$.  The sum is taken over all values of $\vec{q}$ for which particles end up at a given $\vec{x}$. Assuming a uniform starting density $\rho_0$, the local density $\rho$ of the distribution then evolves in time as
\begin{equation}
\rho(\vec{x},t) = \sum_{\vec{q} \to \vec{x}} \frac{\rho_0}{\left|d\vec{x}/d\vec{q}\right|} = \sum_{\vec{q} \to \vec{x}}  \frac{\rho_0}{\left| \delta_{ik} + t \frac{\partial v_{0,i}}{\partial q_k}\right|}
\label{eq:rho-x-t}
\end{equation}
Caustics arise when the determinant in the denominator is zero.

To randomly sample a simple one-dimensional caustic along $x$ in our 3D distribution, we can choose a uniform distribution in $y$ and $z$:
\begin{eqnarray}
y &=& q_y\\
z &=& q_z
\end{eqnarray}
and an extremely simple displacement rule for x that includes a constant initial velocity $v_{0,x}(q_x)$ with a small random component $v_{\mathrm{th}}(q_x)$ along the dimension of the caustic and no further interactions:
\begin{equation}
\label{eq:qtox-with-disp}
x = q_x + t \left(v_{0,x}(q_x)  + v_{\mathrm{th}}(q_x)\right)
\end{equation}
$v_{\mathrm{th}}(q_x)$ is a random variable drawn from a distribution $f(v_{\mathrm{th}})$ for each particle in the system labeled by a unique $q_x$.  This ``thermal velocity" is included so that $\Gamma$ will not diverge at the caustic surface.  Equation \eqref{eq:qtox-with-disp} is not invertible since the $v_{\mathrm{th}}$ are random.  Furthermore the particle ordering at $t>0$ is unknown because of these random velocities, so the density at some point may now include many streams, since the initial velocity determines when the particles cross the caustic.  To make sure we include all the streams, we express the sum over streams as an integral over a delta function (dropping the subscripts on $q$ for clarity):
\begin{equation}
\label{eq:sum-over-streams}
\rho[v_{\mathrm{th}}(q)](x,t) dx = \rho_0 \int_{-\infty}^{\infty} dq\ \delta\left[ x-q-v(q) t \right] dx
\end{equation}
where $v(q) \equiv v_0(q) + v_{\mathrm{th}}(q)$ includes both the uniform and random parts of the initial particle velocity.  The delta function and integral over $q$ form a sum over all particles that arrive at the location $x$ from any $q$ at a time $t$.

We choose to represent the thermal velocities with a Gaussian (Maxwell) distribution with one-dimensional dispersion $\sigma$:
\begin{equation}
f(v_{\mathrm{th}}) dv_{\mathrm{th}}  = \frac{1}{\sqrt{2\pi \sigma^2}} e^{-v_{\mathrm{th}}^2 / 2\sigma^2} dv_{\mathrm{th}}
\end{equation}
$v_{\mathrm{th}}(q)$ is thus a function of $q$ in the sense that for each particle, labeled by its $q$, a different random velocity is assigned.  The dispersion $\sigma$ could easily be time-dependent, but again we take the simplest case and assume it is a constant.  We refer to the $v_{\mathrm{th}}$ as a thermal component because systems in thermal equilibrium have velocities described by the same distribution, so it is often associated with temperature since $\sigma^2 \sim T$ in that case.

To make the integral in the expression for $\rho(x)$ easier to evaluate, we can also replace the delta function with its limit definition in terms of a Gaussian:
\begin{equation}
\delta\left[ x-q-v(q) t \right] = \lim_{\epsilon \to 0} \frac{1}{\sqrt{2\pi \epsilon^2}} e^{-\left[ x-q-v(q) t \right]^2 / 2\epsilon^2}
\end{equation}
Combining, we find
\begin{equation}
\rho[v_{\mathrm{th}}(q)](x,t) = \rho_0 \int_{-\infty}^{\infty} dq \lim_{\epsilon \to 0} \frac{1}{\sqrt{2\pi \epsilon^2}} e^{-\left[ x-q-v(q) t \right]^2 / 2\epsilon^2} 
\end{equation}

To get the distribution $\rho(x,t)$ we must take the ensemble average over $v_{\mathrm{th}}$:
\begin{equation}
\rho(x,t) = \left< \rho[v_{\mathrm{th}}(q)](x,t) \right> = \int_{-\infty}^{\infty} dv_{\mathrm{th}} \rho[v_{\mathrm{th}}(q)](x,t) f(v_{\mathrm{th}}) 
\end{equation}
We reintroduce $v(q) \equiv v_0(q) + v_{\mathrm{th}}(q)$ and pull out the parts of the exponent containing $v_{\mathrm{th}}$:
\begin{equation}
\label{eq:sum-and-average}
\rho(x,t) = \rho_0 \int_{-\infty}^{\infty} dq \lim_{\epsilon \to 0} \frac{1}{\sqrt{2\pi \epsilon^2}} e^{-\left[ x-q-v_0(q) t \right]^2 / 2\epsilon^2}  \int_{-\infty}^{\infty} dv_{\mathrm{th}}  \frac{1}{\sqrt{2\pi \sigma^2}} e^{-v_{\mathrm{th}}^2 / 2\sigma^2} e^{-\left\{ t^2 v_{\mathrm{th}}^2 - 2 v_{\mathrm{th}} t\left[ x-q-v_0(q) t \right] \right\}/2\epsilon^2}
\end{equation}
For notational simplicity, we temporarily define $\left[ x-q-v_0(q) t \right] \equiv \Delta$.  Regrouping terms, we find a quadratic expression in the exponent in the inner integral:
\begin{equation}
\rho(x,t) = \rho_0 \int_{-\infty}^{\infty} dq \lim_{\epsilon \to 0} \frac{1}{2\pi \sigma\epsilon} e^{-\Delta^2 / 2\epsilon^2} \int_{-\infty}^{\infty} dv_{\mathrm{th}} \exp\left\{-\left[ v_{\mathrm{th}}^2 \left(\frac{1}{2\sigma^2} + \frac{t^2}{2\epsilon^2}\right) - \frac{t \Delta}{\epsilon^2} v_{\mathrm{th}} \right]\right\}
\end{equation}
The integral over $v_{\mathrm{th}}$ can be evaluated by completing the square on the quantity in curly brackets.  Setting 
\begin{equation}
a^2 \equiv \left(\frac{1}{2\sigma^2} + \frac{t^2}{2\epsilon^2}\right)
\end{equation}
we find that the inner integral becomes
\begin{equation}
I \equiv \int_{-\infty}^{\infty} dv_{\mathrm{th}} \exp \left[ - a^2 \left(v_{\mathrm{th}} -  \frac{t  \Delta}{2\epsilon^2}\right)^2 + \frac{t^2 \Delta^2}{4 \epsilon^4 a^2} \right]
\end{equation}
Using the substitution $u = a(v_{\mathrm{th}} -  t \Delta/2\epsilon^2)$, we can immediately evaluate the integral to be:
\begin{equation}
I(x,t) = \exp\left({\frac{t^2 \Delta^2}{4 \epsilon^4 a^2}}\right) \sqrt{\frac{\pi}{a^2}}
\end{equation}
Replacing this into the equation for the density, we find that
\begin{equation}
\rho(x,t) = \rho_0 \int_{-\infty}^{\infty} dq \lim_{\epsilon \to 0} \frac{1}{\sqrt{4\pi \sigma^2 a^2 \epsilon^2}} \exp \left[-\Delta^2 \left( \frac{1}{2\epsilon^2} - \frac{t^2 }{4 \epsilon^4 a^2} \right) \right]
\end{equation}
Replacing $a^2$ with its definition and simplifying makes the limit easy to take:
\begin{eqnarray}
\rho(x,t) &=& \rho_0 \int_{-\infty}^{\infty} dq \lim_{\epsilon \to 0} \frac{1}{\sqrt{2\pi\left(\epsilon^2 + \sigma^2 t^2\right)}} e^{-\Delta^2/2\left(\epsilon^2 + \sigma^2 t^2\right)}
\end{eqnarray}
and finally we are down to the last integral:
\begin{equation}
\rho(x,t) = \frac{\rho_0 }{\sqrt{2\pi\sigma^2 t^2}} \int_{-\infty}^{\infty} dq\  e^{-\left[ x-q-v_0(q) t \right]^2/2\sigma^2 t^2}
\label{eq:nice-single-integral}
\end{equation}
As $\sigma$ or $t$ approaches zero, this expression approaches $\delta\left[ x-q-v_0(q) t \right]$, and we recover the density in the case of zero dispersion.  

To perform the integral over $q$ (that is, to sum over streams) we must choose a form for $v_0(q)$.  The simplest form that creates a localized caustic surface is $v_0(q) = -\alpha q^2$:
\begin{equation}
\label{eq:single-integral-alpha}
\rho(x,t) = \frac{\rho_0 }{\sqrt{2\pi\sigma^2 t^2}} \int_{-\infty}^{\infty} dq\  e^{-\left[ x-q + \alpha q^2 t \right]^2/2\sigma^2 t^2}
\end{equation}
For this form of $v_0$ we can evaluate the integral 
\begin{equation}
I(x,t)= \int_{-\infty}^{\infty} dq\  e^{-\left[ x-q + \alpha q^2 t \right]^2/2\sigma^2 t^2}
\end{equation}
in terms of Bessel functions if we make a slightly suspect change of variables.  First we complete the square in $q$ inside the brackets:
\begin{equation}
I(x,t) = \int_{-\infty}^{\infty} dq\  \exp\left\{-\frac{\alpha^2}{2\sigma^2}\left[ \frac{x}{\alpha t}-\frac{1}{4 \alpha^2 t^2}+\left(q - \frac{1}{2 \alpha t}\right)^2\right]^2\right\}
\end{equation}
Then we make the substitution
\begin{equation}
u \equiv q - \frac{1}{2 \alpha t}
\end{equation}
This change of variables is certainly suspect as $t\to 0$, but we make it anyway, hoping that since we know the answer at $t=0$ we can check the result and verify that it gives $\rho_0$ everywhere.  With this substitution
\begin{equation}
I(x,t) = \int_{-\infty}^{\infty} du\  \exp\left[-\frac{\alpha^2}{2\sigma^2}\left( \frac{x}{\alpha t}-\frac{1}{4 \alpha^2 t^2}+u^2\right)^2\right]
\end{equation}
We now identify the caustic position $x_c$.  In the case of perfectly cold initial conditions, the location of the caustic is determined by setting the denominator of Equation \eqref{eq:rho-x-t} to zero and solving for $x_c$, with the help of Equation \eqref{eq:eul-as-lag}.  We use this same quantity to describe the ``position" of the caustic in the warm case, since although the caustic now has a width, the peak density will still occur near $x_c$.  For our choice of $v_0(q)$ we find that this corresponds to
\begin{equation}
\label{eq:xcDefined}
\frac{x_c}{\alpha t}-\frac{1}{4 \alpha^2 t^2} = 0
\end{equation}
so that $x_c = 1/(4\alpha t)$ and
\begin{equation}
\frac{x}{\alpha t}-\frac{1}{4 \alpha^2 t^2} = \frac{1}{\alpha t}(x-x_c) \equiv \bar{x}
\end{equation}
for notational simplicity when performing the integral.  So now we have a factored quartic in the exponent:
\begin{equation}
I(x,t) = \int_{-\infty}^{\infty} du\  \exp\left[-\frac{\alpha^2}{2\sigma^2}\left(\bar{x} +u^2\right)^2\right]
\label{eq:sum-over-streams-quartic}
\end{equation}
Expanding this expression and pulling out a constant term leaves us with an integral that can be evaluated in terms of Bessel functions:
\begin{equation}
I(\bar{x},t) = e^{-\alpha^2\bar{x}^2/2\sigma^2 } \int_{-\infty}^{\infty} du\  \exp\left(-\frac{\alpha^2}{2\sigma^2}u^4-\frac{\bar{x}\alpha^2}{\sigma^2}u^2\right)
\label{eq:sum-over-streams-final-form}
\end{equation}
Performing the integral, replacing $\bar{x}$ with its definition, and reintroducing the prefactor, we find that 
\begin{equation}
\label{eq:rhoasy-disp}
\rho(x,t) = \frac{\rho_0}{\sqrt{2\pi\sigma^2}} \sqrt{\frac{\left|x-x_c\right|}{ \alpha t^3}}\ e^{-\left(x-x_c\right)^2/4\sigma^2t^2}\ \mathcal{B}\left[\frac{(x-x_c)^2}{4 \sigma^2 t^2}\right]
\end{equation}
with
\begin{equation}
\mathcal{B}(u) = 
\left\{
\begin{array}{cc}
\frac{\pi}{2} \left[ \mathcal{I}_{-1/4}(u) +  \mathcal{I}_{1/4}(u) \right]  & x \le x_c  \\
\frac{\pi}{2} \left[ \mathcal{I}_{-1/4}(u) -  \mathcal{I}_{1/4}(u) \right]  &  x > x_c 
\end{array}
\right.
\end{equation}
where $ \mathcal{I}$ is a modified Bessel function of the first kind.  Taking the limit $t\to 0$ recovers $\rho_0$ everywhere, and validates our change of variables.  This form is the same as that obtained by \citet{2006MNRAS.366.1217M} for the density profile of a caustic with Gaussian velocity dispersion, with the substitutions $\alpha_k \sigma_v \to \sigma t$, $\Delta x \to x-x_c$, and $A_k = \rho_0 / \sqrt{\alpha t}$.  That is, it describes the same shape as a caustic formed by secondary self-similar infall from an initial population with Gaussian velocity dispersion, but whose normalization varies as $1/\sqrt{t}$ and whose position varies linearly with time.  In fact, this density profile is, with a few slight changes, universal for any initial population with a Gaussian velocity dispersion, regardless of the equation of motion of those particles.  This is because the shape of the phase space distribution of particles near the caustic can always be approximated by a tilted quadratic with curvature $\kappa$.  For the equation of motion used to derive Equation \eqref{eq:rhoasy-disp}, $\kappa = \alpha t^3$; for motion in a gravitational potential, $\kappa$ depends on the radial gravitational force at the caustic:  
\begin{equation}
\label{eq:kappaFromGravity}
\kappa = - \half \left. \frac{d^2r}{dv_r^2}\right|_{r_c} = -\frac{1}{2 g(r_c)}
\end{equation}
where $g(r)$ is the gravitational field or radial derivative of the potential, $\partial V/\partial r$.  

Inspection of Equation \eqref{eq:rhoasy-disp} shows that the curvature $\kappa$ influences the height of the caustic relative to the initial density, while the product $\sigma t$ determines its sharpness.  In a more general case, the spreading of the particles in phase space is not linear in time and the initial velocity dispersion, and $\sigma t$ can be replaced everywhere in the density profile by a generic parameter representing the width of the caustic, $\delta r$.  The quantity $\rho_0/\sigma \equiv f_0$ in the normalization of the caustic indicates that the physical density at time $t$ depends on the initial phase space density, with the caveat that in cases where a caustic forms in one dimension but motion occurs in more than one dimension, $\sigma$ remains the one-dimensional velocity dispersion while $\rho_0$ must be determined via mass conservation, since the density profile does not account for the behavior of the population in the neglected dimensions or for the Jacobian associated with the volume element in non-Cartesian metrics (for example, spherical coordinates).

The caustic width $\delta r$ can be estimated using energy conservation.  Although each particle turns around at apocenter where $v_r = 0$, the caustic itself can have a net positive velocity because its location at a given time depends on the relative periods of particles on neighboring orbits.  Thus the energy at the caustic surface, assuming spherical symmetry, is:
\begin{equation}
E = V(r) + \half v_r^2
\end{equation}
For two particles on neighboring orbits, their energy difference $\Delta E$ is found by subtracting their energies, so
\begin{equation}
\Delta E = \half ( v_2^2 - v_1^2 ) +  V(r_2) - V(r_1) 
\end{equation}
Particles at the very face of the caustic are close in $r$ and far from $r=0$, so we can write $r_2 = r_1 + \Delta r$, taking $\Delta r/r_1$ to be much less than 1.  We cannot also write $v_2 = v_1 + \Delta v$ and $\Delta v/v_1 \ll 1$,  since the average velocity of particles in the caustic may not be much larger than zero.  Using the condition on $r_1$, the energy difference can be rewritten in terms of the local gravity (defined earlier):
\begin{equation}
\Delta E = \half ( v_2^2 - v_1^2 ) - g(r_1) \Delta r 
\end{equation}
In reality many particles will comprise the caustic surface; one can use this expression by taking its variance, realizing that $r$ and $v$ are correlated and assuming $g(r)/ \Delta r \gg dg/dr$:  
\begin{equation}
\label{eq:energyConsvVariances}
\sigma_E^2 = \half \sigma_{v^2}^2 - g(r) \textrm{cov}(r, v^2) + g(r)^2 \sigma_r^2
\end{equation}
The size of the covariant term is related to the amount of curvature over the region used to calculate the variances.  It can be minimized by taking a region around the caustic small enough that the curvature is smaller than the radial thickness, or $\delta v \leq \delta r/ \sqrt{\kappa}$.  Then the covariance term can be ignored and the equation solved to give an estimate of the width $\sigma_r$:
\begin{equation}
\label{eq:EstimatingDeltaR}
\sigma_r = \frac{\sqrt{\sigma_E^2 - \half \sigma_{v^2}^2}}{g(r_c)}
\end{equation}

\begin{figure}[htbp]
\begin{center}
\plotone{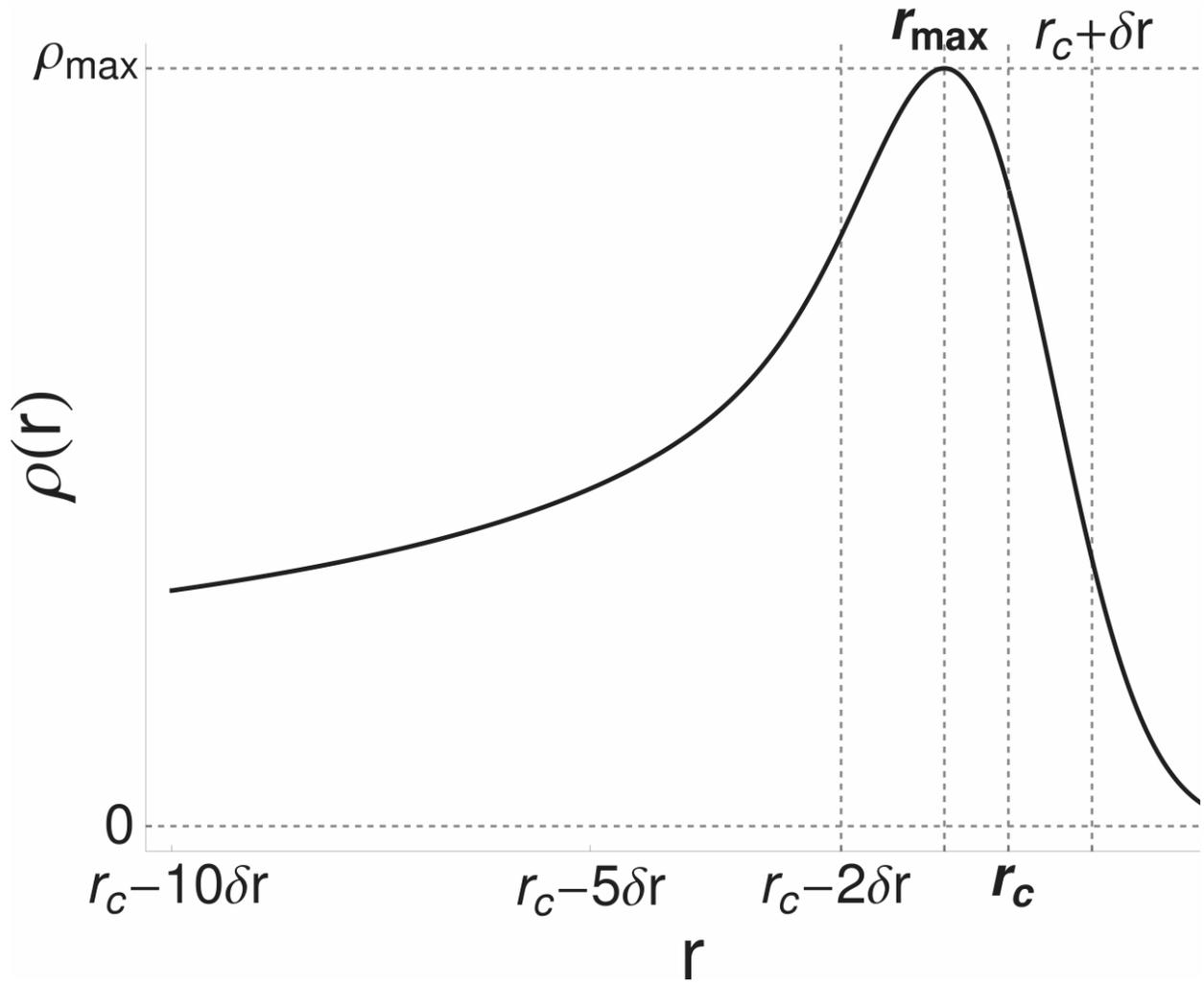}
\caption{Universal caustic form for radial infall from a population with initial Gaussian velocity dispersion.  Formulae for the peak radius and density are given in the text. \label{fig:exampleCaustic}}
\end{center}
\end{figure}

With these physically motivated generalizations, the density profile
\begin{equation}
\label{eq:generalDP}
\rho(x,t) = \frac{f_0}{\sqrt{2\pi}} \sqrt{\frac{\left|x-x_c\right|}{ \kappa}}\ e^{-\left(x-x_c\right)^2/4\delta r^2}\ \mathcal{B}\left[\frac{(x-x_c)^2}{4 \delta r^2}\right]
\end{equation}
will universally fit the projected radial density profile of any shell near its peak.  Using this relation, we can find the location of the peak by solving $d\rho/dr = 0$.  The equation reduces to a linear combination of various Bessel functions, which can be solved numerically to find $r_{\mathrm{max}}$.  Here we give the result to 10 decimal places, but more are easily available if necessary simply by increasing the number of iterations in the root finder.
\begin{equation}
r_{\mathrm{max}} = r_c - 0.7649508674\delta r 
\end{equation}
Substitution into the density profile gives the maximum density in terms of the physical parameters.  Within a few percent the numerical coefficient is unity:
\begin{equation}
\rho_{\mathrm{max}} = 1.0211365847 f_0 \sqrt{\frac{\delta r}{\kappa}}
\end{equation}
The peak density occurs about one characteristic width behind the nominal caustic surface $r_c$ (Figure \ref{fig:exampleCaustic}).

\subsection{Rate}
Armed with an analytic expression for the density, we can now calculate the rate $\Gamma$ for comparison with estimates from random realizations.  We must integrate the density with finite limits on $q$ and $x$, since we will use the density estimator on a finite-sized sample of the distribution.  Since the caustic is only in the $x$ dimension,
\begin{equation}
\Gamma(t) =  4 L_y L_z \int_{x_{-}}^{x_{+}} \rho^2(x, t) dx
\end{equation}
with $x_{\pm}$ the lower and upper edges of the box in the $x$ direction, and $\pm L_y$ and $\pm L_z$ the box dimensions in the $y$ and $z$ directions, respectively.  Furthermore, we are now considering only particles that originated in some range $[q_{-}, q_{+}]$, so we must also recalculate the density function integral, Equation \eqref{eq:sum-over-streams-quartic}, over a finite range:
\begin{equation}
\rho(\bar{x},t) = \frac{\rho_0 }{\sqrt{2\pi\sigma^2 t^2}} \int_{u_{-}}^{u_{+}} du\  \exp\left[-\frac{\alpha^2}{2\sigma^2}\left(\bar{x} +u^2\right)^2\right]
\end{equation}
where 
\begin{equation}
u_{\pm} \equiv q_{\pm} - \frac{1}{2\alpha t}
\end{equation}

Technically, we should use this expression for $\rho^2$ in the rate integral, so that
\begin{equation}
\Gamma(t) = \frac{2 L_y L_z \alpha \rho_0^2}{\pi \sigma^2 t} \int_{\bar{x}_{-}}^{\bar{x}_{+}}  \left[  \int_{u_{-}}^{u_{+}} du\  \exp\left[-\frac{\alpha^2}{2\sigma^2}\left(\bar{x} +u^2\right)^2\right] \right]^2 d\bar{x},
\label{eqn:rate-numerical-version}
\end{equation}
with
\begin{equation}
\bar{x}_{\pm} \equiv \frac{x_{\pm} - x_c}{\alpha t}
\end{equation}
the endpoints of the rate integral about the caustic.

Both these integrals must be evaluated numerically, so we would like to borrow our analytic solution, Equation 
\eqref{eq:rhoasy-disp}, instead if possible.  We can see that this is sufficient by examining the form of Equation \eqref{eq:rhoasy-disp}.  The exponential and Bessel function terms are functions of the quantity
\begin{equation}
\frac{(x-x_c)}{2\sigma t} \equiv \bar{x}_d
\end{equation}
which compares the distance from the caustic to the characteristic length scale for dispersion, $\sigma t$.  For distances much larger than $\sigma t$, the exponential piece approaches zero.  The modified Bessel function $\mathcal{K}_{1/4} \equiv \frac{\pi}{\sqrt{2}} \left[ \mathcal{I}_{-1/4}(u) -  \mathcal{I}_{1/4}(u) \right]   $ also has an exponential cutoff for $(x-x_c)\gg \sigma t$, or $u \to \infty$, as seen in its asymptotic series:
\begin{equation}
K_{1/4}(u) = \sqrt{\frac{\pi}{2}} e^{-u^2} \left[ \frac{1}{u} - \frac{3}{32} \frac{1}{u^3} + \mathcal{O}\left(\frac{1}{u^5}\right)\right]
\end{equation}
The sum of modified Bessel functions $\mathcal{I}_{\pm 1/4}$ has an exponentially increasing term, shown in its asymptotic series:
\begin{equation}
  \mathcal{I}_{-1/4}(u) + \mathcal{I}_{1/4}(u) = e^{u^2} \sqrt{\frac{2}{\pi }} \left[\frac{1}{u}+\frac{3 }{32}\frac{1}{u^3}+\mathcal{O}\left(\frac{1}{u^5}\right)\right]+e^{-u^2} \frac{i}{\sqrt{\pi }}\left[\frac{1}{u}-\frac{3 }{32 }\frac{1}{u^3}+\mathcal{O}\left(\frac{1}{u^5}\right)\right]
\end{equation}
that is cancelled by the exponential piece of the equation, leaving power-law convergence to zero for $(x-x_c)\gg \sigma t$ in the asymptotic series:  
\begin{equation}
e^{-u^2} \left(  \mathcal{I}_{-1/4}(u) + \mathcal{I}_{1/4}(u) \right) = \sqrt{\frac{2}{\pi }} \left[\frac{1}{u}+\frac{3 }{32}\frac{1}{u^3}+\mathcal{O}\left(\frac{1}{u^5}\right)\right]+e^{-2u^2} \cdot  \mathcal{O}\left(\frac{1}{u}\right)
\end{equation}
From examining these limits, we find that for $x>x_c$ the density is completely negligible for distances more than a few times $\sigma t$ from the caustic, and for $x<x_c$ it becomes negligible like a power law.  Thus it is unsurprising that substituting the infinite-range expression for the finite-range one gives the right answer. 

Thus, as long as $x-x_c \gg \sigma t$ at the endpoints, we can safely write
\begin{equation}
\Gamma(t) = 4 L_y L_z \frac{\rho_0^2}{2 \pi \sigma^2 t^2} \frac{1}{t} \int_{x_{-}}^{x_{+}} \left|x-x_c\right| e^{-2(x-x_c)^2/4\sigma^2t^2} \mathcal{B}^2\left[\frac{(x-x_c)^2}{4\sigma^2t^2}\right] dx
\end{equation}
Practically speaking, the approximation is sufficient for our needs as long as $\bar{x} \gtrsim 500$ at the endpoints.  If this condition isn't satisfied we must do the two numerical integrals in Equation \eqref{eqn:rate-numerical-version}.

We change variables to $\bar{x}_d$.  Inserting the form for $\mathcal{B}$ breaks the integral up into two parts:
\begin{equation}
\Gamma(t) =  \frac{4 L_y L_z \rho_0^2}{ t} \left\{ \int_{\bar{x}_{d_-}}^0 |\bar{x}_d| e^{-2\bar{x}_d^2} \left[ \mathcal{I}_{-1/4}\left(\bar{x}_d^2 \right) + \mathcal{I}_{1/4} \left(\bar{x}_d^2 \right) \right]^2 d\bar{x}_d  + \int^{\bar{x}_{d_+}}_0 |\bar{x}_d| e^{-2\bar{x}_d^2} \left[ \mathcal{I}_{-1/4}\left(\bar{x}_d^2 \right) - \mathcal{I}_{1/4} \left(\bar{x}_d^2 \right) \right]^2 d\bar{x}_d \right\}
\end{equation}
where we have assumed that we are integrating across the caustic so that the new endpoints,
\begin{equation}
\bar{x}_{d_{\pm}} \equiv \frac{x_{\pm} - x_c}{2 \sigma t},
\end{equation}
are such that $\bar{x}_{d_{-}} <0$ and $\bar{x}_{d_{+}} >0$.  This integral must be performed numerically, but it's a single integral rather than a two-step process.

\end{document}